\documentclass{ar-astro2e}
\usepackage{ARAstroBib}
\usepackage{amsfonts}
\usepackage{amssymb}
\usepackage{booktabs,caption,fixltx2e}
\usepackage[flushleft]{threeparttable}
\usepackage{amsmath}
\def\HI{\hbox{H\,$\scriptstyle\rm I$}}

\def\CIV{\hbox{C\,$\scriptstyle\rm IV$}}
\def\CIII{\hbox{C\,$\scriptstyle\rm III$}}
\def\CII{\hbox{C\,$\scriptstyle\rm II$}}

\def\OVI{\hbox{O\,$\scriptstyle\rm VI$}}

\def\SiIII{\hbox{Si\,$\scriptstyle\rm III$}}
\def\SiIV{\hbox{Si\,$\scriptstyle\rm IV$}}
\def\Lya{Ly$\alpha$}

\def\etal{et~al.\ }
\def\msun{\,{\rm M_\odot}}
\def\zsun{\,{Z_\odot}}
\def\mdens{\,{\rm M_\odot\,Mpc^{-3}}}

\def\ndotunits{\,{\rm s^{-1}\,Mpc^{-3}}}
\def\sfrd{\,{\rm M_\odot\,year^{-1}\,Mpc^{-3}}}
\def\lum{\,{\rm erg\,s^{-1}\,Hz^{-1}}}

\def\sfr{\,{\rm M_\odot\,year^{-1}}}

\def\spose#1{\hbox to 0pt{#1\hss}}
\def\lta{\mathrel{\spose{\lower 3pt\hbox{$\mathchar"218$}}
     \raise 2.0pt\hbox{$\mathchar"13C$}}}
\def\gta{\mathrel{\spose{\lower 3pt\hbox{$\mathchar"218$}}
     \raise 2.0pt\hbox{$\mathchar"13E$}}}

\def\pasp{{\it Pub. Astron. Soc. Pac. \,}}
\def\apj{{\it Ap. J. \,}}
\def\apjl{{\it Ap. J. Let. \,}}
\def\aj{{\it Astron. J. \,}}

\def\apjs{{\it Ap. J. Suppl. \,}}
\def\mnras{{\it MNRAS \,}}
\def\aa{{\it Astron. Astrophys. \,}}
\def\aap{{Astron. Astrophys. \,}}

\def\araa{{\it Annu. Rev. Astron. Astrophy. \,}}
\def\nat{{\it Nature \,}}

\begin{document}

\input epsf.def   
\input psfig.sty

\jname{Annu. Rev. Astron. Astrophys.}
\jyear{2013}
\jvol{}
\ARinfo{1056-8700/97/0610-00}

\title{Cosmic Star-Formation History}

\markboth{P.\ Madau \& M.\ Dickinson}{Cosmic Star-Formation History}

\author{Piero Madau
\affiliation{Department of Astronomy and Astrophysics, University of California, Santa Cruz, \\ California 95064; email: pmadau@ucolick.org}
Mark Dickinson
\affiliation{National Optical Astronomy Observatory, Tucson, Arizona 85719; email: \\ med@noao.edu}
}

\begin{abstract}
Over the past two decades, an avalanche of data from multiwavelength imaging and spectroscopic surveys has revolutionized our view of galaxy formation and evolution. 
Here we review the range of complementary techniques and theoretical tools that allow astronomers to map the cosmic history of star formation, heavy element production,
and reionization of the Universe from the cosmic ``dark ages'' to the present epoch. A consistent picture is emerging, whereby the star-formation rate density peaked 
approximately 3.5 Gyr after the Big Bang, at $z \approx 1.9$, and declined exponentially at later times, with an e-folding timescale of 3.9 Gyr. Half of the stellar 
mass observed today was formed before a redshift $z = 1.3$. About 25\% formed before the peak of the cosmic star-formation rate density, and another 25\% formed after 
$z = 0.7$. Less than $\sim 1$\% of today's stars formed during the epoch of reionization. Under the assumption of a universal initial mass function, the global stellar 
mass density inferred at any epoch matches reasonably well the time integral of all the preceding star-formation activity. The comoving rates of star formation and 
central black hole accretion follow a similar rise and fall, offering evidence for co-evolution of black holes and their host galaxies. The rise of the mean metallicity 
of the Universe to about 0.001 solar by $z = 6$, one Gyr after the Big Bang, appears to have been accompanied by the production of fewer than ten hydrogen Lyman-continuum 
photons per baryon, a rather tight budget for cosmological reionization.
\end{abstract}


\maketitle

\section{INTRODUCTION}
\label{sec:introduction}

The origin and evolution of galaxies are among the most intriguing and complex 
chapters in the formation of cosmic structure, and observations in this field have 
accumulated at an astonishing pace. Multiwavelength 
imaging surveys with the {\it Hubble} (HST) and {\it Spitzer} space telescopes and ground-based facilities, 
together with spectroscopic follow-up with 8-m-class telescopes, have led to the 
discovery of galaxies with confirmed redshifts as large as $z = 7.5$ (Finkelstein \etal 2013), as well as 
compelling photometric candidates as far back as $z \approx 11$ (Coe \etal 2013)
when the Universe was only 3\% of its current age.
Following the seminal work of Steidel \etal (1995), color-selection criteria that are 
sensitive to the presence of intergalactic \HI\ absorption features in the spectral energy distribution (SED)
of distant sources have been used to build increasingly large samples of star-forming galaxies at $2.5\lta z \lta 9$ 
(e.g., Madau \etal 1996, Steidel \etal 2003, Giavalisco \etal 2004a, Bouwens \etal 2011b). 
Infrared (IR)-optical color selection criteria efficiently isolate both actively 
star-forming and passively evolving galaxies at $z \approx 2$ (Franx \etal 2003, Daddi \etal 2004).
Photometric redshifts have become an unavoidable tool for placing faint galaxies onto 
a cosmic timeline.  {\it Spitzer,} {\it Herschel}, and submillimeter telescopes 
have revealed that dusty galaxies with star-formation rates (SFRs) of order 
100$\,\msun$~year$^{-1}$ or more were abundant when the Universe was only 2--3 Gyr old 
(Barger \etal 1998, Daddi \etal 2005, Gruppioni \etal 2013). Deep near-infrared (NIR) observations 
are now commonly used to select galaxies on the basis of their optical rest-frame light and to 
chart the evolution of the global stellar mass density (SMD) at $0<z<3$ (Dickinson \etal 2003).
The {\it Galaxy Evolution Explorer} (GALEX) satellite has quantified the ultraviolet galaxy luminosity 
function (LF) of galaxies in the local Universe and its evolution at $z\lta 1$.   
Ground-based observations and, subsequently, UV and IR data from {GALEX} and {Spitzer}
have confirmed that star-formation activity was significantly 
higher in the past (Lilly \etal 1996, Schiminovich \etal 2005, Le~Floc'h \etal 2005).
In the local Universe, various galaxy properties (colors, surface mass densities, 
and concentrations) have been observed by the {Sloan Digital Sky Survey} (SDSS) 
to be ``bimodal'' around a transitional stellar mass of $3\times 10^{10}\,\msun$ 
(Kauffmann \etal 2003), showing a clear division between faint, blue, active galaxies 
and bright, red, passive systems. The number and total stellar mass of blue galaxies 
appear to have remained nearly constant since $z\sim 1$, whereas those of red galaxies 
(around $L^\ast$) have been rising (Faber \etal 2007).  At redshifts $0 < z < 2$ at least, 
and perhaps earlier, most star-forming galaxies are observed to obey a relatively tight 
``main-sequence'' correlation between their SFRs and stellar masses 
(Brinchmann \etal 2004, Noeske \etal 2007, Elbaz \etal 2007, Daddi \etal 2007).
A minority of starburst galaxies have elevated SFRs above this
main sequence as well as a growing population of quiescent galaxies that fall 
below it.

With the avalanche of new data, galaxy taxonomy has been enriched by the addition of new acronyms such as
LBGs, LAEs, EROs, BzKs, DRGs, DOGs, LIRGs, ULIRGs, and SMGs. Making sense of it all and fitting it together into 
a coherent picture remains one of astronomy's great challenges, in part because of the observational difficulty 
of tracking continuously transforming galaxy sub-populations across cosmic time and in part    
because theory provides only a partial interpretative framework. The key 
idea of standard cosmological scenarios is that primordial density 
fluctuations grow by gravitational instability driven by cold, collisionless dark
matter, leading to a ``bottom-up'' $\Lambda$CDM (cold dark matter) scenario of structure formation (Peebles 1982). Galaxies 
form  hierarchically: Low-mass objects (``halos'') collapse earlier and merge to form increasingly larger systems over time -- 
from ultra-faint dwarfs to clusters of galaxies (Blumenthal \etal 1984). Ordinary matter in the Universe follows the dynamics 
dictated by the dark matter until radiative, hydrodynamic, and star-formation processes take over (White \& Rees 1978). The ``dark side'' of galaxy
formation can be modeled with high accuracy and has been explored in detail through $N$-body 
numerical simulations of increasing resolution and size (e.g., Davis \etal 1985, Dubinski \& Carlberg 1991, 
Moore \etal 1999, Springel \etal 2005, 2008, Diemand \etal 2008, Stadel \etal 2009, Klypin \etal 2011). However, the same does not hold for the baryons. 
Several complex processes are still poorly understood, for example, baryonic dissipation inside evolving CDM halos, the transformation of cold gas into stars, 
the formation of disks and spheroids, the chemical enrichment of gaseous material on galactic and intergalactic scales, 
and the role played by ``feedback'' [the effect of the energy input from stars, supernovae (SNe), and 
massive black holes on their environment] in regulating star formation and generating galactic outflows. 
The purely phenomenological treatment of complex physical processes that is at the core of semi-analytic schemes of galaxy formation
(e.g., White \& Frenk 1991, Kauffmann \etal 1993, Somerville \& Primack 1999, Cole \etal 2000) and -- at a much 
higher level of realism -- the ``subgrid modeling'' of star formation and stellar feedback that must be implemented even in the more accurate 
cosmological hydrodynamic simulations (e.g., Katz \etal 1996, Yepes \etal 1997, Navarro \& 
Steinmetz 2000, Springel \& Hernquist 2003, Keres al. 2005, Ocvirk \etal 2008, Governato \etal 2010, Guedes \etal 2011, Hopkins \etal 2012, 
Kuhlen \etal 2012, Zemp \etal 2012, Agertz \etal 2013)
are sensitive to poorly determined parameters and suffer from various degeneracies, a weakness that has 
traditionally prevented robust predictions to be made in advance of specific observations. 

Ideally, an in-depth understanding of galaxy evolution would encompass the full sequence of events 
that led from the formation of the first stars after the end of the cosmic dark ages to 
the present-day diversity of forms, sizes, masses, colors, luminosities, metallicities, and 
clustering properties of galaxies. This is a daunting task, and it is perhaps not surprising 
that an alternative way to look at and interpret the bewildering variety of galaxy data has 
become very popular in the past two decades. The method focuses on the emission
properties of the galaxy population as a whole, traces the evolution with cosmic time 
of the galaxy luminosity density from the far-UV (FUV) to the far-infrared (FIR), and offers the
prospect of an empirical determination of the global history of star formation and heavy element production 
of the Universe, independently of the complex evolutionary phases of individual 
galaxy subpopulations. The modern version of this technique relies on some basic properties of 
stellar populations and dusty starburst galaxies: 

\begin{enumerate}

\item The UV-continuum emission in all but the oldest galaxies is dominated by
short-lived massive stars. Therefore, for a given stellar initial mass function (IMF) 
and dust content, it is a direct measure of the instantaneous star-formation rate density (SFRD).

\item The rest-frame NIR light is dominated by near-solar-mass evolved stars that
make up the bulk of a galaxy's stellar mass and can then be used as a tracer
of the total SMD. 

\item Interstellar dust preferentially absorbs UV light and re-radiates it in the
thermal IR, so that the FIR emission of dusty starburst galaxies can be a sensitive tracer of young stellar populations  
and the SFRD. 

\end{enumerate}

By modeling the emission history of all stars in the Universe at UV, optical, and IR wavelengths from the
present epoch to $z\approx 8$ and beyond, one can then shed light on some key questions in
galaxy formation and evolution studies: Is there a characteristic cosmic epoch of the formation
of stars and heavy elements in galaxies?  What fraction of the luminous baryons
observed today were already locked into galaxies at early times? Are the data consistent
with a universal IMF? Do galaxies reionize the Universe at a redshift greater than 6? 
Can we account for all the metals produced by the global star-formation activity from the Big Bang to the present? 
How does the cosmic history of star formation compare with the history of mass accretion onto massive black holes as traced by luminous quasars?

This review focuses on the range of observations, methods, and theoretical tools that are allowing astronomers to map 
the rate of transformation of gas into stars in the Universe, from the cosmic dark ages to the present epoch.
Given the limited space available, it is impossible to provide a thorough survey of such a huge community effort without leaving out significant 
contributions or whole subfields. We have therefore tried to refer only briefly to earlier findings, and present recent observations in more detail,
limiting the number of studies cited and highlighting key research areas. In doing so, we hope to provide a manageable overview of how the field 
has developed and matured in line with new technological advances and theoretical insights, and of the questions with which astronomers still struggle nowadays.

The remainder of this review is organized as follows. The equations of cosmic chemical evolution that govern the consumption of gas into stars and 
the formation and dispersal of heavy elements in the Universe as a whole are given in Section \ref{sec:chemev}. We turn to the topic of measuring mass from light, and
draw attention to areas of uncertainty in Section \ref{sec:massfromlight}. Large surveys, key data sets and the analyses thereof are highlighted in Section 
\ref{sec:surveys}. 
An up-to-date determination of the star-formation history (SFH) of the Universe is provided and its main implications are discussed in Serction \ref{sec:obs_to_para}. 
Finally, we summarize our conclusions in Section \ref{sec:conclusion}.  Unless otherwise stated, all results presented here
will assume a ``cosmic concordance cosmology'' with parameters $(\Omega_M, \Omega_\Lambda, \Omega_b, h)=(0.3, 0.7,$ $0.045, 0.7)$.

\section{THE EQUATIONS OF COSMIC CHEMICAL EVOLUTION}
\label{sec:chemev}

To pursue and cast light into a quantitative form of the idea of a history of cosmic star formation and metal enrichment 
-- not of any particular type of galaxy but of the Universe as a whole -- 
it is useful to start by generalizing the standard equations of galaxy evolution (Tinsley 1980) over all 
galaxies and intergalactic gas in the Universe. In a representative cosmological comoving volume with 
density $\rho_\ast$ in long-lived stars and stellar remnants (white dwarfs, neutron stars, black 
holes) and gas density $\rho_g$ and in which new stars are formed at the rate $\psi$, 
the equations of cosmic chemical evolution can be written as   
\begin{eqnarray}
{{\rm d}\rho_\ast \over {\rm d}t} & = & (1-R)\psi \nonumber \\
{{\rm d}\rho_g \over {\rm d}t} & = & - {{\rm d}\rho_\ast\over {\rm d}t}  \label{chem_ev} \\
\rho_g {{\rm d}Z \over {\rm d}t} & = & y(1-R)\psi \nonumber .
\end{eqnarray}
Here, $Z$ is the metallicity in the gas and newly born stars, $R$ is the ``return fraction'' or the 
mass fraction of each generation of stars that is put back into the interstellar medium (ISM) and 
intergalactic medium (IGM), and $y$ is the net metal yield or the mass of new heavy elements 
created and ejected into the ISM/IGM by each generation of stars per unit mass locked into
stars. The above equations govern the formation, destruction, and distribution of heavy elements as they 
cycle through stars and are ultimately dispersed into the ISM/IGM. By treating all galaxies as a single  
stellar system and all baryons in the ISM/IGM as its gas reservoir, their solution enables the 
mean trends of galaxy populations to be calculated with the fewest number of free parameters. 
The equations state that, for every new mass element locked forever into long-lived stars
and stellar remnants, $\Delta \rho_\ast$, the metallicity of the ISM/IGM increases as 
$\Delta Z=y\Delta \rho_\ast/\rho_g$, whereas the mass of heavy elements in the ISM/IGM changes 
as $\Delta (Z\rho_g)=(y-Z)\Delta \rho_\ast$. The latter expression is a consequence of metals being  
released into the gas from mass loss during post-main-sequence stellar evolution as well as 
being removed from the ISM/IGM when new stars condense out. However, compared with the source term, the metal sink 
term can be neglected at early epochs when $Z\ll y$. 

At redshift $z$, Equation \ref{chem_ev} can be integrated to give the following:
\begin{enumerate}

\item The total mass density of 
long-lived stars and stellar remnants accumulated from earlier episodes of star formation,
\begin{equation}
\rho_\ast(z)=(1-R)\int_0^{t(z)}\psi {\rm d}t=(1-R)\int_z^\infty \psi {{\rm d}z'\over H(z')(1+z')}, 
\label{eq:rhostar}
\end{equation}
where $H(z')=H_0[\Omega_M(1+z')^3+\Omega_\Lambda]^{1/2}$ is the Hubble parameter in a flat cosmology.

\item The total mass density of gas, 
\begin{equation}
\rho_g(z)=\rho_{g,\infty}-\rho_\ast(z),
\label{eq:rhogas}
\end{equation}
where $\rho_{g,\infty}$ is the comoving density of gas at some suitable high redshift where there are no stars or heavy elements. 

\item The total mass density of heavy elements in the ISM/IGM, 
\begin{eqnarray}
Z(z)\rho_g(z) & = & y(1-R)\int_0^{t(z)}\psi {\rm d}t -(1-R)\int_0^{t(z)}Z\psi {\rm d}t \nonumber \\
& \equiv & [y- \langle Z_\ast(z)\rangle] \rho_\ast(z) 
\label{eq:rhoZ}
\end{eqnarray}
where the term $\langle Z_\ast \rangle \rho_\ast$ is the total metal content of stars and remnants at that redshift. Note that the instantaneous total metal ejection 
rate, $E_Z$, is the sum of a recycle term and a creation term (Maeder 1992),
\begin{equation}
E_Z=ZR\psi +y(1-R)\psi, 
\end{equation}
where the first term is the amount of heavy elements initially lost from the ISM when stars formed that are now being re-released, and the 
second represents the new metals synthesized by stars and released during mass loss.

\end{enumerate}

For a given universal stellar IMF, the quantities $R$ and $y$ can be derived using the following formulas:
\begin{eqnarray}
R & = & \int_{m_0}^{m_u} (m-w_m)\phi(m){\rm d}m \\
y (1-R) & = & \int_{m_0}^{m_u} my_m \phi(m){\rm d}m ,
\label{eq:yR}
\end{eqnarray}
where $m$ is mass of a star, $w_m$ is its remnant mass, $\phi(m)$ is the IMF 
[normalized so that $\int_{m_l}^{m_u} m\phi(m){\rm d}m=1$], and $y_m$ is the stellar yield, i.e., the 
fraction of mass $m$ that is converted to metals and ejected. In this review, the term 
``yield'' generally indicates the net yield $y$ of a stellar population as defined in Equation \ref{eq:yR}; instead we 
explicitly speak of ``stellar yields'' to indicate the $y_m$ resulting from nucleosynthesis calculations. 
The above equations have been written under the simplifying assumptions of ``instantaneous recycling''
(where the release and mixing of the products of nucleosynthesis by all stars more massive than $m_0$ occur on 
a timescale that is much shorter than the Hubble time, whereas stars with $m<m_0$ live forever), ``one zone'' 
(where the heavy elements are well mixed at all times within the volume under 
consideration), ``closed box'' (flows of gas in and out the chosen volume are negligible), and 
``constant IMF and metal yield.'' 

Recall now that the main-sequence timescale is shorter than 0.6 Gyr (the age of the Universe at $z=8.5$) for stars
more massive than $2.5\,\msun$, whereas stars less massive than $0.9\,\msun$ never evolve off the main sequence. [Stellar evolutionary
models by Schaller \etal (1992) show that, for $m<7\,\msun$, solar-metallicity stars have longer lifetimes than their metal-poor counterparts, 
whereas the opposite is true for $m>9\,\msun$.] So over the redshift range of interest here, the instantaneous recycling approximation may break down in the    
limited mass range $0.9<m<2.5\,\msun$. For illustrative purposes, in the following we adopt the initial-final mass values for white 
dwarfs tabulated by Weidemann (2000), which can be fit to few-percent accuracy over the interval $1\,\msun<m<7\,\msun$   
as $w_m=0.444+0.084m$. We also assume that all stars with $8\,\msun<m<m_{\rm BH}=40\,\msun$ return all but a $w_m=1.4\,\msun$ remnant,
and stars above $m_{\rm BH}$ collapse to black holes without ejecting material into space, i.e., $w_m=m$. Few stars form with masses
above $40\,\msun$, so the impact of the latter simplifying assumption on chemical evolution is minimal. Thus, taking $m_0=1\,\msun$ as the 
dividing stellar mass for instantaneous recycling and a Salpeter (1955) IMF 
with $\phi(m)\propto m^{-2.35}$ in the range $m_l=0.1\,\msun<m<m_u=100\,\msun$, one derives a return fraction of $R=0.27$. 
Under the same assumptions, a Chabrier (2003) IMF,
\begin{equation}
\phi(m)\propto 
\begin{cases} e^{-(\log m -\log m_c)^2/2\sigma^2}/m~~~~~ & (m<1\,\msun)\\
m^{-2.3}~~~~~ & (m>1\,\msun)
\end{cases}
\end{equation}
(with $m_c=0.08\,\msun$ and $\sigma=0.69$) is more weighted toward short-lived massive stars and yields a larger 
return fraction, $R=0.41$.  In the instantaneous recycling approximation, the fraction of ``dark'' stellar remnants
formed in each generation is 
\begin{equation}
D=\int_{m_0}^{m_u} w_m\phi(m){\rm d}m.
\end{equation}
The two IMFs produce a dark remnant mass fraction of $D=0.12$ and $D=0.19$, respectively. 
The stellar nucleosynthetic yields depend on metallicity, rotation, and the mass limit for black hole formation $m_{\rm BH}$. 
By integrating over the IMF the subsolar metallicity stellar yields (where the effect of mass loss is negligible) tabulated by Maeder (1992) from $10\,\msun$ to 
$m_{\rm BH}=40,\msun$, we obtain $y=0.016$ for a Salpeter and $y=0.032$ for a Chabrier IMF. When integrated to $m_{\rm BH}=60\,\msun$, 
the same tabulation implies $y=0.023$ (with $R=0.29$) and $y=0.048$ (with $R=0.44$) for a Salpeter and Chabrier IMF, respectively. 
Notice that some of the uncertainties associated with the IMF and the mass cutoff $m_{\rm BH}$ become smaller when computing the term 
$y(1-R)$ in the equations (Equation \ref{chem_ev}). For massive stars at solar metallicities, stellar winds eject large amount 
of helium and carbon into the ISM before these are processed into heavier elements, but the effect on the integrated metal yields ($10-40\,\msun$) is
weak (Maeder 1992). Total stellar yields (including the wind and pre-SN contributions) obtained from rotating stellar models at solar metallicity 
have been presented by Hirschi \etal (2005). Over the same range $10-40\,\msun$, in this case we  derive $y=0.019$ for Salpeter and $y=0.038$ 
for Chabrier. For comparison, the zero-metallicity stellar yields of Chieffi \& Limongi (2004) imply $y=0.015$ for Salpeter and $y=0.030$ for Chabrier.

Although disfavored by many observations, a Salpeter IMF in the mass range $0.1-100\,\msun$ is used as a reference throughout the rest of this 
review. Similarly, for consistency with prior work, we assume the canonical metallicity scale where solar metallicity is 
$\zsun=0.02$, rather than the revised value $\zsun=0.014$ of Asplund \etal (2009).

\section{MEASURING MASS FROM LIGHT}
\label{sec:massfromlight}

\begin{figure}
\centerline{\psfig{figure=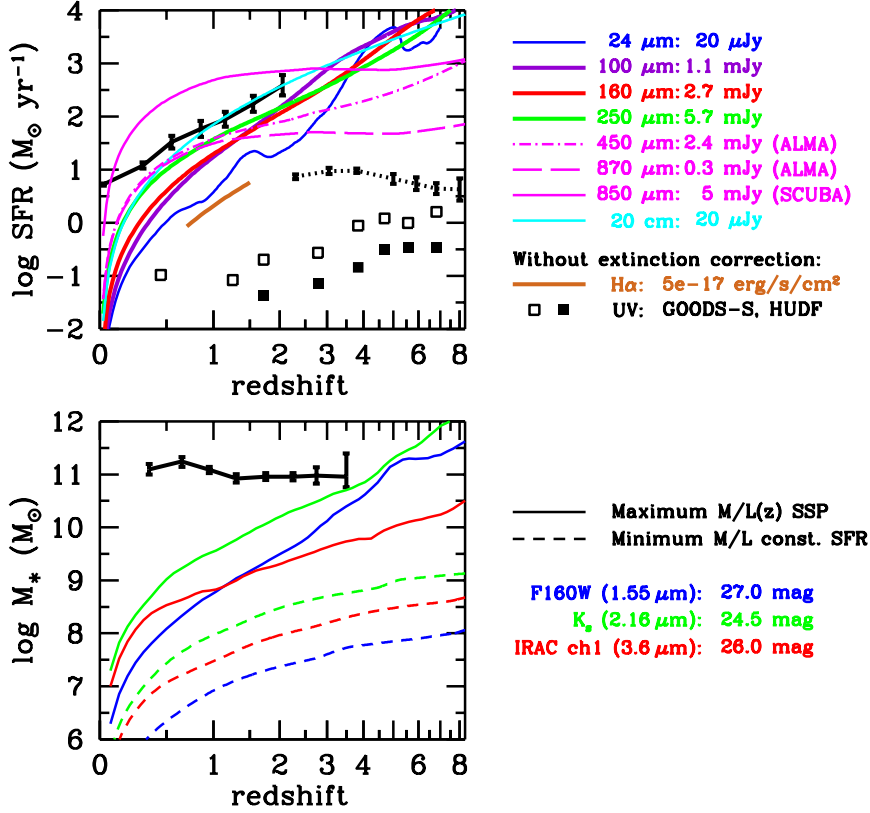,width=0.9\textwidth}}
\caption{\footnotesize 
({\it Top panel}) Sensitivity to star formation versus redshift for deep survey data at various wavelengths.
The key at right indicates the wavelengths and the corresponding flux limits.   Most of these limits were computed 
using data available in the GOODS fields or are simply chosen to be representative values for very deep 
surveys.  For ALMA, we use projected sensitivities of the completed interferometer
for modest 10-min integration times, noting that the small ALMA primary beam at 870$\,\mu$m and 450$\,\mu$m
requires approximately 50 and 170 pointings, respectively, to map 1 square arcminute with uniform sensitivity.   All conversions
to SFR assume a Salpeter IMF from 0.1 to 100 M$_\odot$. The limits for mid-IR ({\it Spitzer}), far-IR 
({\it Herschel}), submillimeter, and radio data use 
bolometric corrections from the observed wavelength based on spectral energy distribution templates by Magdis \etal (2012).
Open square points show rest-frame 1,500-\AA\ sensitivities assuming {no extinction};  in practice, dust attenuation 
can significantly increase these limits. The open squares are based on GOODS data from {GALEX}, ground-based 
$U$-band (Nonino \etal 2009), and {HST} ACS and WFC3 (CANDELS), whereas the filled points show {HST} ACS 
and WFC3 limits for the HUDF (Bouwens \etal 2011b). The H$\alpha$ curve shown here represents the typical limiting 
sensitivity of the 3D-HST IR grism spectroscopic survey (Brammer \etal 2012), again assuming {no extinction}, 
with conversion to SFR from Kennicutt (1998). The data points (shown as error bars) connected by a solid line 
show the SFR corresponding to the characteristic infrared luminosity $L^\ast({\rm IR})$ from Magnelli \etal (2013), 
and the points connected by a dotted line show the SFR corresponding to the characteristic FUV luminosity $L^\ast({\rm FUV})$ 
for Lyman break galaxies at $2 < z < 8$, uncorrected for extinction (Reddy \& Steidel 2009, Bouwens \etal 2012b).
({\it Bottom panel}) Sensitivity to stellar mass versus redshift, for flux limits at several wavelengths, 
as indicated at right.  The solid curves assume a passively evolving simple stellar population with the age 
of the Universe, solar metallicity, a Salpeter IMF, and no extinction, computed using the models of Conroy \etal (2009).
This approximates the maximum mass-to-light ratio potentially visible at any redshift and, hence, provides an upper bound to
the mass completeness limit for a given survey.  Less massive galaxies can easily be detected, however, if they 
have young stellar populations. The dashed curves show sensitivities for an unreddened galaxy with a constant SFR 
and age of $10^7$~years, when $M/L$ at these wavelengths reaches a minimum value.  The IR sensitivity limits
(given in AB magnitudes) are chosen to be representative of deep surveys such as GOODS and CANDELS [e.g., 
$K_s$ data from Retzlaff \etal (2010) or Wang \etal (2010)], but are not specific to a particular data set.
The data points (shown as error bars) connected by a solid line show the characteristic stellar mass $M_\ast$ 
at redshifts $0.2 < z < 4$ (Ilbert et al. 2013). 
}
\label{fig1}
\vspace{+0.3cm}
\end{figure}

Fundamentally, deriving the history of star formation in galaxies involves inferring mass from light.
We observe the emission from galaxies at various wavelengths, and from those measurements we try 
to infer either the rates at which the galaxies are forming stars or their integrated stellar masses.
Figure~\ref{fig1} illustrates the sensitivity of today's premier multiwavelength surveys to the SFRs
and stellar masses of galaxies at high redshift.  Rest-frame UV, IR, submillimeter and radio emission, as well as nebular lines such as H$\alpha$
are all used to measure SFRs and are discussed in this section.  In the absence of
extinction, UV measurements are more sensitive than current IR or radio data by orders of magnitude,
but in practice dust attenuation is often severe.  Long-wavelength data are essential to gain
a comprehensive picture of cosmic star formation, but are limited by current instrumental 
sensitivities, although ALMA (Acatama Large Millimeter Array) enables dramatic improvements at submillimeter wavelengths
that are particularly valuable at higher redshifts.  NIR to mid-infrared (MIR) measurements are 
critical for deriving stellar masses.  Their sensitivity to stellar mass depends critically 
on the mass-to-light ratio of the stellar population in a distant galaxy, hence on its age, SFH, 
and extinction.  Figure 1b illustrates two limiting
cases:  a maximum-$M/L$ model defined as a passively evolving stellar population as old as the 
Universe, and a minimum-$M/L$ model defined as a very young, unreddened, actively star-forming 
galaxy.  In principle, surveys should be mass-complete to the maximum-$M/L$ limits.  Much less 
massive galaxies with young, low-$M/L$ stellar populations can easily be detected, but 
observations would miss dusty or evolved galaxies with lower masses.  The {HST} WFC3 camera 
has significantly improved NIR sensitivity compared with most ground-based imaging, but it samples only
optical rest-frame light at $z < 3$.  {\it Spitzer}'s IRAC remains the premier 
resource for deriving stellar masses at higher redshifts, and {\it James Webb Space Telescope} (JWST) will provide a major advance.

The conversions from light to mass are derived or calibrated using stellar population synthesis 
models, which encode our knowledge of stellar evolution and of the SEDs 
of stars, and compute the emergent spectrum for a galaxy with given properties.  This knowledge 
is imperfect, although astronomers have made great progress developing population synthesis models 
and improving the libraries of empirical and theoretical stellar spectra that they use (for a recent 
review, see Conroy 2013).  

A galaxy (or the Universe as a whole) consists of stars that
span a wide range of masses, ages, and metal abundances.  The light from those
stars may be attenuated by dust before it emerges from the galaxy;  the dust dims
and generally reddens the galaxy spectrum, and the heated dust re-emits energy
in the IR. A galaxy spectrum arises from a composite stellar population
whose true distribution of properties is generally unknown.  For nearby galaxies,
resolved color-magnitude diagrams can reveal the actual distributions of stellar properties,
but for most galaxies we can observe only their integrated light, and properties
of the emergent spectrum (particularly broadband colors) are often degenerate to
different intrinsic properties.  An often-noted example is the degeneracy between
age, metallicity, and dust attenuation, all of which can redden the spectrum
of a galaxy.  Observations at higher spectral resolution, for example, of individual
spectral lines, can help to resolve some degeneracies (e.g., to constrain stellar
metallicities, population ages from absorption line strengths, or reddening from
emission line ratios), but never all: The inherently composite nature of stellar
populations requires that we make simplifying assumptions when interpreting the light,
assumptions that generally cannot be uniquely tested for individual galaxies.
Examples of such assumptions include the form of the IMF, 
the stellar metallicity distribution, the wavelength dependence of dust attenuation,
or the precise SFH of the galaxy.  The hope is that these
assumptions can be made as reasonably as possible, that their impact on derived
masses or SFRs can be estimated, and that ultimately they may 
be tested or constrained by observations in various ways.


The IMF underlies the relation between mass, light, and stellar population age.
It controls the ratio of hot, bright stars that dominate the light to cool,
faint stars that usually dominate the mass. It regulates the luminosity and color 
evolution of the integrated stellar population, as stars with 
different masses evolve at different rates. It also affects the time evolution 
of the integrated stellar mass, which changes as more massive 
stars lose gas to the ISM via winds or detonate as SNe.

It is essentially impossible to constrain the IMF from photometric
measurements of the integrated light from galaxies: The color of a galaxy does not
uniquely reveal its underlying IMF, as there are too many degeneracies to permit useful
constraints.  Even detailed spectroscopy does not usually offer strong constraints
on the IMF overall, although certain spectral features can be useful diagnostics of the
number of stars in a given mass range (e.g., Leitherer \etal 1999). The most direct constraints on the IMF
come from counting stars as a function of mass in resolved, nearby stellar populations,
but they must be very nearby (within our Galaxy and its satellites) to
detect sub-solar dwarf stars that dominate the mass of a stellar system.  The next-best 
constraints come from integrated measurements of the mass-to-light ratio for star
clusters or galaxies, using kinematics (velocity dispersions or rotation curves) to
derive a mass for comparison to the luminosity.  However, these measurements are
difficult to make for faint galaxies at high redshift, and require careful modeling
to account for the role of dark matter and many other effects.

For lack of better information, astronomers often assume that the IMF is universal,
with the same shape at all times and in all galaxies. Although the IMF of various stellar populations
within the Milky Way appears to be invariant (for a review, see Bastian \etal 2010), 
recent studies suggest that the low-mass IMF slope may be a function of the global galactic
potential, becoming increasingly shallow (bottom-light) with decreasing galaxy velocity 
dispersion (Conroy \& van Dokkum 2012, Geha \etal 2013).  
It is still unknown, however, how galaxy to galaxy variations may affect the 
``cosmic'' volume-averaged IMF as a function of redshift. 
In Section \ref{sec:obs_to_para}, we see how a universal IMF can provide a reasonably
consistent picture of the global SFH.
The exact shape of the IMF at low stellar masses is fairly unimportant for deriving
{relative} stellar masses or SFRs for galaxies.  Low-mass stars
contribute most of the mass but almost none of the light, and do not evolve over a
Hubble time.  Therefore, changing the low-mass IMF mainly rescales the mass-to-light ration 
$M/L$ and, hence, affects both stellar masses and SFRs derived from photometry to a similar
degree.  Changes to the intermediate- and high-mass region of the IMF, however, can
have significant effects on the luminosity, color evolution, and the
galaxy properties derived from photometry.   It is quite common to adopt the simple
power-law IMF of Salpeter (1955), truncated over a finite mass range (generally,
0.1 to 100~$\msun$, as adopted in this review).  However, most observations show that 
the actual IMF turns over from the Salpeter slope at masses $< 1\,\msun$, resulting 
in smaller $M/L$ ratios than those predicted by the Salpeter IMF.  Some common versions of such an
IMF are the broken power-law representation used by Kroupa (2001) and the log-normal
turnover suggested by Chabrier (2003).


Dust extinction is another important effect that must often be assumed or inferred, rather than directly measured. 
The shape of the extinction law depends on the properties of the dust grains causing the extinction.
For observations of a single star, photons
may be absorbed by dust or scattered out of the observed sightline.  However, galaxies
are 3D structures with mixed and varying distributions of stars and dust.
Photons may be scattered both into and out of the sightline, and the optical depth of
dust along the line of sight to the observer will be different for every star in the
galaxy.  These effects are generally lumped together into the simplifying assumption
of a net dust attenuation curve, and such relations have been derived for local galaxy
samples both empirically (e.g., Calzetti \etal 2000) and using theoretical modeling
(Charlot \& Fall 2000).  However, all galaxies are not equal, and no net attenuation law is equally appropriate for all galaxies.
There can always be stars that are completely obscured behind optically thick dust such that little
or none of their light emerges directly from the galaxy, except re-radiated as dust
emission.  Although this may not be a significant factor for many galaxies, there are
certainly some starburst galaxies in which huge and bolometrically dominant star-formation 
activity takes place in regions screened by hundreds of magnitudes of dust
extinction. UV/optical measurements will never detect this light, but 
the star formation can be detected and measured at other wavelengths, e.g., with FIR or radio data.


In order to derive SFRs or stellar masses for galaxies using stellar population synthesis models, astronomers typically assume 
relatively simple, parameterized SFHs. However, the SFHs of individual galaxies are unlikely to be smooth and simple;
they may vary on both long and short timescales.  The fact that young stars are more luminous 
than older stars leads to the problem of ``outshining'' (e.g., Papovich \etal 2001,  
Maraston \etal 2010) -- the light from older stars can be lost in the glare of more
recent star formation and contributes relatively little to the observed photometry from
a galaxy, even if those stars contribute significantly to its mass. SED model fits to
galaxies with recent star formation tend to be driven largely by the younger, brighter
starlight, and may not constrain the mass (or other properties) of older stars that
may be present.

For the Universe as a whole there is one ``cosmic'' IMF that
represents the global average at a given time or redshift, regardless of whether the IMF varies
from one galaxy to another.  Similarly, there is a ``cosmic'' distribution of metallicities,
a ``cosmic'' net attenuation of starlight by dust at a given wavelength, and the Universe
as a whole obeys one ``cosmic'' SFH that, moreover, was probably
relatively smooth over time -- i.e., any stochasticity or ``burstiness'' averages out
when considered for the Universe as a whole.  In principle, these facts can simplify
the determination of the cosmic SFH, particularly when it is derived from
measurements of integrated light averaging over all galaxies.  In practice, however,
astronomers often derive SFRs and stellar masses for individual galaxies
in their deep surveys, and then sum them to derive comoving volume averages.
In which case, some of the advantages of the ``cosmic averaging'' are reduced.

\subsection{Star-Formation Rates}
\label{sec:sfrates}

There are many ways in which to infer SFRs from observations of the
integrated light from galaxies. Kennicutt (1998) and Kennicutt \& Evans (2012) have presented 
extensive reviews of this topic, and here we recap only points that
are especially relevant for measurements of the global SFH, particularly at high redshift. 
Virtually all observational tracers of star formation fundamentally 
measure the rate of massive star formation, because massive stars emit most of the energy from
a young stellar population.  However, different observational tracers are sensitive to
different ranges of stellar masses: hence, they respond differently as a function of stellar
population age.  For example, H$\alpha$ emission arises primarily from HII regions
photoionized by O stars with lifetimes shorter than 20~Myr, whereas the UV continuum
is produced by stars with a broader mass range and with longer lifetimes.
The time-dependence of different indicators can complicate efforts to derive
accurate SFRs for individual galaxies, especially if their SFRs may be rapidly
changing (e.g., during a starburst event), but they should average out when summing
over a whole population of galaxies.

\subsubsection{UV light}
\label{sec:sfrates_uv}

\begin{figure}[ht]
\centerline{\psfig{figure=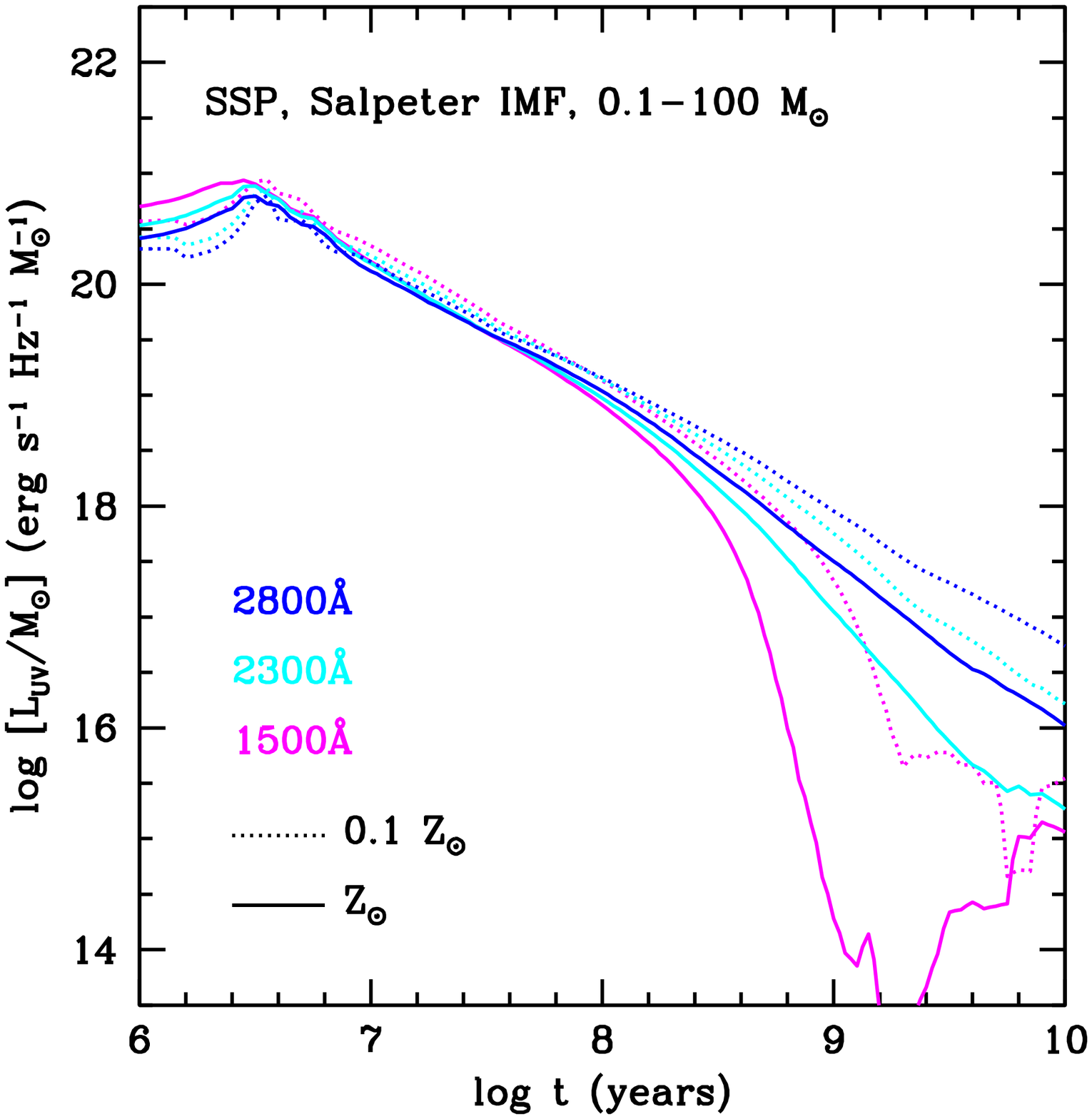,width=0.49\textwidth}
            \psfig{figure=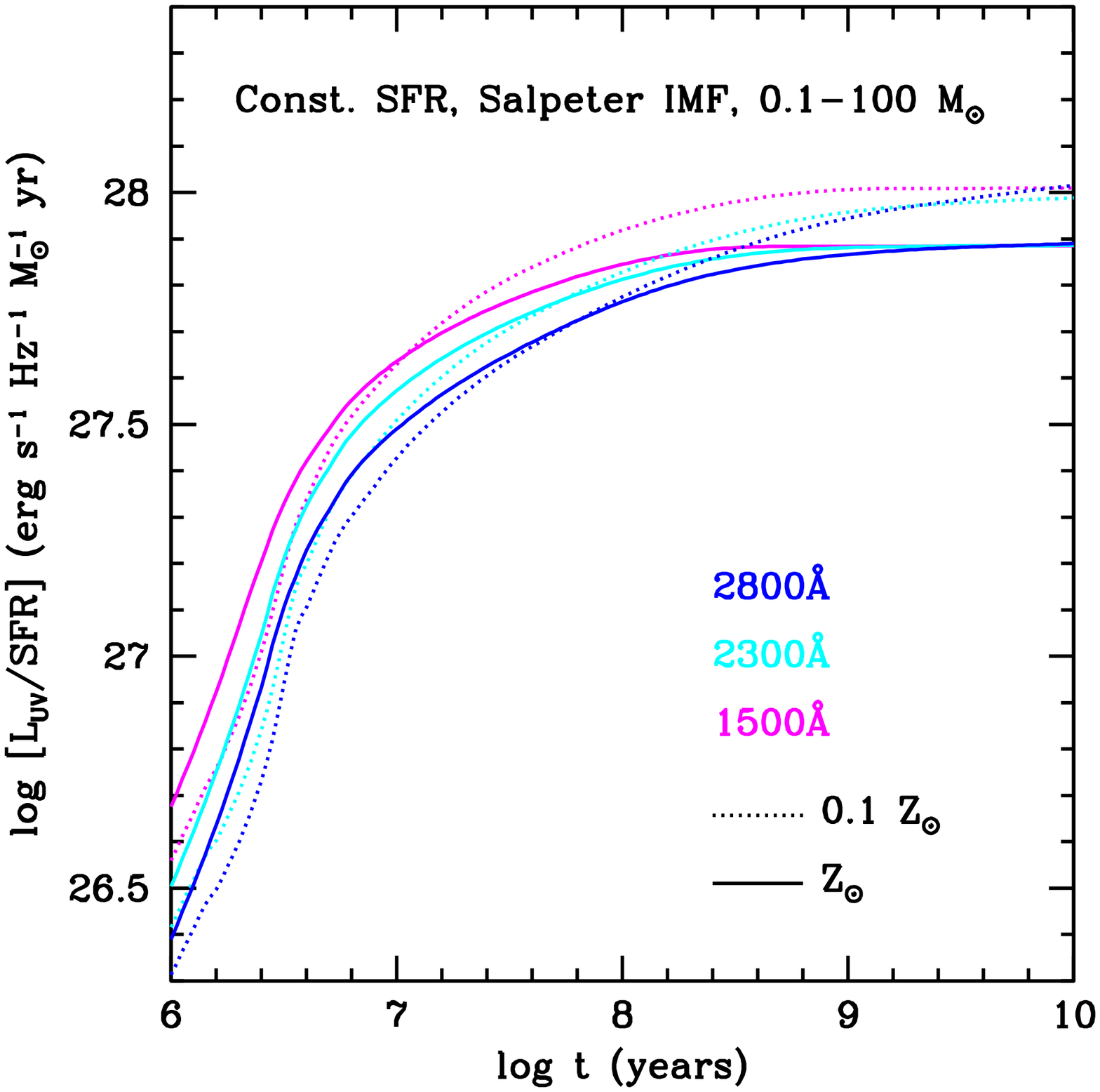,width=0.49\textwidth}}
\caption{\footnotesize{
({\it Left panel}) Time dependence of the UV luminosity of an SSP 
with initial mass 1 $\msun$, formed with a Salpeter IMF in the range $0.1-100\,\msun$,
measured at three wavelengths, 1,500\,\AA, 2,300\,\AA, and 2,800\,\AA,
and computed with the stellar population synthesis models of Conroy \etal (2009):
({\it solid lines}) solar metallicity ($Z_\ast = 0.02$), ({\it dashed lines}} 1/10th solar metallicity ($Z_\ast = 0.002$).
({\it Right panel}) Time dependence of the UV luminosity per unit SFR 
for a model with a constant SFR, shown for the same UV wavelengths and metallicities as in the left panel.
}
\label{fig2}
\end{figure}

Newly-formed stellar populations emit radiation over a broad spectrum.  For a normal IMF, 
low-mass stars dominate the mass integrated over the whole stellar population, but at young 
ages the luminosity is dominated by ultraviolet emission from massive stars.  These stars 
have short lifetimes, so the UV emission fades quickly.  For a Salpeter IMF, the 1,500-\AA\ 
luminosity from an evolving simple stellar population (SSP) (i.e., an ensemble of stars formed
instantaneously and evolving together) with solar metallicity fades by a factor of 100 after 
$10^8$~years, and by factors of $10^3$ to $10^6$ after 10$^9$~years, depending on metallicity
(Figure~\ref{fig2}).   Bolometrically, at least half of the luminous energy that an SSP 
produces over a 10-Gyr cosmic lifetime emerges in the first 100~Myr, mostly in the UV, 
making this a natural wavelength from which to infer SFRs. 

For a galaxy forming stars at a constant rate, the 1,500-\AA\ luminosity stabilizes
once O-stars start to evolve off the main sequence. For solar metallicity, by an age 
of $10^{7.5}$ years, the 1,500-\AA\ luminosity has reached 75\% of its asymptotic value, although convergence
is somewhat slower at lower metallicity (Figure~\ref{fig2}).  For these reasons, 
the UV luminosity at wavelengths of $\sim$ 1,500\,\AA\ (wavelengths from 1400\,\AA\ to 1700\,\AA\ have been 
used in the literature for both local and high redshift studies) 
is regarded as a good tracer of the formation rate of massive stars, provided that the 
timescale for significant fluctuations in the SFR is longer than a few~$10^7$ years.
For shorter bursts or dips in the SFR, changes in the UV continuum flux may lag those in
the SFR and smooth over such variations.

Although the 1,500-\AA\ rest frame is readily accessible with ground-based
optical observations of galaxies at redshifts $z \gtrsim 1.4$, measurements
at lower redshifts require space-based UV data (e.g., from {GALEX} or {HST})
or are limited to longer UV wavelengths. Mid-UV reference wavelengths that have been 
used in the literature include 2,300\,\AA\ (the approximate central 
wavelength of the {GALEX} near-UV passband) and 2,800\,\AA\ (used, e.g., by Lilly \etal 1996).
The mid-UV emission from a galaxy can have a larger contribution from longer-lived, 
lower-mass stars, particularly at later ages, and the time evolution of the luminosity 
is more gradual.  This is particularly true after $\sim 250$~Myr, when the 1,500-\AA\ 
luminosity of an SSP drops off sharply, whereas the 2,800-\AA\ luminosity continues
to fade at an approximately exponential rate (Figure~\ref{fig2}).  For a constant 
SFR, the UV spectral slope reddens moderately with time, as the 
1,500\,\AA\ luminosity reaches a steady-state level earlier, while longer-lived 
(B and A) stars continue to build up and contribute to the 2,800\,\AA\ luminosity.  
This complicates the conversion from luminosity to SFR, as well as any
correction for dust extinction based on the UV spectral slope.  
Still, for young ages, both shorter and longer UV wavelengths usefully trace 
the SFR and have been used extensively in the literature.
Moreover, longer UV wavelengths are subject to somewhat lesser dust attenuation.
Wavelengths shorter than that of \Lya\ (1,216 \AA) are rarely used to estimate SFRs, 
particularly at high redshift where absorption from neutral hydrogen 
in the IGM is strong.

The UV luminosity output by a stellar population also depends on its metallicity,
which affects stellar temperatures and line blanketing.  Generally speaking, less-metal-rich 
stars produce more UV light.   The amplitude of this effect is 
not insignificant, and depends on the details of the SFH. 
From a Salpeter IMF and constant SFR, the range of FUV luminosity per 
unit SFR for stars spanning a factor of 100 in metallicity (from $Z = 0.0003$ to 0.03)
is less than 0.24~dex, or 70\%.  These variations are larger at higher metallicities and older ages; therefore, 
we may expect significant evolution in the $L_{\rm FUV}$ to the
SFR conversion factor as the global metallicity of galaxies evolves.

We express the conversion factor between the intrinsic FUV-specific luminosity $L_\nu$(FUV)
(before extinction, or corrected for extinction) and the ongoing SFR as
\begin{equation}
{\rm SFR} = {\cal K}_{\rm FUV} \times L_\nu({\rm FUV}),
\label{eq:KFUV}
\end{equation}
where $L_\nu({\rm FUV})$ is expressed in units of $\lum$ and SFR in units of $\sfr$.
The precise value of the conversion factor ${\cal K}_{\rm FUV}$ is sensitive to the recent 
SFH and metal-enrichment history as well as to the choice of the IMF.  
It is relatively insensitive to the exact FUV wavelength, as the UV spectrum of a galaxy
with a constant SFR is quite flat in $f_\nu$ units, at least
for ages much longer than $10^7$~years.  Generally in this review, we use FUV to refer to 1,500-\AA\ 
emission or are explicit when we refer to other UV wavelengths.  
For a Salpeter IMF in the mass range $0.1-100\,\msun$ and constant SFR, the 
{flexible stellar population synthesis} (FSPS) models of Conroy \etal (2009) 
yield ${\cal K}_{\rm FUV}=(1.55,1.3,1.1,1.0) \times 10^{-28}$ for $\log Z_\ast/\zsun=(+0.2,0,-0.5,-1.0)$ 
at age $\gta 300$ Myr.  The GALAXEV models of Bruzual \& Charlot (2003) yield 
values of ${\cal K}_{\rm FUV}$ that are $\sim $5\% smaller.  

\begin{figure}[ht]
\centerline{\psfig{figure=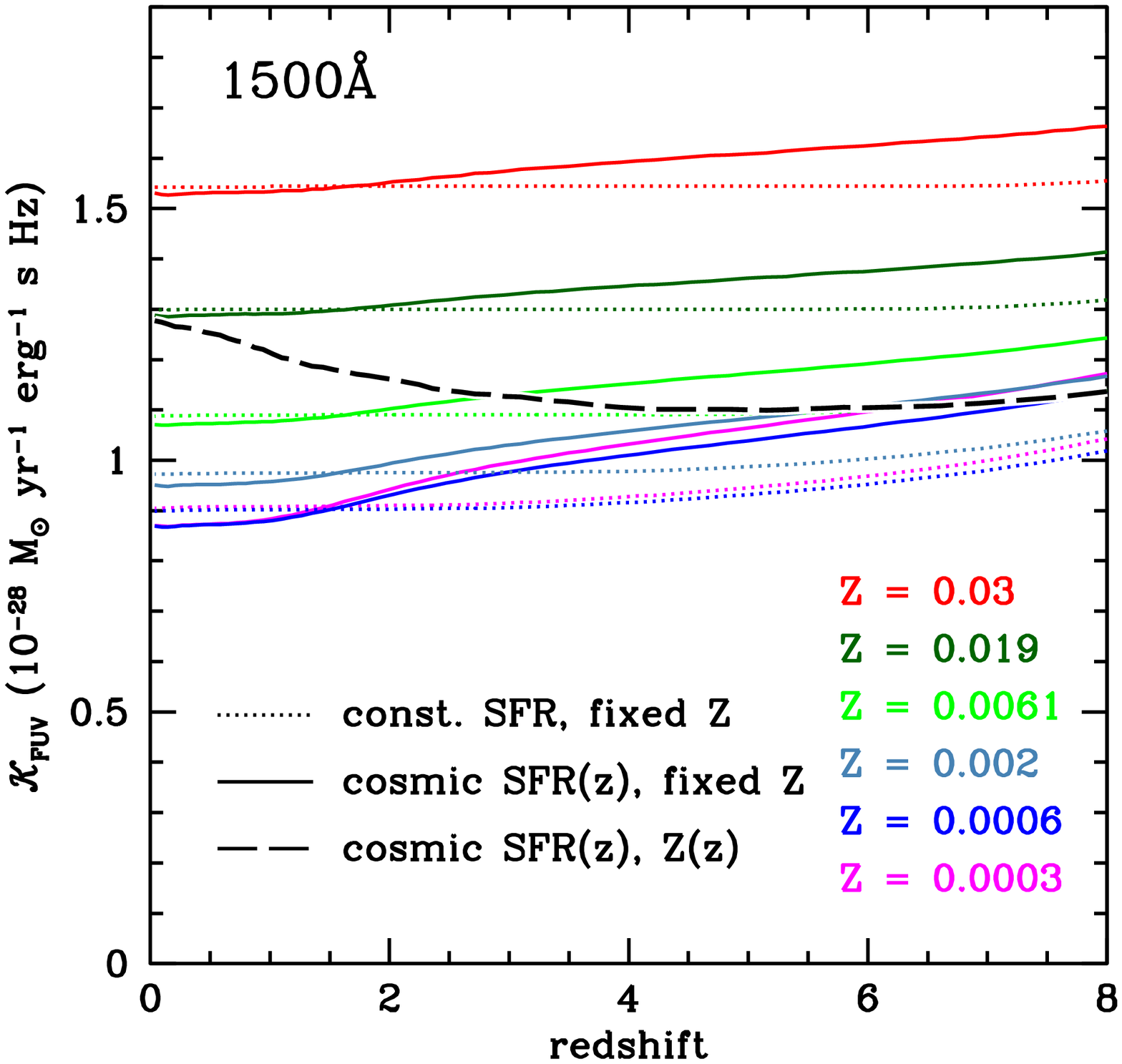,width=0.49\textwidth}
            \psfig{figure=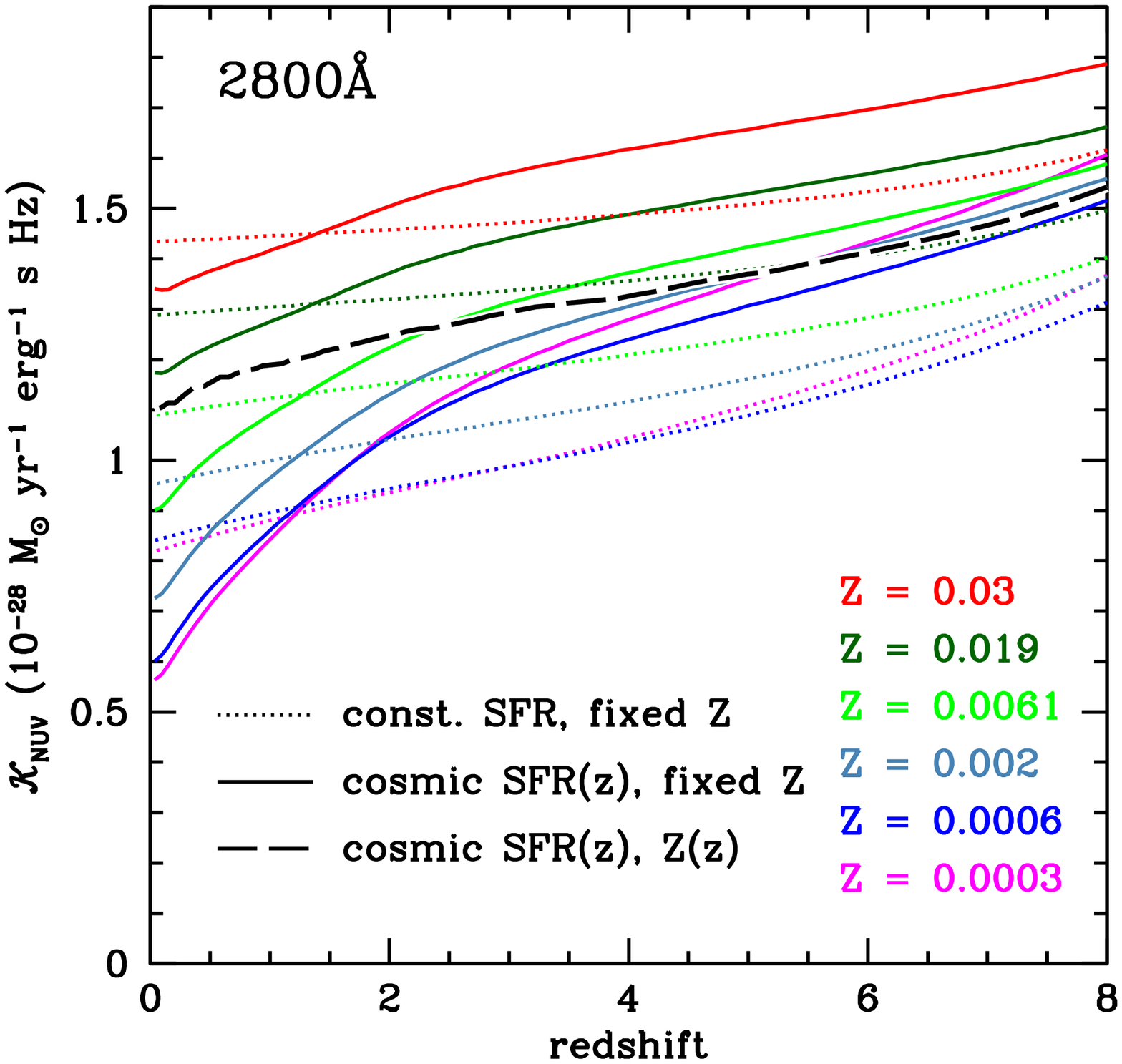,width=0.49\textwidth}}
\vspace{-0.0cm}
\caption{\footnotesize Age and metallicity dependence of the SFR to UV luminosity ratio 
($={\cal K}$) (Equation \ref{eq:KFUV}) for a stellar population with a Salpeter IMF in the range $0.1-100\,\msun$, 
using the spectral population synthesis models of Conroy \etal (2009):  
({\it dotted lines}) ${\cal K}_{\rm FUV}(z)$ assuming constant SFR (starting at $z = 12$) and fixed stellar metallicity;
({\it solid lines}) same assuming the functional form for cosmic SFR density given in Equation \ref{eq:sfrd} (again starting at $z = 12$) and 
fixed stellar metallicity; ({\it dashed lines}) same assuming the functional form for cosmic SFR density given in Equation \ref{eq:sfrd} 
(again starting at $z = 12$) and a stellar metallicity evolving as $Z_\ast=\zsun 10^{-0.15z}$ (Kewley \& Kobulnicky 2007).
The relation is shown at ({\it a} 1,500\,\AA\ and ({\it b}) 2,800\,\AA, respectively. 
}
\label{fig3}
\end{figure}

Figure~\ref{fig3} illustrates the combined effects of the evolution of the global SFR and metal density 
on the global mean UV-to-SFR conversion factor as a function of redshift, on the basis of the FSPS models. 
Concentrating on the FUV behavior at 1,500\,\AA, for constant $\psi(z)$, the conversion factor is nearly constant, though slightly 
elevated at the highest redshifts as the cosmic age gets young, particularly for lower metallicity models. A SFH that increases 
with time from $z=12$ to 1.7, so that the UV-emitting population is on average younger over that redshift range, leads to a more gradually declining 
trend in ${\cal K}_{\rm FUV}(z)$ with time. The figure also illustrates one scenario for a global change in the metallicity
of the star-forming population, evolving as $Z_\ast=\zsun 10^{-0.15z}$ (Kewley \& Kobulnicky 2007).
This particular evolution is only moderately well constrained at lower redshifts
and should be taken as illustrative only, but we may certainly expect metallicities
to be lower on average at higher redshift. The effects of metallicity dominate over
those of age in this scenario, but the two counterbalance each other other to a certain degree,
so that ${\cal K}_{\rm FUV}(z)$ changes by less than 20\%.  At 2,800\,\AA, 
the redshift dependence of ${\cal K}_{\rm NUV}(z)$ is stronger, particularly at $z < 2$
as the global SFRD declines with time, although this may be partially canceled
by the effects of metallicity evolution.   This is an example of why shorter FUV 
wavelengths should be preferred for deriving galaxy SFRs.

In this review, we adopt a constant FUV conversion factor 
${\cal K}_{\rm FUV} = 1.15 \times 10^{-28}\,\msun$~year$^{-1}$ erg$^{-1}$~s~Hz
(we typically leave out the units) as a compromise value based on the evolutionary scenario from Figure~\ref{fig3}.
The widely used value from Kennicutt (1998) (and based on the calibration by Madau \etal 1998b),
${\cal K}_{\rm FUV}=1.4 \times 10^{-28}$, is 20\% larger than our calibration.  Other recent analyses 
based on the GALAXEV libraries have also found lower mean conversion factors,
both for low- and high-redshift galaxy populations (e.g., Salim \etal 2007, Haardt \& Madau 2012).
The FUV conversion tabulated in Kennicutt \& Evans (2012) (from Murphy \etal 2011), if rescaled
from the Kroupa to Salpeter IMF, is very close to the $z = 0$, solar metallicity value
of ${\cal K}_{\rm FUV}(z)$ in Figure~\ref{fig3}, but our somewhat smaller value should 
be more representative for the peak era of cosmic star formation at high redshift.
Some authors express FUV luminosity as $L_{\rm FUV} = \nu L_\nu$ in solar units.
In that case, ${\cal K}_{\rm FUV} = 2.2 \times 10^{-10}\,\msun {\rm year}^{-1} L_\odot^{-1}$
at 1,500\AA, and the conversion factor will depend on the wavelength.

\begin{figure}[ht]
\centerline{\psfig{figure=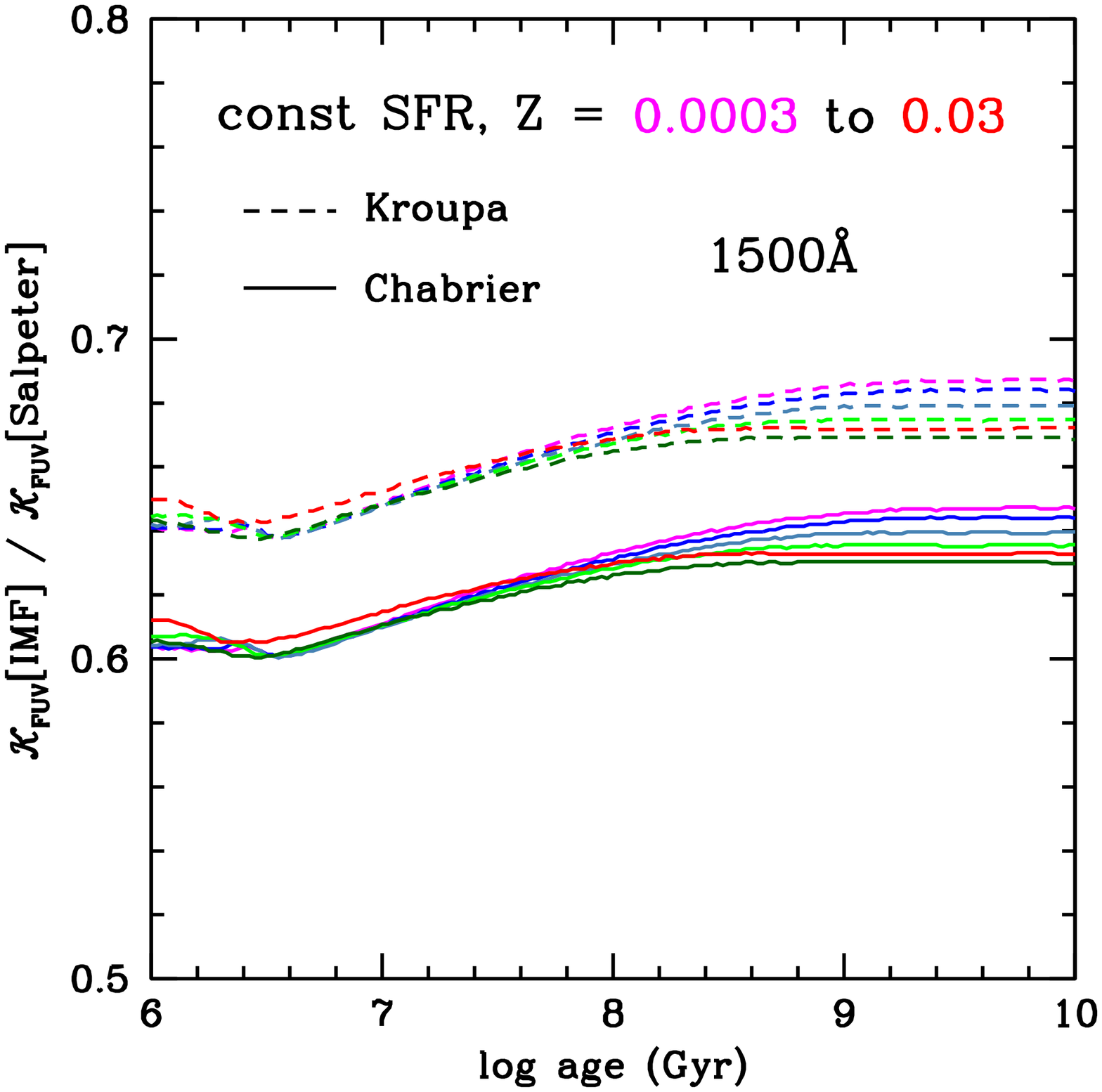,width=0.49\textwidth}
            \psfig{figure=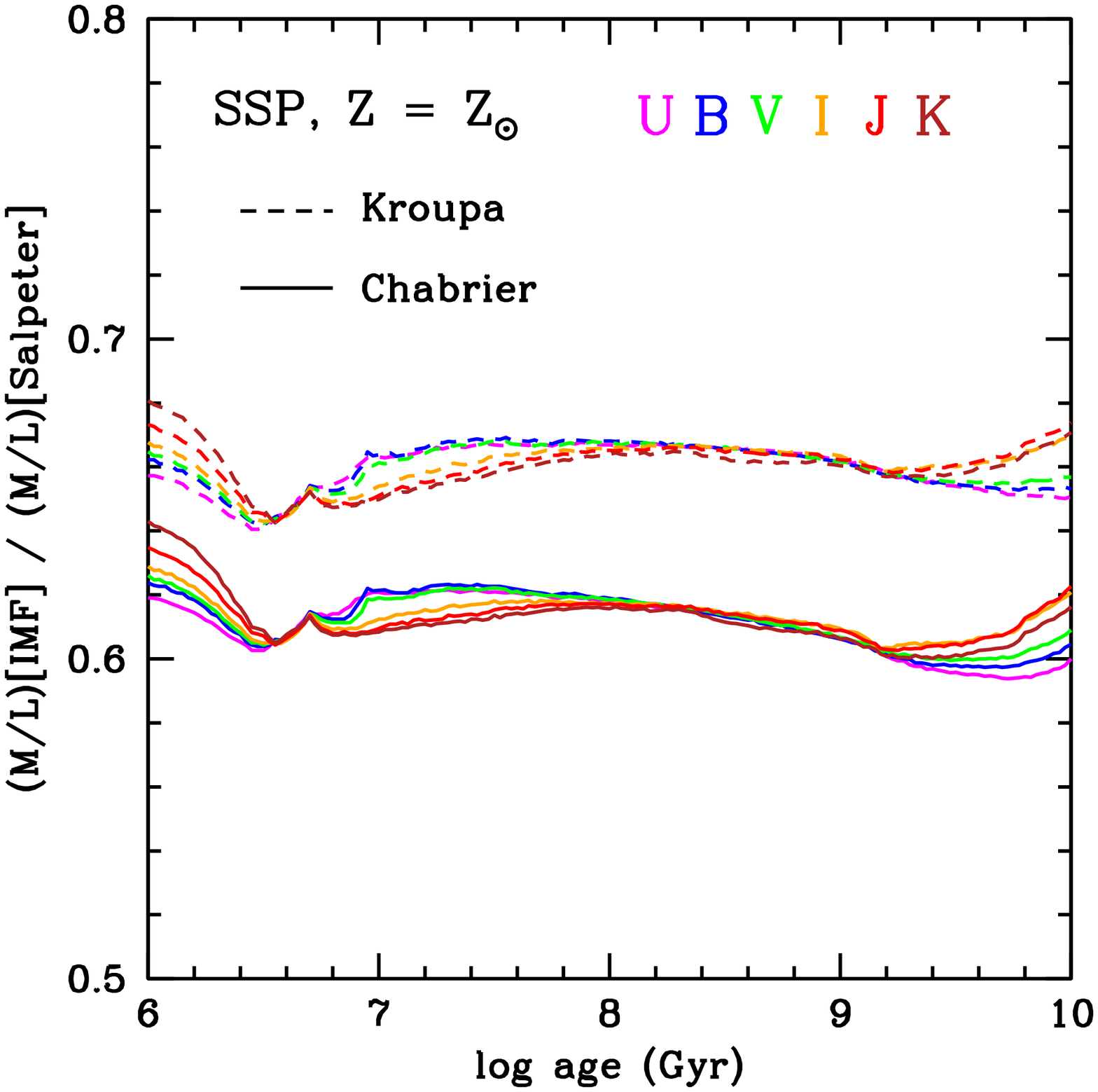,width=0.49\textwidth}}
\caption{\footnotesize ({\it Left panel}) The ratio of SFR/$L_{\rm FUV}$ conversion factors ${\cal K_{\rm FUV}}$
(Equation ~\ref{eq:KFUV}) for Chabrier or Kroupa IMFs to that for a Salpeter IMF (0.1 to 100~$\msun$), for a constant 
SFR and for various metallicities, computed with the FSPS models of Conroy \etal (2009).  
The difference in this conversion factor varies only slightly with age or metallicity over the ranges shown here.
({\it Right panel}) Ratio of mass-to-light ratios for a simple stellar population as a function of age,
for various bandpasses from the near-UV through near-IR, comparing values for Chabrier and Kroupa IMFs
to Salpeter. This factor also has only a small dependence on age, bandpass, or metallicity (the last
not shown here).
}
\label{fig4}
\end{figure}

Figure~\ref{fig4} shows the ratio of ${\cal K}_{\rm FUV}$ for the Chabrier
or Kroupa IMFs to that for the Salpeter IMF as a function of age for a constant SFR calculated using FSPS.
This ratio is nearly constant, varying by only 5\% with age and 3\% over a factor of 100
in metallicity.   Where necessary to convert SFRs from the literature from Chabrier or Kroupa IMFs to the
Salpeter IMF, we divide by constant factors of 0.63 (Chabrier) or 0.67 (Kroupa).  Similarly,
Figure~\ref{fig4} examines mass-to-light ratios for an SSP as a function of age,
in various bandpasses, comparing values for the Chabrier or Kroupa IMFs to the Salpeter IMF.  Again, these
ratios are fairly constant with age and have very little dependence on the bandpass. In other words,
the color evolution for an SSP with Chabrier or Kroupa IMFs is very similar to that for the Salpeter IMF, showing
a roughly constant offset in $M/L$. [The similar time dependence of $M/L$ for the ``bottom-light'' Chabrier or Kroupa IMFs to 
that for the Salpeter IMF is something of a coincidence (or a conspiracy). 
The rate of luminosity evolution for an SSP depends on the logarithmic IMF slope at
masses greater than 1~$\msun$, and is faster for the flatter $x = 1.3$ (Kroupa or Chabrier) 
than for the Salpeter value $x = 1.35$.  However, the evolution of the recycled mass
fraction is also faster for the Kroupa and Chabrier IMFs because their low--mass turnovers 
give them smaller mass fractions of long-lived stars.  These two effects roughly cancel for 
$x = 1.3$, resulting in a time dependence for $M/L$ that is nearly the same as that for the 
Salpeter IMF.  For an SSP with a ``bottom-light'' IMF with a Salpeter slope $x = 1.35$, 
the ratio of $M/L$ compared with that for a Salpeter IMF would decrease by $\sim$ 16\% over $\sim 5$ Gyr, 
and a constant IMF rescaling factor for derived stellar masses would be inappropriate.]  
The dependence on metallicity (not shown) is very weak.
To rescale stellar masses from Chabrier or Kroupa to Salpeter IMF, we divide by constant factors
0.61 and 0.66, respectively.

The greatest drawback for UV measurements of star formation is the obscuring effect of dust.
Extinction is strong in the UV, so even modest amounts of dust can dramatically suppress
the emerging UV flux.  Dust re-emits the absorbed energy in the IR, which we 
discuss in the next section.  A reliable measurement of SFRs from UV light
must either correct for the effects of dust absorption, or measure the absorbed energy
directly through IR emission.  We will return to the relation between UV dust
attenuation and IR emission in Section \ref{sec:UVEIRE} below.

\subsubsection{Infrared emission}
\label{sec:sfrates_ir}

The energy that dust absorbs from the UV is re-radiated at MIR and FIR 
wavelengths, making IR observations another important tool for measuring SFRs. 
The effect of dust extinction at FIR wavelengths is generally
regarded as negligible, although in the MIR extinction can still be relevant
for the most deeply buried star formation and active galactic nuclei (AGN).  The total IR luminosity 
($L_{\rm IR}$, usually defined as being integrated over the wavelength range 8--1000$\,\mu$m) is
a measurement of the energy that was absorbed by dust, mainly at UV wavelengths.
Because most UV emission comes from star formation, the IR luminosity is often
interpreted as being directly proportional to the absorbed fraction of the energy
from star formation.   However, active nuclei can also produce strong UV emission,
often in dusty environments, and may contribute to IR emission by heating
dust in the torus and clouds surrounding the AGN.   Older stellar populations can
also heat dust that is present in the ISM of a galaxy,
contributing to the FIR emission.  This is important particularly for ``mature''
galaxies with low current SFRs in the nearby Universe: For a galaxy such as our Milky Way, 
perhaps half of the FIR emission comes from dust heated
by older stars, not from young star-forming regions (Lonsdale Persson \& Helou 1987).
However, for very actively star-forming galaxies without AGN it is generally assumed 
that most of the IR emission arises from new star formation.  Ideally, a galaxy's
total IR luminosity would be measured by fitting a dust emission model to 
observations at several wavelengths, hopefully spanning the peak of dust emission.
In practice, however, such multiwavelength data are often unavailable, and astronomers 
frequently use an SED template that is often derived from observations of local galaxies 
to extrapolate from a single observed flux density at some MIR or FIR wavelength,
not necessarily close to the dust emission peak, to a total $L_{\rm IR}$. Thus, variations 
in the dust emission properties from galaxy to galaxy can lead to significant 
uncertainties in not only this bolometric correction, but also in the estimation of 
SFRs.

Arising from various components heated to different temperatures, 
the spectrum of dust emission is fairly complex. Most of the dust mass in a galaxy is usually in
the form of relatively cold dust (15-60 K) that contributes strongly to the emission
at FIR and submillimeter wavelengths (30-1,000$\,\mu$m).  Dust at several
different temperatures may be present, including both colder grains in the ambient
ISM and warmer grains in star-forming regions.  Emission from still hotter, small-grain
dust in star-forming regions, usually transiently heated by single photons and not
in thermal equilibrium, can dominate the MIR continuum ($\lambda < 30\,\mu$m)
and may serve as a useful SFR indicator (e.g., Calzetti \etal 2007).
The MIR spectral region (3--20$\,\mu$m) is both spectrally and physically
complex: It has strong emission bands from polycyclic aromatic hydrocarbons and absorption bands primarily 
from silicates.  The strength of emission from polycyclic aromatic hydrocarbons can 
depend strongly on ISM metallicity and radiation field intensity 
(e.g., Engelbracht \etal 2005, 2008, Smith \etal 2007). Strong silicate absorption 
features are seen when the column density of dust and gas is particularly large toward 
obscured AGN and perhaps even nuclear starburst regions.  AGN may contribute 
strong continuum emission from warm dust, and can dominate over star formation at MIR wavelengths. 
By contrast, in the FIR, their role is less prominent.

The {\it Infrared Space Observatory} (ISO) and the {\it Spitzer Space Telescope}
were the first telescopes with MIR sensitivities sufficient
to detect galaxies at cosmological redshifts.  In particular, {\it Spitzer} observations
at 24$\,\mu$m with the MIPS instrument are very sensitive and capable of detecting
``normal'' star-forming galaxies out to $z \approx 2$ in modest integration times.
{\it Spitzer} is also very efficient for mapping large sky areas. It has a 24-$\mu$m beam size that is
small enough (5.7 arcsec) to reliably identify faint galaxy counterparts to
the IR emission.  However, only a fraction of the total IR luminosity
emerges in the MIR. As noted above, it is a complicated spectral region that leads 
to large and potentially quite uncertain bolometric corrections
from the observed MIR flux to the total IR luminosity.
At $z \approx 2$, where 24-$\mu$m observations sample rest-frame wavelengths
around 8$\,\mu$m, where the strongest polycyclic aromatic hydrocarbon bands are found, spectral templates
based on local galaxies span more than an order of magnitude in the ratio
$L_{\rm IR} / L_{8\mu {\rm m}}$ (e.g., Chary \& Elbaz 2001, Dale \& Helou 2002, 
Dale \etal 2005).  More information about the type of galaxy being observed is needed to 
choose with confidence an appropriate template to convert the
observed MIR luminosity to $L_{\rm IR}$ or a SFR. 

The FIR thermal emission is a simpler and more direct measurement of star-formation energy. 
Partly owing to their large beam sizes that resulted
in significant confusion and blending of sources and in difficulty localizing galaxy
counterparts, ISO and {\it Spitzer} offer only relatively limited FIR
sensitivity for deep observations.  The {\it Herschel Space Observatory} dramatically improved such
observations:  Its 3.5-m mirror diameter provided a point spread function FWHM (full width half maximum) small 
enough to minimize confusion and to identify source counterparts in observations from 70 to 250$\,\mu$m. However, 
at the longest wavelengths of the {\it Herschel} SPIRE instrument, 350 and 500$\,\mu$m,
confusion becomes severe. {\it Herschel} observations can directly detect galaxies near the peak of their FIR
dust emission: Dust SEDs typically peak at 60-100$\,\mu$m in the rest frame,
within the range of {\it Herschel} observations out to $z < 4$.   Temperature variations
in galaxies lead to variations in the bolometric corrections for observations
at a single wavelength, but these differences are much smaller than for MIR 
data, generally less then factors of 2.

Despite {\it Herschel}'s FIR sensitivity, deep {\it Spitzer} 24-$\mu$m observations, in general, 
still detect more high-$z$ sources down to lower limiting IR luminosities or SFRs.
At $z \approx 2$, the deepest {\it Herschel} observations only barely reach
to roughly $L^\ast_{\rm IR}$ [the characteristic luminosity of the ``knee'' of the IR luminosity function (IRLF)], 
leaving a large fraction of the total cosmic SFRD undetected, at least for individual sources,
although stacking can be used to probe to fainter levels.   Deep {\it Spitzer} 24-$\mu$m observations detect 
galaxies with SFRs several times lower, and many fields were surveyed to faint
limiting fluxes at 24-$\mu$m during {\it Spitzer}'s cryogenic lifetime.  Therefore, 
there is still value in trying to understand and calibrate ways to measure
star formation from deep MIR data, despite the large and potentially
uncertain bolometric corrections.

In practice, observations of IR-luminous galaxies detected at high redshift with both
{\it Spitzer} and {\it Herschel} have demonstrated that the IR SEDs for
many galaxies are well-behaved and that variations can be understood at least in part.
Several pre-{\it Herschel} studies (Papovich \etal 2007, Daddi \etal 2007,
Magnelli \etal 2009, 2011) compared 24-$\mu$m observations of distant galaxies with those of 
other SFR tracers, including {\it Spitzer} FIR measurements (either individual
detections or stacked averages) and radio emission.   On average, the MIR to FIR
flux ratios for galaxies at $z \lesssim 1.3$ match those predicted by local IR SED
templates such as those of Chary \& Elbaz (2001), implying that 24-$\mu$m-derived SFRs
should be reliable.  However, at higher redshift, $1.3 < z < 2.5$, the 24-$\mu$m
fluxes were brighter than expected relative to the FIR or radio data, i.e., SFRs derived from 24-$\mu$m 
data using local SED templates may be systematically
overestimated at $z \approx 2$.  This result was upheld by early {\it Herschel} studies
(Nordon \etal 2010, Elbaz \etal 2010).  In a joint analysis of the IR SED
properties of both nearby and high-redshift IR-luminous galaxies, Elbaz \etal (2011)
provided an explanatory framework for these observations in terms of the distinction
between a majority population of galaxies obeying a ``main-sequence'' correlation
between their SFRs and stellar masses and a minority ``starburst''
population with substantially higher SFRs per unit mass (or sSFR).
Locally, starburst galaxies have more compact, high surface density star forming
regions, whereas normal disk galaxies on the star-forming main sequence have star
formation distributed on larger scales with lower surface density.  Starbursts
also have warmer average dust temperatures and a significantly larger ratio
between their FIR and 8-$\mu$m rest-frame luminosities than those of the main-sequence disk 
galaxies.   Locally, most luminous and ultraluminous IR 
galaxies (LIRGs and ULIRGs, with $L_{\rm IR} > 10^{11} L_\odot$ and $> 10^{12} L_\odot$,
respectively) are merger-driven starbursts, but at $z \approx 2$ where 
the SFRs and sSFRs of galaxies are globally much larger, the majority of
LIRGs and ULIRGs are ``normal'' main-sequence galaxies.  Their IR SEDs are
more similar to those of ordinary, local star-forming spiral galaxies, and have 
smaller bolometric corrections from observed 24-$\mu$m data
(rest frame $\lambda \approx 8\,\mu$m) than those predicted by SED templates
designed to match local LIRGs and ULIRGs.   Elbaz \etal (2011) constructed a 
``universal'' main-sequence SED from the ensemble of high-$z$ {\it Spitzer} and
{\it Herschel} photometry for galaxies in the Great Observatories Origins Deep Survey (GOODS) 
fields at $0.3 < z < 2.5$. This SED leads to consistent total IR luminosities for the large majority of galaxies
over that redshift range.  Although no single template can be used to accurately
derive $L_{\rm IR}$ or SFR from MIR observations for all galaxies, we now have
a better understanding of how this can be done on average, which may be sufficient
for deriving the global redshift evolution of the IR luminosity density or
its corresponding SFRD.  Rodighiero \etal (2011) (see also Sargent \etal 2012) showed 
that starbursts (whose IR SEDs deviate
significantly from those of the main-sequence population) account for only 10\% of
the global SFRD at $z \approx 2$.   With the data now available from {\it Herschel}
and {\it Spitzer}, a broad understanding of the evolving IRLF and IR
luminosity density, at least at $0 < z < 2.5$, seems within reach.

MIR and FIR observations require space-based telescopes, but at
submillimeter and millimeter wavelengths, observations can once again be made
from the ground within certain atmospheric transmission windows.    The advent
of submillimeter bolometer array cameras such as SCUBA on the JCMT revolutionized
the field, and led to the first detections of a large population of ULIRGs
at high redshift (e.g., Smail \etal 1997, Hughes \etal 1998, Barger \etal 1998).
Until recently, only the most luminous high-$z$ objects could be readily detected,
but the new ALMA interferometer will improve detection sensitivities by more than
an order of magnitude, albeit over small fields of view.  As noted above, 
submillimeter observations measure emission beyond the peak of dust
emission, where flux is declining steeply with wavelength in the Rayleigh-Jeans
part of the SED. This leads to a negative $K$ correction
so strong that it cancels out the effects of distance:  A galaxy with a given
IR luminosity will have roughly constant submillimeter flux if it is observed
at any redshift $1 < z < 10$.  By contrast, the
bolometric corrections from the observed submillimeter wavelengths to the
total IR luminosities are large and depend strongly on dust temperature.
This can lead to significant uncertainties interpreting submillimeter fluxes
from high-redshift sources, and a bias toward detecting galaxies with
the coldest dust emission.

\begin{figure}[ht]
\vspace{-1cm}
\centerline{\psfig{figure=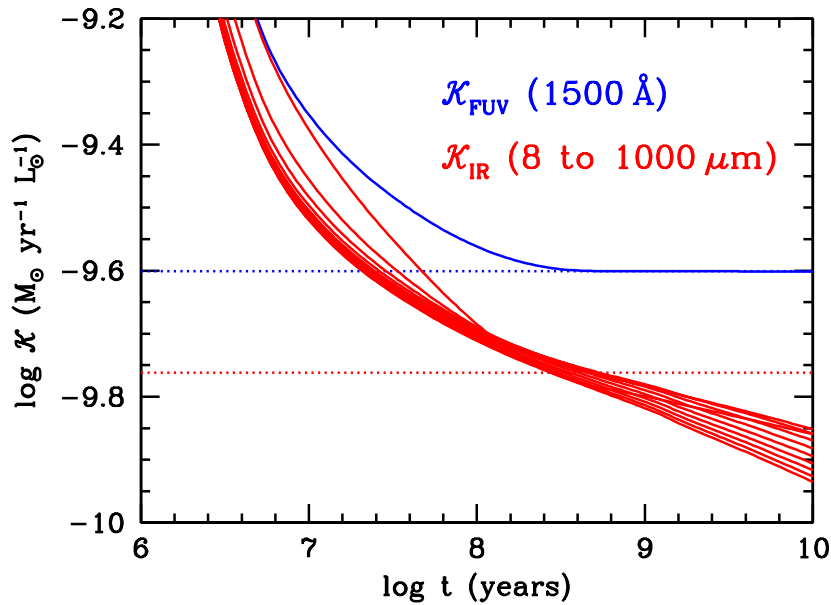,width=0.9\textwidth}}
\vspace{-1cm}
\caption{\footnotesize The SFR to luminosity ratio ${\cal K}$ for dusty galaxies, 
in solar luminosity units, calibrated with the FSPS models of Conroy \etal (2009), assuming a constant SFR, 
a Salpeter IMF, and solar metallicity.  The blue curve shows the FUV conversion factor ${\cal K}_{\rm FUV}$
(see Section \ref{sec:sfrates_uv}), and the blue dotted line indicates its asymptotic value.  The red curves
show the FIR conversion factor ${\cal K}_{\rm IR}$ computed as per Equation ~\ref{eq:KIR2}, for 1,500-\AA\ 
extinction ranging from 0.055 to 5.5 magnitudes.  The dotted red line indicates the value of 
${\cal K}_{\rm IR}$ from Kennicutt (1998), which we also adopt here.
}
\label{fig5}
\end{figure}

By analogy with Equation \ref{eq:KFUV}, we express the conversion from IR luminosity ($L_{\rm IR}$) to ongoing SFR as
\begin{equation}
{\rm SFR_{\rm IR}} = {\cal K_{\rm IR}} \times L_{\rm IR},
\label{eq:KIR}
\end{equation}
where $L_{\rm IR}$ is the IR luminosity integrated over the wavelength range from 
8 to 1,000\,$\mu$m.  Here, it is assumed that the IR emission is entirely due to recent
star formation, but in practice, AGN and older stars can contribute
to dust heating.  Furthermore, if the net dust opacity to young star-forming regions in
a galaxy is not large, and if a significant amount of UV radiation emerges, then the SFR  
derived from the IR luminosity will represent only a fraction of the total.  Hence, we write
${\rm SFR_{\rm IR}}$ in Equation \ref{eq:KIR} to indicate that this is only the dust-obscured 
component of the SFR. For this reason, some authors advocate summing the SFRs 
derived from the observed IR and UV luminosity densities, the latter {uncorrected} for extinction.
Once again, we calibrate the conversion factor ${\cal K}_{\rm IR}$ using the FSPS models of 
Conroy \etal (2009), which also incorporate dust attenuation and re-emission.  We assume 
simple foreground-screen dust attenuation from Calzetti \etal (2000), although the details 
of the dust absorption model matter relatively little.  The luminosity integrated from 
8 to 1000$\,\mu$m depends only mildly on the detailed dust emission parameters (essentially, 
the dust temperature distribution) for a broad range of reasonable values.  Because the dust 
luminosity is primarily reprocessed UV emission from young star formation, the conversion 
factor ${\cal K}_{\rm IR}$ also depends on the details of the SFH 
and on metallicity.  In practice, we may expect that galaxies with substantial extinction
and bolometrically dominant dust emission are unlikely to have low metallicities;  here 
we assume solar metallicity for our calibration.  We modify Equation \ref{eq:KIR} to account
for both the FUV and FIR components of star formation:
\begin{equation}
{\rm SFR_{tot}} = {\cal K}_{\rm FUV} L_{\rm FUV} + {\cal K}_{\rm IR} L_{\rm IR},
\label{eq:KIR2}
\end{equation}
where $L_{\rm FUV}$ is the observed FUV luminosity at 1,500\,\AA with no correction 
for extinction.  We use FSPS models with a Salpeter IMF, solar metallicity, and constant SFR 
to compute $L_{\rm FUV}$ and $L_{\rm IR}$ as a function of age for various levels of dust attenuation:
we then solve for ${\cal K}_{\rm IR}$.  Figure~\ref{fig5} shows the result of this calculation: SFR is 
expressed in units of $\sfr$, and both the FUV and IR luminosities are expressed in solar units
(with $L_{\rm FUV} = \nu L_\nu$) to display both on the same scale.  As shown
in Section \ref{sec:sfrates_uv}, the FUV emission reaches a steady state after $\sim 300$ Myr,
and for this calculation, we use the asymptotic value 
${\cal K_{\rm FUV}} = 2.5\times 10^{-10} \msun\, {\rm year}^{-1} L_\odot^{-1}$
(equivalently, ${\cal K}_{\rm FUV} = 1.3 \times 10^{-28}\,\msun$~year$^{-1}$ erg$^{-1}$~s~Hz).
Instead, $L_{\rm IR}$ increases slowly (hence, $K_{\rm IR}$ decreases) as the optical rest-frame
luminosity of longer-lived stars continues to build, some fraction of which is then absorbed by 
dust and re-emitted. This model with constant SFR and constant dust attenuation results in a modest effect
of $\sim 0.1$ dex in $\log {\cal K}_{\rm IR}$ per dex in $\log t$. However, in practice, older stars will 
likely have lower dust extinction than younger stars, thus further reducing this trend.  At ages of a few~$10^8$~years, 
$K_{\rm IR}$ depends very little on the total extinction.  Kennicutt (1998) proposed a calibration 
factor ${\cal K_{\rm IR}} = 1.73\times 10^{-10}\,\msun\, {\rm year}^{-1} L_\odot^{-1}$, which
is fully consistent with the models shown in Figure ~\ref{fig5} for an age of 300~Myr.  We adopt
that value for this review.  For luminosities measured in cgs units, we can write
${\cal K}_{\rm IR} = 4.5 \times 10^{-44}\,\msun$~year$^{-1}$ erg$^{-1}$~s.

\subsubsection{UV extinction and IR emission}
\label{sec:UVEIRE}

As noted above, dust can substantially attenuate UV emission, not only compromising its utility for measuring 
SFRs, but also producing IR  emission, which is a valuable tracer of star-formation activity.  Considerable effort
has been invested in understanding the physics and phenomenology of extinction
in galaxies (for a review, see Calzetti 2001).  In principle, the best way to account
for the effect of dust attenuation is to directly measure the energy emitted at
both UV and IR wavelengths, i.e., both the luminosity that escapes the galaxy
directly and that which is absorbed and re-radiated by dust.  This provides a
``bolometric'' approach to measuring SFRs. In practice, however,
data sensitive enough to measure FIR luminosities of high-redshift galaxies
are often unavailable. {\it Herschel} greatly advanced these sorts of observations,
but its sensitivity, although impressive, was sufficient to detect only galaxies with high 
SFRs $> 100\,\msun$~yr$^{-1}$, at $z > 2$.

For star-forming galaxies with moderate extinction at $z>1$, optical photometry
measuring rest-frame UV light is obtained much more easily than are suitably deep
FIR, submillimeter or radio data. Current observations of UV light are also typically 
much more sensitive to star formation than are those at other wavelengths (Figure ~\ref{fig1}). 
As a result, trying to infer SFRs from rest-frame UV observations alone it tempting, but this requires 
reliable estimates of dust extinction corrections. For example, Lyman break galaxies (LBGs) are a 
UV-selected population of star-forming high-redshift
galaxies.  Their selection would favor relatively low extinction, but even LBGs are quite
dusty: Reddy \etal (2012) used {\it Herschel} observations to determine that, on average,
80\% of the FUV emission from typical ($\sim L^\ast_{\rm FUV}$) LBGs at $z \approx 2$
is absorbed by dust and re-radiated in the FIR.  Many more massive galaxies
with high SFRs have greater extinction.  So-called dust-obscured galaxies (Dey \etal 2008) 
have MIR to UV flux density ratios $> 1,000$ (typically corresponding to $L_{\rm IR} / L_{\rm FUV}>100$) (Penner \etal 2012)
and are quite common, contributing 5--10\% of the SFRD at
$z \approx 2$ (Pope \etal 2008); many of these are nearly or entirely invisible
in deep optical images.

Nevertheless, the widespread availability of rest-frame UV data for high-redshift
galaxies encourages their use for measuring the cosmic SFH. 
Presently, at $z \gg 2$, there is little alternative:  Even the deepest {\it Spitzer},
{\it Herschel}, radio, or submillimeter surveys can detect only the rarest and most ultraluminous
galaxies at such redshifts. By contrast, deep optical and NIR surveys
have now identified samples of thousands of UV-selected star-forming galaxies
out to $z \approx 7$ and beyond.

Attempts to measure and correct for dust extinction in high-$z$ galaxies have generally
used the ultraviolet spectral slope (designated $\beta$) as a measure of UV reddening,
and have adopted empirical correlations between UV reddening and UV extinction.
Calzetti \etal (1994, 2000) used ultraviolet and optical spectroscopy to derive an empirical,
average dust attenuation curve for a sample of local UV-bright star-forming galaxies.
Meurer \etal (1999) (later updated by Overzier \etal 2011) used UV and
FIR data for a similar local sample to empirically calibrate the relation between UV
reddening ($\beta$) and UV extinction ($\mathrm{IRX} \equiv L_{\rm IR} / L_{\rm FUV}$, which can
be directly related to $A_{\rm FUV}$).  The reasonably tight IRX--$\beta$ relation
obeyed by the local UV-bright galaxies is broadly consistent with the Calzetti attenuation
law, hence reinforcing its popularity.  However, other local studies showed clearly that
some galaxies deviate from these relations.  Goldader \etal (2002) found that
nearby ULIRGs deviate strongly from the Meurer IRX--$\beta$ relation; these ULIRGs have very large values 
of IRX but often with relatively blue UV spectral index $\beta$.  This was interpreted to 
mean that that the observed UV light from local ULIRGs is relatively unreddened star formation
in the host galaxy that is unrelated to the bolometrically dominant star formation,
which is entirely obscured from view at UV-optical wavelengths, and detected only
in the FIR.   Instead, observations of ordinary spiral galaxies (Kong \etal 2004,
Buat \etal 2005) measured redder values of $\beta$ for a given IRX.
This is generally taken as evidence that light from older and less massive stars
contributes significantly to the near-UV emission, leading to redder UV colors
for reasons unrelated to extinction.  In general, different relative distributions
of stars and dust can lead to different net attenuation properties.  Extinction
can easily be patchy: Winds from star-forming regions can blow away dust
on certain timescales, whereas other regions that are younger or more deeply
embedded in the galaxy's ISM remain more heavily obscured.  Dust heating also
depends on geometry, leading to different distributions of dust temperatures
and different emission spectra at IR and submillimeter wavelengths.

At high redshift there are only relatively limited tests of the relation between
UV reddening and extinction.  Reddy \etal (2004, 2006, 2010, 2012) have compared
various SFR tracers (including radio, {\it Spitzer} 24-$\mu$m, and {\it Herschel} 100--160-$\mu$m
emission) to show that Calzetti/Meurer UV extinction laws are broadly appropriate for
the majority of $L^\ast$ LBGs at $z \approx 2$. However, they found evidence
for systematic deviations for galaxies with the largest SFRs ($> 100\,\msun$~year$^{-1}$),
which, similar to local ULIRGs, show ``grayer'' effective attenuation (i.e., less UV
reddening for their net UV extinction). They also found evidence for systematic deviations for the 
youngest galaxies, which show stronger reddening for their net FUV extinction, perhaps because of the 
metallicity or geometric effects that steepen the wavelength dependence
of the UV reddening function compared with results from the Calzetti law.  Assuming Calzetti attenuation, 
Daddi \etal (2007) and Magdis \etal (2010) also found broad consistency between UV-based and IR-
or radio-based SFR measurements for samples at $z \approx 2$--3 (although, see Carilli \etal 2008).  However,
studies that have selected galaxies primarily on the basis of their IR emission have
tended to find significant deviations from Meurer/Calzetti attenuation. In general, these deviations 
indicate that UV-based SFRs using Meurer/Calzetti UV slope corrections
significantly underestimate total SFRs (e.g., Chapman \etal 2005, Papovich \etal 2007).
Such studies have also found that differently-selected populations 
may obey systematically different net dust attenuation relations depending 
on the properties of the galaxies (Buat \etal 2012, Penner \etal 2012).

Therefore, we must remain cautious about SFRs derived from UV
data alone, even when estimates of UV reddening are available. Current
evidence suggests that these may work well on average for UV-bright LBGs
with relatively low reddening but may fail for other galaxies including the
most IR-luminous objects that dominate the most rapidly star-forming galaxy
population.  Star formation that is obscured by too much dust, e.g., in compact
starburst regions, will be unrecorded by UV observations, and can
be measured directly only with deep IR, submillimeter, or radio measurements.

\subsubsection{Other indicators: nebular line, radio, and X-ray emission}
\label{sec:sfr_other}

Star formation also produces nebular line emission from excited and ionized gas
in HII regions.  Recombination lines of hydrogen such as H$\alpha$ and \Lya\ 
are often used to measure SFRs, because they have a close relation
to photoionization rates that are mainly due to intense UV radiation from OB stars.
Hence, they trace massive star formation quite directly, although the presence of AGN
can also contribute to these lines.  Other lines from heavier elements such as
[OII]~3,727 \AA\ or [OIII]~5,007 \AA have been used, but they tend to have more complex dependence
on ISM conditions such as metallicity or excitation.  Emission lines are also 
subject to absorption by dust in the star-forming regions.  This is particularly true 
for \Lya, which is a resonance line, scattered by encounters with neutral hydrogen atoms. Such encounters 
can greatly increase the path length of travel for \Lya, and hence increase the 
likelihood that it may encounter a dust grain and be absorbed.  Overall, H$\alpha$
is regarded as the most reliable among the easily accessible nebular SFR tracers
(e.g., Moustakas \etal 2006).  Weaker but less extinguished hydrogen 
lines in the NIR, like Paschen~$\alpha$, can be very useful 
for measuring SFRs in dusty galaxies, but they are generally accessible 
only at very low redshift, although the JWST
will open the possibility for measuring these for significant numbers of galaxies 
at cosmological distances.

Radio emission is also correlated with star formation, as SN--accelerated
electrons emit non-thermal radiation at centimeter wavelengths;  thermal (free-free)
emission from electrons in HII regions can also contribute, particularly at higher frequencies
($>$5~GHz).   The physics is somewhat complicated and not entirely understood,
but a remarkably tight correlation is observed between radio emission and FIR 
emission in local galaxies spanning many orders of magnitude in luminosity (e.g., Condon 1992, 
Yun \etal 2001).  Radio emission is free from dust extinction, and thus offers a relatively
unbiased tracer of star formation.  However, it is difficult to obtain radio observations 
deep enough to detect ordinary star-forming galaxies at high redshift, although recent 
upgrades to the Karl G.\ Jansky VLA have significantly improved its sensitivity.  AGN can 
also contribute to radio emission, occasionally dominating for
radio-loud AGN (which are a minority population).   Radio emission
should also be suppressed at earlier cosmic epochs, as electrons should lose energy by
inverse Compton scattering off microwave background photons whose energy density increases
at high redshift.  Recent studies have found little evidence for redshift evolution 
in the FIR to radio correlation (Appleton \etal 2004; Ivison \etal 2010a,b; Sargent \etal 2010a,b; 
Mao \etal 2011).

Even X-rays have been used to trace SFRs.  X-rays are typically regarded
as a quintessential signature of AGN activity in galaxies, but they are also produced by young stellar
populations, notably by X-ray binaries.  In the absence of an AGN, X-ray emission may
be measured from individual star-forming galaxies out to $z \approx 1$ in the deepest 
{\it Chandra} fields, and stacking measurements have been used to reach fainter fluxes
in studies of UV-selected galaxies, with detections at $1 < z < 4$ and upper limits 
at higher redshifts  (Reddy \& Steidel 2004; Lehmer \etal 2005; Laird \etal 2005, 2006; 
Basu-Zych \etal 2013).  However, the proportionality between X-ray luminosity and SFR 
may vary with stellar population age and other parameters that could affect 
the mix of low- and high-mass X-ray binaries present in a galaxy; various calibrations
that differ significantly have been published (e.g., Ranalli \etal 2003, Persic \etal 2004).
Overall, because most of the cosmic X-ray background arises from AGN (for a review, see Brandt \& Hasinger 2005), 
the value of using X-rays to measure the cosmic SFH seems limited (we do not discuss this method further).

\subsection{``Weighing'' Stellar Mass}
\label{sec:stellarmass}

Whereas hot young stars emit most of their energy at UV wavelengths, the cooler low-mass stars 
that dominate the stellar mass of a galaxy emit most of their light
at red optical and NIR wavelengths.   If we examine the SED 
of an evolving SSP from ages older than $10^9$~years,
the bulk of the luminosity (in $\lambda f_\lambda$ energy units) is emitted in a broad
plateau between 0.4 and 2.5$\,\mu$m, peaking at $\sim 1\,\mu$m for ages
$> 2$~Gyr. (In $f_\nu$ flux density or AB magnitude units, the SED peak
is at approximately 1.6$\,\mu$m, where there is a minimum in the H$^-$ opacity of cool 
stellar atmospheres.) The effects of dust extinction are also greatly
reduced at NIR wavelengths: For Calzetti attenuation, the
extinction (in magnitudes) $A$ in the $K$-band is 10 times smaller than
that in the $V$-band and 25 times smaller than that at 1,600 \AA.

The luminosity, and hence the mass-to-light ratio, of a stellar population
evolves very steeply with time at UV and blue wavelengths: Young stars evolve quickly 
off the main sequence but more slowly at red and NIR
wavelengths.  Therefore, observations in the NIR rest frame more closely 
trace the integrated stellar mass of a galaxy, but we cannot neglect the effects of evolution: 
The flux at 1$\,\mu$m still changes by more than an order of
magnitude as a stellar population ages from 0.1 to 10~Gyr (see, e.g., figure ~9 from
Bruzual \& Charlot 2003).  Therefore, we need to do more than simply measure 
the NIR luminosity to infer a stellar mass.

In effect, astronomers use the colors or SED of
a galaxy to infer the expected mass-to-light ratio at some wavelength
(preferably in the red or NIR) and then multiply the observed
luminosity by $M/L$ to estimate the stellar mass ($M_\ast$).  The most common method
is to fit spectral templates generated by stellar population synthesis models
to broadband photometry in whatever bands are available that span
rest-frame UV to NIR wavelengths, where stellar photospheric emission
dominates the galaxy light.  Generally speaking, researchers generate a large suite of models 
that span a wide range of stellar population parameters, including
the past SFH, age, metallicity, and dust attenuation.
The IMF is typically fixed, because there is almost no photometric signature
that can usefully constrain it.   The suite of models is redshifted to match
a galaxy of interest. The models are then convolved by the filter bandpasses to generate synthetic 
broadband fluxes that are fit to the photometry, allowing the luminosity
normalization to vary and minimizing $\chi^2$ or some other likelihood parameter.
The unnormalized models have a specified unit mass; therefore, the 
normalization of the best-fitting model provides the best estimate of
the galaxy's stellar mass, given the range of input parameters that were allowed.

In principle, this method can be used to constrain other stellar population
parameters such as the galaxy's age, SFRs, or the degree of extinction
that is present.  In practice, the fitting results for various parameters are
often quite degenerate.  For example, age, extinction, and metallicity all
affect the integrated colors of a galaxy. As a result, the derived values of
these parameters tend to be highly covariant: A galaxy may be red because
it is old, dusty, or very metal rich.  With very good photometry, particularly
spanning a large range of wavelength and with many bandpasses that can more
accurately sample the detailed spectral shape (e.g., measuring relatively
sharp age-sensitive features such as the Balmer or 4,000-\AA\ breaks), these
constraints can be improved, but it is hard to avoid significant degeneracies.
Careful practitioners may consider joint probability distributions for models
that fit with acceptable likelihood.  The stellar mass tends to be the best-constrained
parameter, largely because the degeneracies in other parameters all tend to affect
the net $M/L$ of the model in similar ways.  Redder colors from age,
dust, or metallicity all tend to affect $M/L$ to a similar (but not identical)
degree.  Whereas parameters such as age or reddening may be individually uncertain,
the net $M/L$ of acceptable models does not vary so much. Thus, the total
mass is well constrained. Many papers have discussed stellar population modeling uncertainties 
in estimating galaxy masses; these are very thoroughly reviewed by Conroy (2013). 

Other than the choice of the IMF, the largest uncertainty that affects the derived stellar 
mass is usually the necessarily imperfect knowledge of the galaxy's past SFH. 
Fundamentally, more recently formed stars can easily outshine
older stars and dominate the observed light, even at red wavelengths.  The observed
photometry may be dominated by the younger starlight, even though the actual galaxy mass
may be dominated by older stars that are lost in the glare of the younger stars (``outshining'', e.g., 
Papovich et al. 2001, Maraston et al. 2010)
and thus have little impact on the choice of the best-fitting models. Therefore, the model fitting 
often underestimates the age of the galaxy or the potential
contribution of older stars to the mass, and it may also underestimate the mass.
If the actual SFH were well known (which is almost never the case in practice) this might not be a problem.
For example, the models used to fit
the photometry are often assumed to have smoothly-varying SFRs, 
but the actual SFHs of real galaxies can be complex and nonmonotonic, fluctuating with time and perhaps punctuated by 
short-duration bursts. Even if very large suites of models with complex SFHs are
considered, outshining tends to ensure that recently formed stars drive the model fitting, whereas
the mass in older stars is poorly constrained. This effect generally leads to underestimation of galaxy stellar 
masses (Pforr et al. 2012). Realistically constraining the distribution of allowable
past SFHs for real galaxies, especially at high redshift, remains a basic limitation when deriving stellar masses.

In practice, these SFH degeneracies are largest for galaxies with recent star formation. For galaxies that 
have not formed stars in a long while (say $>1$~Gyr) or for which the current SFR is small compared with 
the stellar mass (often quantified by the sSFR), the outshining is small and, thus, so is the resulting systematic 
uncertainty on $M/L$.  Therefore, stellar masses for present-day elliptical galaxies, which are old with little or no 
ongoing star formation, or for ordinary spiral galaxies such as the Milky Way tend
to be reasonably well-constrained, whereas those for very actively star-forming
galaxies are less certain.   As an example, Papovich \etal (2001) fit models
to {HST} WFPC2 and NICMOS photometry for faint LBGs at $z \approx 2.5$ in the Hubble Deep Field North (HDF-N).  
When using models with smoothly-varying SFHs, they found stellar mass uncertainties
to be $\sigma(\log M_\ast) < 0.5$~dex, with typical uncertainties of 0.25~dex,
i.e., less than a factor of 2.  However, if they considered ``maximal $M/L$''
models, which allowed for as much older stellar mass as possible within the
$\chi^2$ fitting constraints, formed at $z = \infty$, the masses could in
principle be as much as 3 to 8 times larger. In practice, such extreme models
seem unlikely. Moreover, the early work of Papovich \etal (2001) used photometry only out to the $K$-band
or rest-frame wavelengths $\sim 6,000$ \AA\ at $z = 2.5$.  Today, deep {\it Spitzer}
IRAC photometry routinely measures fluxes for high-$z$ galaxies at redder rest-frame
wavelengths and can significantly improve stellar mass constraints. Nevertheless, even with
the best data from {\it Spitzer} (or the JWST in the future) the effects of outshining
fundamentally limit our certainty about stellar mass estimates for individual
objects. These effects can be reduced only if reasonable prior assumptions can more
tightly constrain the range of allowable SFHs. 

Interestingly, at very high redshifts some of these SFH uncertainties are reduced,
simply because the Universe is much younger.  At $z > 6$, the Universe is less than
1~Gyr old, and the oldest stars in the galaxies must be younger than that;  this sets a cap
on $M/L$ for a hypothetical unseen old population and thus on its possible contribution
to the total stellar mass.  Curtis-Lake \etal (2013) provided a recent and detailed 
discussion of stellar population modeling uncertainties for galaxies at $z \sim 6$.

Additionally, the practitioners who create stellar population models have not reached complete consensus: 
Questions regarding evolutionary tracks, the contributions of certain stellar sub-populations, and the behavior of stellar
populations at low and high metallicities remain topics of debate or are poorly calibrated by observations.  One 
widely recognized example of such uncertainties
was highlighted by Maraston (2005), whose models featured significantly greater
contributions of emission from thermally pulsating asymptotic giant branch
(TP-AGB) stars to the red and NIR rest-frame light at SSP ages
between a few hundred million years and $\sim 2$ Gyr.   The enhanced red luminosity led
to lower $M/L$ at these wavelengths and redder colors, with potentially quite 
significant effects (factors of 2 or more) in derived stellar masses for galaxies
dominated by stars in this age range.  Although such populations may not dominate
in most present-day galaxies, at $z \approx 2$--4 when the Universe 
was only a few billion years old their role must be accurately modeled to ensure proper estimates of stellar masses.  Maraston \etal (2006)
found that this could reduce derived stellar masses by $\sim 60$\% on average for
$K$-band-selected star-forming galaxies at $z \approx 2$ compared with results computed using the popular models of Bruzual \& 
Charlot (2003). Although Bruzual \etal (2013) released new models in 2007 that featured enhanced TP-AGB emission, they have 
argued in recent conference presentations for weaker TP-AGB emission more similar to that in the older models. 
Given the lack of completely satisfactory way to compute this contribution on theoretical principles, a lot hangs on the 
sparseness of data available to empirically calibrate the emission and evolution of TP-AGB stars.

\section{TRACING THE GALAXY EMISSION HISTORY WITH LARGE SURVEYS}
\label{sec:surveys}

\begin{figure}[ht]
\centerline{\psfig{figure=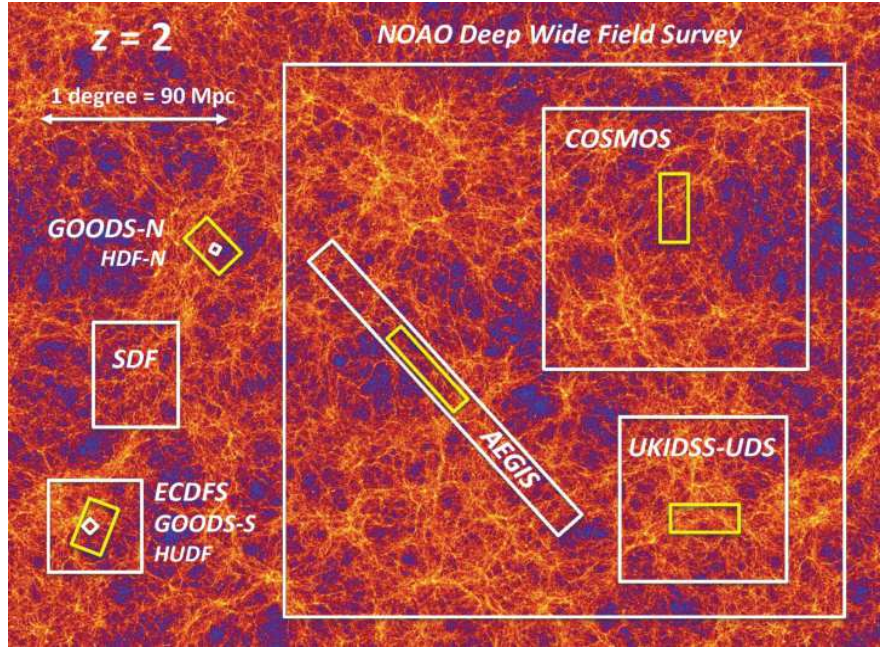,width=0.9\textwidth}}
\vspace{0.3cm}
\caption{\footnotesize Relative sizes of the regions on the sky observed in several important surveys of the distant Universe. 
The two Great Observatories Origins Deep Survey (GOODS) fields, the Subaru Deep Field (SDF) and the Extended Chandra Deep Field 
South (ECDFS), are shown on the left. Very-deep surveys such as the Hubble Deep Field North (HDF-N) and the Hubble Ultradeep Field (HUDF)
[Advanced Camera for Surveys (ACS) area shown], which are embedded within the GOODS fields, 
can detect fainter galaxies, but cover only very tiny regions on the sky. Other surveys such as the Cosmic Evolution Survey (COSMOS), 
the UK Infrared Deep Sky Survey (UKIDSS), the Ultradeep Survey (UDS), the All-Wavelength Extended Groth Strip International Survey (AEGIS) 
and the National Optical Astronomy Observatory (NOAO) Deep Wide Field Survey cover wider regions of the sky, usually to shallower 
depths, i.e., with less sensitivity to very faint 
galaxies. However, they encompass larger and perhaps more statistically representative volumes of the Universe.  The yellow boxes indicate 
the five fields from the Cosmic Assembly Near-Infrared Deep Extragalactic Legacy Survey (CANDELS), each of which is embedded within another 
famous survey area.  The image in the background shows a cosmological $N$-body simulation performed within the MultiDark project 
(see http://www.multidark.org/MultiDark/), viewed at $z = 2$, more than 10 Gyr ago. The colors represent the matter density 
distribution in a slice 43-Mpc thick, or $\Delta(z) = 0.03$ at that redshift, and all lengths are given in comoving units for $h = 0.7$. 
Small surveys may sample under- or over-dense regions, whereas larger surveys can average over density variations, but may not be sensitive 
to the ordinary, relatively faint galaxies that are most numerous in the Universe.  Averaging over redshift intervals that are greater 
than that shown in the background figure will smooth over density variations, but for any redshift binsize cosmic variance will be smaller 
for wider surveys or when a survey is divided into fields sampling multiple, independent sightlines.
}
\label{fig6}
\end{figure}

Over the past 18 years, a sea of published measurements of the
cosmic SFRD or SMD at many different redshifts have used different data sets and methods.  
Much of the observational work has been carried out in deep survey fields that have 
accumulated outstanding multiwavelength data for this purpose and cover different angular 
scales to different depths (Figure ~\ref{fig6}).  We dot not attempt a comprehensive review
of this literature. Instead, we highlight key data sets and analyses from roughly the 
first decade of cosmic SFR measurements (1996-2006) and then focus 
on the most important or extensive recent measurements that we use in our analysis
in Section \ref{sec:obs_to_para}.  Hopkins (2004) and Hopkins \& Beacom (2006) provided an extensive 
compilation of SFRD measurements up to 2006, whereas Wilkins \etal (2008a) summarized SMD 
derivations through 2007.  Other authors have also compiled these data more
recently (e.g., Behroozi \etal 2013).

The number of papers that present measurements of the cosmic SFR 
history vastly exceeds the number of different data sets that have been used for this purpose 
because certain well-trodden surveys such as the HDF-N, the Hubble Ultradeep Field (HUDF), GOODS, and COSMOS have been 
used repeatedly by many groups or by the same groups who continue to refine their analyses 
or add new observational information.  One should thus be cautious compiling results from 
many different studies: Although the analyses are independent, the data used and 
the actual galaxies measured may not be different.  For example, the SFH of GOODS-South 
and COSMOS are particularly well studied, but true cosmic variance due to clustering 
in those fields will not cancel out from one analysis to another.

\subsection{UV Surveys}
\label{sec:surveys_uv}

The largest number of analyses of the cosmic SFRD have used 
rest-frame ultraviolet continuum measurements, largely because the method is quite 
sensitive (see Figure ~\ref{fig1}) and can be applied over a very broad range of redshifts.
Rest-frame FUV (1,500\AA) SFRs at $1.4 < z < 6$ can be measured using optical 
photometry that is (relatively) easily obtained with ground-based or HST imaging.
The heavy use of photometric redshifts in recent years has led to much work based
on imaging data alone, with little or no spectroscopy. However, in popular
deep survey fields such as GOODS or COSMOS, the photo-$z$' values are typically well calibrated
thanks to the widespread availability of thousands of spectroscopic measurements.
At $z < 1$, rest-frame UV measurements ideally require space-based telescopes such as  
{GALEX} or {HST} to reach rest-frame wavelengths near 1,500\AA, but several 
studies have used blue or $U$-band imaging to sample mid-UV wavelengths (e.g., 2,800\,\AA) 
at $z < 1$ instead.

The modern era of SFRD measurements arguably began with
the analysis of Lilly \etal (1996), who were the first to combine a large and deep 
(for its time) spectroscopic redshift survey with multiwavelength photometry and to 
derive LFs and luminosity densities at several different rest-frame 
wavelengths, including the rest-frame UV.  The Canada-France Redshift Survey (CFRS)
was carried out using the 4-m {\it Canada-France-Hawaii Telescope} and mainly surveyed the Universe out to $z < 1$.
The available $BVIK$-band photometry permitted direct measurement of 2,800-\AA\ rest-frame
luminosities at $z > 0.5$, and down to $z \approx 0.3$ with modest spectral extrapolation.
Lilly \etal found that the 2,800-\AA\ luminosity density declined by approximately one order
of magnitude from $z=1$ to the present, which they interpreted as a steep decline in
the SFRD. 

Madau \etal (1996) used the then-new HDF observations to extend 
this analysis to much higher redshift. They employed color-selected LBG 
samples at $\langle z \rangle = 2.75$ and 4.  The deep {HST} WFPC2 photometry allowed 
luminosities to be measured at 1,500\AA\ in the rest frame, reaching fainter than 
contemporaneous ground-based LBG data at $z \approx 3$ from Steidel \etal (1996) 
and thus integrating further down the LF. Madau \etal (1996) quoted 
only lower limits for the SFRD, without extrapolation to fainter luminosities 
(relatively small, given the depth of the HDF imaging) and without correction 
for dust absorption (significant, but at the time little known).  Later analyses
(e.g., Sawicki \etal 1997, Madau \etal 1998b, Steidel \etal 1999) fit Schechter LFs 
to the photometric samples to extrapolate to total UV luminosity densities. 
Connolly \etal (1997) and Pascarelle \etal (1998) combined the optical HST imaging of the HDF with ground-based 
NIR data to improve photometric redshift analyses in the ``redshift desert'' at $1 < z < 2$, between the regime 
of Lilly et al.\ (1996) and that of Madau et al.\ (1996)." Taken together, the 
HDF measurements at $z > 1$ and the CFRS measurements at $z < 1$ suggested a general ``rise and fall'' 
picture of a UV luminosity density and, by inference, the cosmic SFH that peaked somewhere between 
$z \approx 1$ and 2.

Various surveys subsequently extended this finding using other data sets or different analyses.
Several groups reanalyzed the HDF (and later the HDF-South) data in various ways or made 
use of deeper spectroscopic surveys with the {\it Keck} telescope.  
Cowie \etal (1999) and Wilson \etal (2002) combined {\it Keck} spectroscopy in several 
fields with deep $U$-band imaging to measure shorter rest-frame UV wavelengths 
(,2000-2,500\AA) at $z < 1$ than were probed in the CFRS analysis of Lilly \etal (1996) 
and derived a shallower rate of decline in the SFRD.  Wolf \etal (2005) used 17-filter 
intermediate and broadband imaging to obtain high-quality photometric redshifts 
at $z < 1.25$ in the Extended Chandra Deep Field South and analyzed the 2,800-\AA\ 
luminosity density evolution.

Ideally, UV rest-frame observations at $z < 1$ should be done from space telescopes
to sample shorter UV wavelengths than those used in the ground-based
studies by Lilly, Wilson, Wolf, and others.  In early work, Treyer \etal (1998) and 
Sullivan \etal (2000) used the FOCA balloon-born UV telescope to measure 2,000-\AA\ 
luminosity densities at $z \approx 0.15$.  {GALEX} has since provided vastly more
FUV data, including both wide-area and deeper surveys.  Wyder \etal (2005) 
combined {GALEX} all-sky imaging survey data with distances from the 
2dF Galaxy Redshift Survey (2dFGRS) over 56~deg$^2$ to measure local 
($z \approx 0.055$) LFs at 1,500\AA\ and 2,300\,\AA. Budav\'{a}ri \etal (2005)
analyzed a similar total sky area using somewhat deeper {GALEX} data and SDSS-derived 
photometric redshifts to compute LFs and densities at $z < 0.25$.
Salim \etal (2007) and Robotham \& Driver (2011) have since analyzed much larger 
{GALEX} data with SDSS data to cover as much as 830 deg$^2$.  Robotham \& Driver presented 
a straightforward derivation of the UV luminosity function (UVLF) and luminosity density, 
whereas Salim \etal used {GALEX} photometry as one ingredient to derive SFRs 
and the total SFRD.

Arnouts \etal (2005) combined much deeper {GALEX} observations with spectroscopy 
from the VIMOS VLT Deep Survey (VVDS) and derived 1,500-\AA\ rest-frame LFs
at $0.2 < z < 1.2$.  Schiminovich \etal (2005) integrated these LFs 
to determine the 1,500-\AA\ luminosity density, measuring evolution 
consistent with $\rho_{{\rm FUV}} \propto (1+z)^{2.5}$.    This remains the most
direct and frequently cited {GALEX} study of FUV luminosity densities at $z < 1$, which is 
somewhat surprising, as there are many more deep {GALEX} observations in 
fields with extensive spectroscopy (COSMOS, AEGIS, GOODS, etc.). Yet, to to our knowledge,
there have been no other published LFs. The Arnouts/Schiminovich
analysis is admirable, but it used only $\sim 1,000$ galaxies with spectroscopic
redshifts over the whole range $\Delta z = 1$ in a single field covering 0.5~deg$^2$. Hence, it may be 
subject to cosmic variance issues.  This area is quite ripe for further exploitation of existing archival data.  

{HST} is the only other modern space telescope with UV capabilities, particularly 
with the UVIS channel of the WFC3 camera.  Until recently, only one 50-arcmin$^2$ field
in GOODS-South has been surveyed to interesting depths to study distant galaxies 
(Hathi \etal 2010, Oesch \etal 2010).  At $z < 1.4$, where these data measure FUV
rest-frame emission, the survey volume and counting statistics are poor.  These data
have also been used for Lyman break color selection at $z \approx 1.5$--2.  
The analyses generally support relatively steep UVLFs with $\alpha < -1.5$ 
with large uncertainties.  Expanded WFC3 UVIS observations in the HUDF and GOODS-North 
field have been recently completed and should improve the existinf measurements somewhat.  
A recent WFC3-UVIS survey of gravitationally lensed galaxies behind the massive cluster 
Abell~1689 (Alavi \etal 2014) has been used for Lyman break color selection to 
unprecedentedly faint FUV luminosities at $z \approx 2$ to $M_{AB} \approx -13$, 
or $\sim 1,000$ times fainter than $L^\ast$ at that redshift. Alavi \etal (2014) found 
no turnover to the LF down to those limits and measured a faint-end 
slope $\alpha = -1.56 \pm 0.13$.

Cucciati \etal (2012) analyzed a larger, deeper and more complete spectroscopic 
sample in the same VVDS survey field studied by Arnouts \etal and Schiminovich \etal. They fit SED 
models to multiband photometry from the $U$- to the $K$-bands to extrapolate flux density 
measurements to FUV rest-frame wavelengths at $z < 1.4$.  They demonstrated consistency with 
the {GALEX} luminosity densities from Schiminovich but did not make direct use of the {GALEX} data.  
Although the largest and deepest spectroscopic sample used to derive UVLFs at $0 < z < 2$,
it is based on only one sightline and requires SED extrapolation to rest frame 1,500\AA\ at lower redshifts.
Tresse \etal (2007) presented a similar, earlier analysis using shallower spectroscopy in two VVDS fields.

Over many years, Steidel and collaborators have carried out an extensive campaign 
of {\it Keck} spectroscopy for Lyman break-selected galaxies, especially at $z \approx 2$
and 3 (Steidel \etal 2003, 2004).  Their survey covers many widely-spread sightlines and provides 
excellent control over cosmic variance.   Among several LF analyses 
from these data, the most recent and definitive are those of Reddy \etal (2008) and
Reddy \& Steidel (2009).  These still rely on deep photometric color-selected samples
to probe the faint-end of the LF, but with a degree of spectroscopic
confirmation and calibration for brighter galaxies that is unmatched by any other 
survey.   The use of UV rest-frame selection means that any LBG-based study will
miss heavily dust-obscured star formation at these redshifts, but as a measure of 
the evolving UV luminosity density the LBG surveys have provided the most robust 
method to date.

At higher redshifts, $4 < z < 7$, deep {HST} observations (discussed below) have dominated 
surveys for LBGs in recent years, but several ground-based imaging programs 
have made significant contributions, particularly surveying wider areas at (relatively)
shallower depths to constrain the bright end of the LF. The {\it Subaru}
telescope and its SuPrime Cam imager have been particularly important in this respect,
although deep IR imaging from {UKIRT} and the {VLT} have also been used.
Notable examples (not an exhaustive list) include Ouchi \etal (2004), Yoshida \etal (2006),
Iwata \etal (2007) and McLure \etal (2009) at $z = 4$ to 6 and Ouchi \etal (2009), 
Castellano \etal (2010a,b), Bowler \etal (2012), and Tilvi \etal (2013) at $z \approx 7$.

The installation of the Advanced Camera for Surveys (ACS) enabled substantially more 
efficient optical {HST} imaging that covers fields much wider than the original HDF.  
ACS also offered significant gains in sensitivity at the reddest wavelengths,
making Lyman break selection practical out to $z \approx 6$. Two major ACS surveys 
led to new derivations of the cosmic SFRD. GOODS (Giavalisco \etal 2004b) observed two independent fields with combined
area $> 60$ times larger than the HDF through four filters.  This 
provided a sample of several thousand Lyman break candidates at $z\approx 4$ and 
of order 1,000 at $z \approx 5$, reaching significantly fainter than $L^\ast$ and 
permitting robust characterization of the luminosity density. LBG selection 
at $z \approx 6$ was less secure from GOODS ACS data alone, as it was based on 
a single color ($i-z$) and could sample only relatively bright galaxies.  The HUDF (Beckwith \etal 
2006) observed a single ACS pointing ($\sim $ 11~arcmin$^2$) located within the GOODS-South region with very long 
exposure times and reaching fainter than the original HDF and with better sensitivity at higher redshifts.
Both GOODS and the HUDF have been repeatedly revisited with new observations from {HST}
over the years to add deeper optical imaging as well as NIR data, first from 
NICMOS (Thompson \etal 2006, Conselice \etal 2011) and later with WFC3 in the HUDF09 
and HUDF12 programs (Bouwens \etal 2011b, Ellis \etal 2013) and the Cosmic Assembly Near-Infrared 
Deep Extragalactic Legacy Survey (CANDELS) of several premier deep survey 
fields including GOODS (Grogin \etal 2011, Koekemoer \etal 2011).
These IR observations make Lyman break selection at $z \approx 6$ far more robust
and extend the method out to $z \approx 8$, with a handful of unconfirmed
candidates identified out to $z \approx 12$.  

Early analyses of the GOODS data (Giavalisco \etal 2004a) found evidence 
for relatively mild evolution of the UV luminosity density from $2 < z < 5$ and 
clear evidence that there were fewer high-luminosity galaxies at $z \approx 6$
(Stanway \etal 2003, Dickinson \etal 2004).   Subsequent studies have 
repeatedly mined the combined GOODS+HUDF observations, using deeper data and more 
rigorous analyses.  At least 20 papers about high-redshift LFs 
using WFC3 data in the HUDF and GOODS/CANDELS have been published since 2010,
augmenting at least a dozen others pre-WFC3. We cannot attempt to review
them all here, but they have convincingly demonstrated that the UVLF evolves 
significantly at $z > 4$.  The current consensus is that this is primarily 
luminosity evolution, at least at $4 < z < 7$, with $L^\ast$ brightening over time
(e.g., Bouwens \etal 2007, Grazian \etal 2011, Bouwens \etal 2012b).  As a result, 
the number density of bright LBGs increases rapidly with time, more quickly than does 
the integrated luminosity density.  Analyses using the CANDELS and HUDF09+12 
NIR imaging point to continued evolution out to $z = 8$ and perhaps beyond 
(Oesch \etal 2012, Yan \etal 2012, Lorenzoni \etal 2013, Schenker \etal 2013), 
although it would be prudent to recall that only a handful of galaxies at 
$z \approx 7$ have reasonably secure spectroscopic confirmation and none at $z > 7.5$.
There is broad agreement that UV spectral slopes for LBGs are bluer at $z > 4$
than at lower redshifts (Bouwens \etal 2012a, Finkelstein \etal 2012a, Dunlop \etal 2013),
which has implications for their dust extinction and total SFRD.
Most of these studies have examined the faint-end slope of the LF, 
measuring very steep values, in some cases approaching 
or exceeding the divergent value $\alpha = -2$ (Bouwens \etal 2012b, McLure \etal 2013).
Several studies have also extended SFRD analyses to $9 < z < 12$, using data
from the HUDF (Bouwens \etal 2011a, Ellis \etal 2013) or from lensing cluster studies 
(Coe \etal 2013). Although still in flux as better data accumulate, these measurements have 
considerable significance for the earliest phases of galaxy 
evolution and for the reionization of the IGM (e.g., Robertson \etal 2013),
but relatively little impact on the global star-formation budget of the 
Universe.   According to current estimates, only $\sim 1\%$ of the cosmic SMD present 
today was formed at $z > 6$.

\subsection{Infrared Surveys}
\label{sec:surveys_ir}

{IRAS} enabled the first measurements of the local FIR luminosity function 
(FIRLF) (Lawrence \etal 1986, Soifer \etal 1987, Saunders \etal 1990, Sanders \etal 2003).
These were typically measured either at 60-$\mu$m observed wavelength or using FIR 
luminosities integrated over a broader wavelength range. They were also generally extrapolated from the
measured {IRAS} fluxes using fitting formulas.  Although various representations of 
FIR luminosity have been adopted, here we consider $L_{\rm IR}$ as the luminosity
integrated over the range 8--1,000 $\mu$m, which encompasses most of the bolometric
luminosity of dust emission from nearly all sources of interest.  The longest wavelength
{IRAS} band was at 100$\,\mu$m, but {AKARI} extended all-sky FIR measurements out to 
160$\,\mu$m to provide more reliable measurements of the bolometric luminosity and reduced
bias against galaxies with cold dust temperatures.  Goto \etal (2011a) and (2011b) reanalyzed 
the local IRLF incorporating {AKARI} data. Despite differences in detail, the results 
are largely consistent with previous {IRAS} measurements in the luminosity range of overlap. 
LFs have also been measured for {IRAS} samples selected at 12 and 25 microns 
(Rush \etal 1993, Shupe \etal 1998).  LFs at longer IR wavelengths 
were measured for (rather small) local samples with {ISO} [90$\,\mu$m (Serjeant \etal 2004); 170$\,\mu$m (Takeuchi \etal 2006)], 
{\it Herschel} (250--500$\,\mu$m) (Dey \etal 2010, Vaccari \etal 2010), and with ground-based (sub)-millimeter 
observations generally for {IRAS}-selected samples [1.2mm (Franceschini \etal 1998); 850$\,\mu$m (Dunne \etal 2000)].  
New, large-area measurements using the largest {\it Herschel} surveys (e.g., H-ATLAS, covering 
550 deg$^2$ with observations from 100 to 500$\,\mu$m) have not yet appeared in the literature.

Nearly all studies found that the bright end of the IRLF cuts off less sharply than does 
the exponential used in the Schechter function.  This has typically been modeled either as 
a double power law (e.g., Lawrence \etal 1986, Sanders \etal 2003) or as a combined 
log-normal and power law (e.g., Saunders \etal 1990).   Locally, the bright end of the 
IRLF is dominated by galaxies with warmer dust temperatures, which tend to be starburst 
galaxies and dusty AGN (Saunders \etal 1990).   Several studies have measured a 
steep faint-end slope $\alpha$ (d$N/$d$L \propto L^\alpha$), 
e.g., $\alpha = -1.6$ (Sanders \etal 2003), $\alpha = - 1.8$ (Goto \etal 2011a), and $\alpha = -1.99$ (Goto \etal 2011b).
However, other studies have found flatter distributions, e.g., $\alpha = -1.2$ to -1.0 (Saunders \etal 1990, Takeuchi \etal 2003, Vaccari \etal 2010).
In practice, the faint-end has not been well-sampled locally except in a few
of the {IRAS} surveys.  Future analysis of the widest {\it Herschel} surveys may help 
resolve this.

In local, relatively quiescent spiral galaxies such as the Milky Way, 
more than half of the FIR luminosity is believed to arise not from dust in 
active star-forming regions, but from dust in the general ISM that is heated by ambient starlight from intermediate- and older-age stellar populations
(Lonsdale Persson \& Helou 1987, Sodroski \etal 1997). The luminosity of the Milky Way is 
typical ($L_{\rm IR} \approx 10^{10} L_\odot$ (Sodroski \etal 1997) compared with 
the knee of the local IRLF at $L^\ast_{\rm IR} = 10^{10.5} L_\odot$ (Sanders \etal 2003).  
This implies that a significant fraction of the local IR luminosity density is not 
the direct result of young star formation. Thus, it may not be a good measure of 
the global SFR today. At higher redshift when the specific SFR of typical galaxies was much larger
and the net dust extinction to star-forming 
regions was, on average, larger (see Section \ref{sec:sfrd}), we may expect the IRLF and its integral
to more reliably trace the total SFRD.  However, Salim \etal (2009) suggested that, even at higher 
redshift ($z \approx 0.7$), intermediate-age stars may significantly
contribute to MIR dust emission observed by {\it Spitzer} at $24\,\mu$m.

The deepest surveys with {ISO} at 15$\,\mu$m detected a few hundred galaxies,
mainly at $z \leq 1$, in the HDF and a few other deep survey regions
where spectroscopic and photometric redshifts were available 
(Rowan-Robinson \etal 1997, Flores \etal 1999, Aussel \etal 1999).
Analyses of these generally agreed that the emission from dusty star formation
increased steeply with redshift, although statistics were generally too poor to 
construct redshift-dependent LFs. 
Chary \& Elbaz (2001) used measurements from {ISO} and SCUBA as well as constraints 
from the FIR background as measured by the {COBE} satellite to constrain a model for the evolution of the 
cosmic SFH. Their model exhibited a sharp
decline in the SFRD by a factor of 10 or more from $z \approx 0.8$ to the present,
with a plateau of nearly constant star formation at $0.8 < z < 2$.   At higher redshifts,
the SFRD was more poorly constrained.  Submillimeter sources placed 
a rough lower bound, whereas the cosmic infrared background (CIRB) set an upper limit.  
Acceptable solutions ranged from flat evolution to an increase by a factor of $\sim 10$ 
from $z = 4.5$ to 2.

{\it Spitzer} greatly enhanced the sensitivity and mapping efficiency
for deep IR observations, particularly at 24$\,\mu$m where the beam size 
(FWHM~$\approx 5^{\prime\prime}.7$) was small enough to enable relatively straightforward 
association with optical counterparts.  {\it Spitzer} also observed in an era when very large 
spectroscopic redshift surveys were available or underway and when photometric redshift 
techniques were well established.  Le~Floc'h \etal (2005) produced an early, 
seminal analysis of 24-$\mu$m sources at $0.3 < z < 1.2$ in the Extended Chandra Deep Field South.  Integrating 
over derived IRLFs, they inferred an evolution of the IR 
luminosity density proportional to $(1+z)^{3.9 \pm 0.4}$, significantly steeper than
the evolving UV luminosity density, $\rho_{\rm FUV} \propto (1+z)^{2.5}$ (Schiminovich \etal 2005). 
With strong luminosity evolution, the fraction of the 
IR luminosity density produced by LIGs and ULIGs evolved 
even more steeply:  Le~Floc'h \etal (2005) found that galaxies with $L_{\rm IR} > 10^{11}\,L_\odot$ 
produced $70\% \pm 15\%$ of the IR luminosity density at $z \approx 1$,
compared with 5--15\% today, depending on the adopted local LF. Several analyses of LFs, mainly at $z < 1$, using shallower
{\it Spitzer} data covering significantly wider areas have also been published. These includes 
Babbedge \etal (2006) and Rujopakarn \etal (2010) at 24$\,\mu$m and 
Patel \etal (2013) at 70$\,\mu$m and 160$\,\mu$m.  The work by Rujopakarn \etal (2010) 
is particularly notable for its combination of (relatively) wide area (9 deg$^2$), 
extensive spectroscopic redshifts (4,047 galaxies with $z \leq 0.65$), and 24-$\mu$m
sensitivity (0.27~mJy, sufficient to reach $\sim L^\ast$ out to $z = 0.65$), making
it arguably the best bridge study to date between local ({IRAS} and {AKARI}) measurements
and deep-field studies at $z \geq 1$.

Several studies extended {\it Spitzer} 24-$\mu$m-based LF measurements 
to higher redshifts, $z \approx 2$ to 2.5 (P\'{e}rez-Gonz\'{a}lez \etal 2005, Caputi \etal 2007,
Rodighiero \etal 2010). Such studies primarily use deeper 24-$\mu$m
data and fainter spectroscopic and photometric redshifts available in the two GOODS 
fields. [P\'{e}rez-Gonz\'{a}lez \etal (2005) used
shallower 24-$\mu$m data, whereas Caputi \etal (2007) and Rodighiero \etal (2010)
employed deeper data from the GOODS team.  Rodighiero \etal (2010) also incorporated 
relatively shallow {\it Spitzer} 24-$\mu$m data covering 0.85~deg$^2$ in one of the VVDS 
redshift survey fields.] Dependenig on the analysis, these studies all found flatter
IR luminosity density evolution at higher redshifts with modestly lower or higher 
$\rho(L_{\rm IR})$ at $z = 2$ than at $z = 1$. All three studies also found that the characteristic IR luminosity $L^\ast_{\rm IR}$ brightened
further at $z > 1$, such that ULIRGs emitted either close to 50\% of the total energy
density at $z \approx 2$ (Caputi \etal 2007, Rodighiero \etal 2010) or 
the majority of it (P\'{e}rez-Gonz\'{a}lez \etal 2005).  

Depending on the data, methodology, and assumptions that are used, IRLFs estimated at high redshift have differed 
at both the faint and bright ends. At the faint end, the available data rarely constrain the slope of the 
LF at high redshift. Indeed, as noted above, there are significant differences in the faint-end slopes that have been measured at $z \approx 0$.
Data with a limiting 24-$\mu$m flux density of 80$\,\mu$Jy (as found for most of the 
earlier {\it Spitzer} studies described above) reach only a factor of a few fainter 
than typical estimates of $L_{\rm IR}^\ast$ at $z \approx 1$, and they barely reach $L_{\rm IR}^\ast$
at $z \approx 2$.   Thus, most analyses are forced to assume a faint-end slope based
on measurements at lower redshifts, making them subject to large (typically $> 100$\%) 
and uncertain extrapolations to total IR luminosity densities.  

Moreover, depending on the SED templates that are adopted,
there are significant differences in the (large) extrapolations from observed 
MIR rest-frame measurements (e.g., 8-$\mu$m rest-frame at $z = 2$) to the 
bolometric IR luminosity and SFR. Indeed, we expect such variations among real galaxies. 
Compared with several other studies, P\'{e}rez-Gonz\'{a}lez \etal (2005) found many more galaxies 
with $L_{\rm IR} > 10^{12} L_\odot$ at $z \approx 2$, in part owing to different assumptions about these bolometric corrections.  
Also, different procedures to account for AGN emission (which can be particularly
strong in the MIR) as well as heavy reliance on photometric redshifts may 
contribute to systematic issues in the IRLF, particularly at the bright end.

As we noted above (Section \ref{sec:sfrates_ir}), several studies (Papovich \etal 2007, Daddi \etal 2007) 
stacked data at longer FIR wavelengths (e.g., 70$\,\mu$m) or in radio and submillimeter data and found 
that standard SED templates such as those of Chary \& Elbaz (2001) tend to overestimate 
typical bolometric corrections from observed 24-$\mu$m data for galaxies 
at $z \approx 2$.  This suggested that true FIR measurements were needed to reliably 
determine luminosities and SFRs at high redshift.  Huynh \etal (2007) made early measurements 
of the IRLF at $z < 1$ using the deepest available {\it Spitzer} 70-$\mu$m data in GOODS-North.
The sample of detected sources was very small, but it was generally consistent 
with the earlier 24-$\mu$m work by Le~Floc'h \etal (2005).  

Magnelli \etal (2009, 2011) used comparably deep 70-$\mu$m data over 
a much wider area from the Far-Infrared Deep Extragalactic Legacy survey.  In addition 
to counting detected sources (mostly at $z < 1.3$, given the depth of the 70-$\mu$m data), 
Magnelli \etal stacked 70-$\mu$m data in bins of 24-$\mu$m flux and redshift
to measure empirically the average conversion between observed MIR and FIR luminosities. Compared with previous studies, 
they also used significantly deeper 24-$\mu$m catalogs, 
extending down to $30\,\mu$Jy in the GOODS fields.  At $z < 1.3$, Magnelli \etal (2009) 
found that the average FIR over MIR flux ratios closely matched predictions from the template library of Chary 
\& Elbaz (2001). They also measured LFs
that were similar to previous measurements, but that extended to fainter luminosities significantly 
below the bend in the IRLF at $z \approx 1$ and with better statistics.  At $z > 1.3$, 
however, Magnelli \etal (2011) confirmed previous suggestions that the average 70-$\mu$m 
to 24-$\mu$m flux ratios deviated systematically from the predictions of standard local
SED templates.  They extrapolated from the averaged 70-$\mu$m fluxes to the bolometric IRLF and found
only a mild increase in $L^\ast$ and the luminosity contribution of
ULIRGs from $z \approx 1$ to 2.   At lower redshifts, the faint-end slope was 
consistent with $\alpha = -1.6$ as measured for local {IRAS} galaxies by Sanders \etal (2003).
By $z = 2$, the data reach only slightly fainter than the IRLF knee, and the slope
is not constrained.  However, extrapolating with a fixed slope $\alpha = -1.6$, 
Magnelli \etal (2011) found that the faint IRLF at $z = 2$ would be quite similar to that 
predicted by Reddy \etal (2008) on the basis of UV-selected galaxies and the dust absorption
predicted from their UV spectral slopes.

The {\it Herschel Space Observatory} significantly improved sensitivity and reduced the beam size
for FIR (70--500$\,\mu$m) observations, and several large programs were dedicated
to surveys of the most important multiwavelength-deep fields.  Even the deepest {\it Herschel}
surveys do not detect as many sources per square arcminute as are found in the deepest 
{\it Spitzer} 24-$\mu$m observations, but direct access to FIR wavelengths is invaluable 
for reliably estimating bolometric luminosities and SFRs at high redshift.
Several analyses presented preliminary LFs out to $z = 2$ to 3 using
data sets obtained early in the {\it Herschel} mission (Eales \etal 2010, Gruppioni \etal 2010,  
Lapi \etal 2011).  

More extensive analyses of larger {\it Herschel} data sets have recently appeared in the 
literature.  Gruppioni \etal (2013) used sources selected in {\it Herschel} PACS observations 
at 70, 100, and 160$\,\mu$m in the two GOODS fields as well as shallower but wider 
observations of the Extended Chandra Deep Field South and COSMOS.  
They fit customized IR SED templates to photometry from both the PACS and SPIRE (250--500$\,\mu$m) instruments 
instruments and computed IR luminosities based on spectroscopic and photometric redshifts. They derived LFs out to $z < 4.2$. 
However, at the highest redshifts ($3 < z < 4.2$), the data are sensitive only to the rarest hyperluminous 
sources. Limiting their analysis to $z<2.3$, Magnelli \etal (2013) restricted their study 
to the deepest 70--160-$\mu$m data available in the GOODS fields, and extracted FIR fluxes to still-fainter 
limits at positions of 24-$\mu$m-detected sources. Note that, as is often the case, the fields analyzed
and the data used in these studies overlap considerably;  even if the methods of analysis
are different, they cannot be considered to be fully independent.  That said, in their 
range of overlap, the two analyses are generally consistent. They find somewhat stronger 
luminosity evolution at $z > 1$ than in the {\it Spitzer} analysis of Magnelli \etal (2011) and, 
hence, demonnstrate a larger contribution of ULIRGs to the total IR luminosity density at 
$z \approx 2$ [but not as large as in some earlier {\it Spitzer} studies, e.g., 
by P\'{e}rez-Gonz\'{a}lez \etal (2005)[.  Magnelli \etal (2011) concluded that this difference 
(compared with their own very similar {\it Spitzer} analysis) is mainly due to better determination
of the total IR luminosities of galaxies using the improved {\it Herschel} FIR 
measurements.  Gruppioni \etal (2013) found that the characteristic luminosity 
$L_{\rm IR}^\ast$ continued to brighten at $z > 2$, albeit at a slower rate.  Neither 
survey reliably measured the faint-end slope of the IRLF at high redshift, and both 
fixed it to values derived locally. Each study adopted distinct values: Magnelli \etal (2011) used $\alpha = -1.6$, 
whereas Gruppioni \etal (2013) used $\alpha=-1.2$. Given these different slopes, it is striking and 
perhaps surprising that these two analyses derive similar values for the total 
IR luminosity density at redshifts $0 < z < 2$.  Broadly speaking, both studies 
find evolution by a factor of $\sim 6$ between $z = 1.1$ and today and comparatively flat evolution at higher redshift
to $\langle z \rangle = 2.05$ (Magnelli \etal 2011) and $\langle z \rangle = 2.75$ (Gruppioni \etal 2013), 
albeit with an increasing range of values allowed within the measurement uncertainties.

\begin{figure}[ht]
\centerline{\psfig{figure=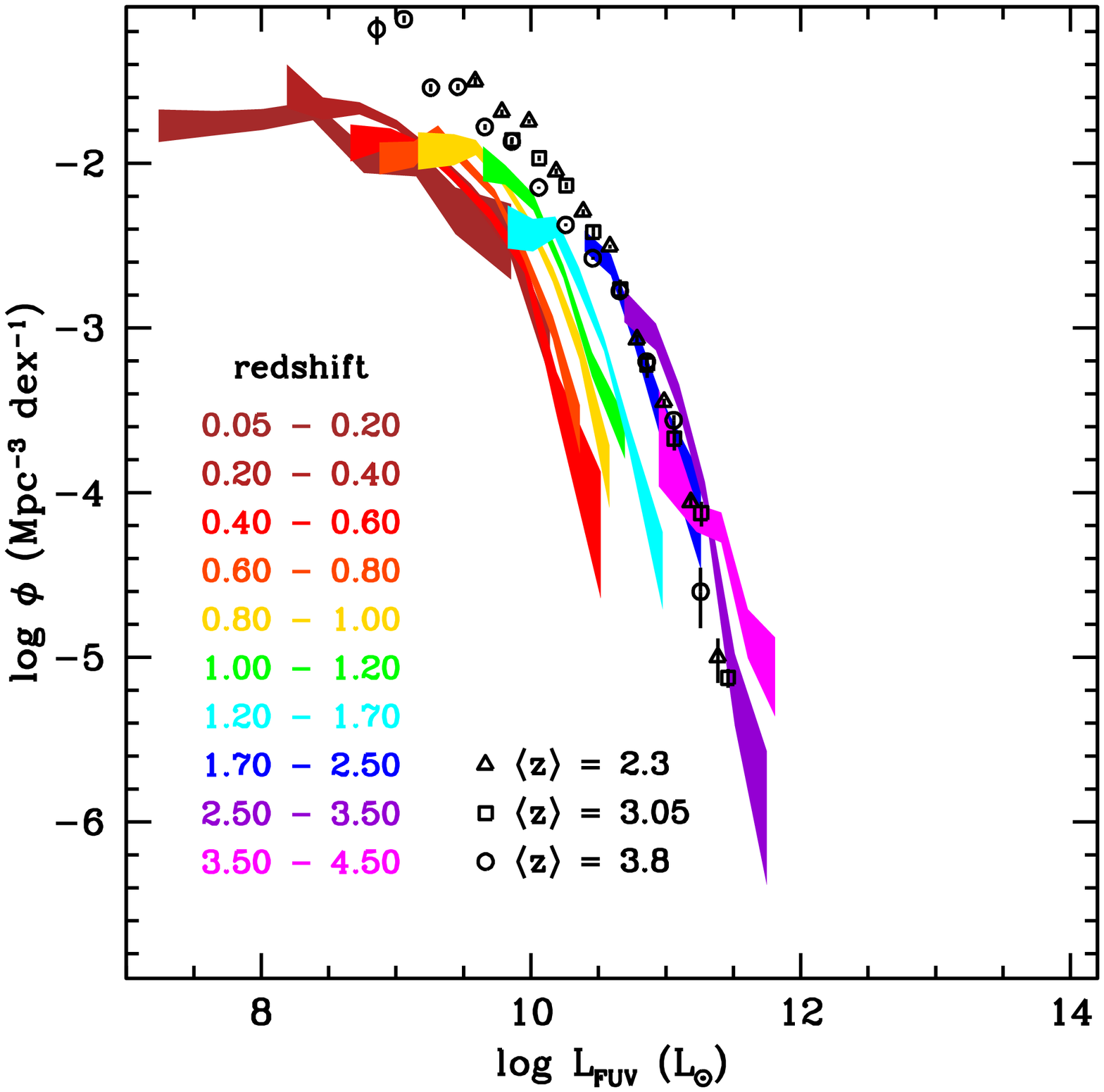,width=0.35\textwidth}
           \psfig{figure=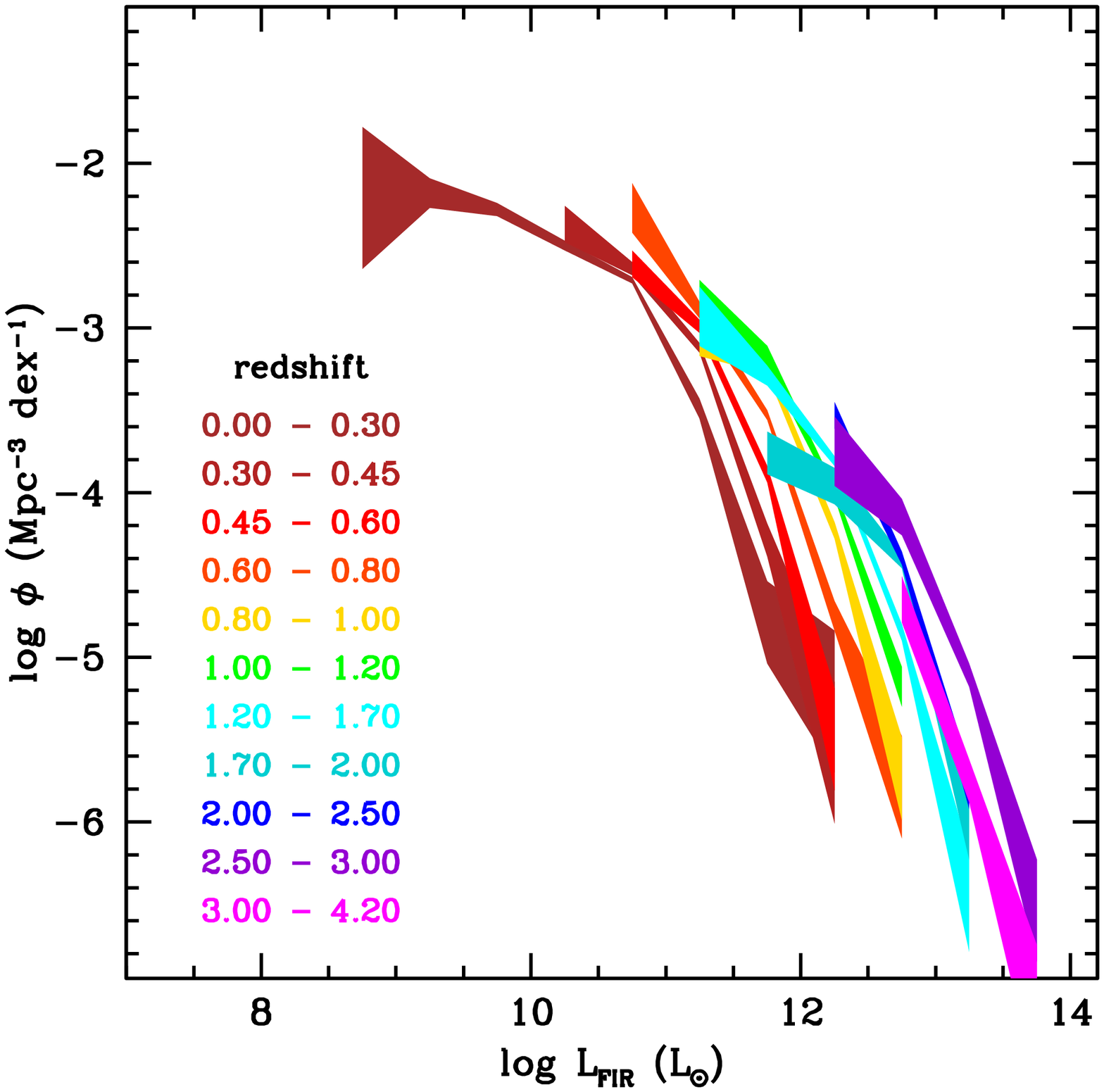,width=0.35\textwidth}
           \psfig{figure=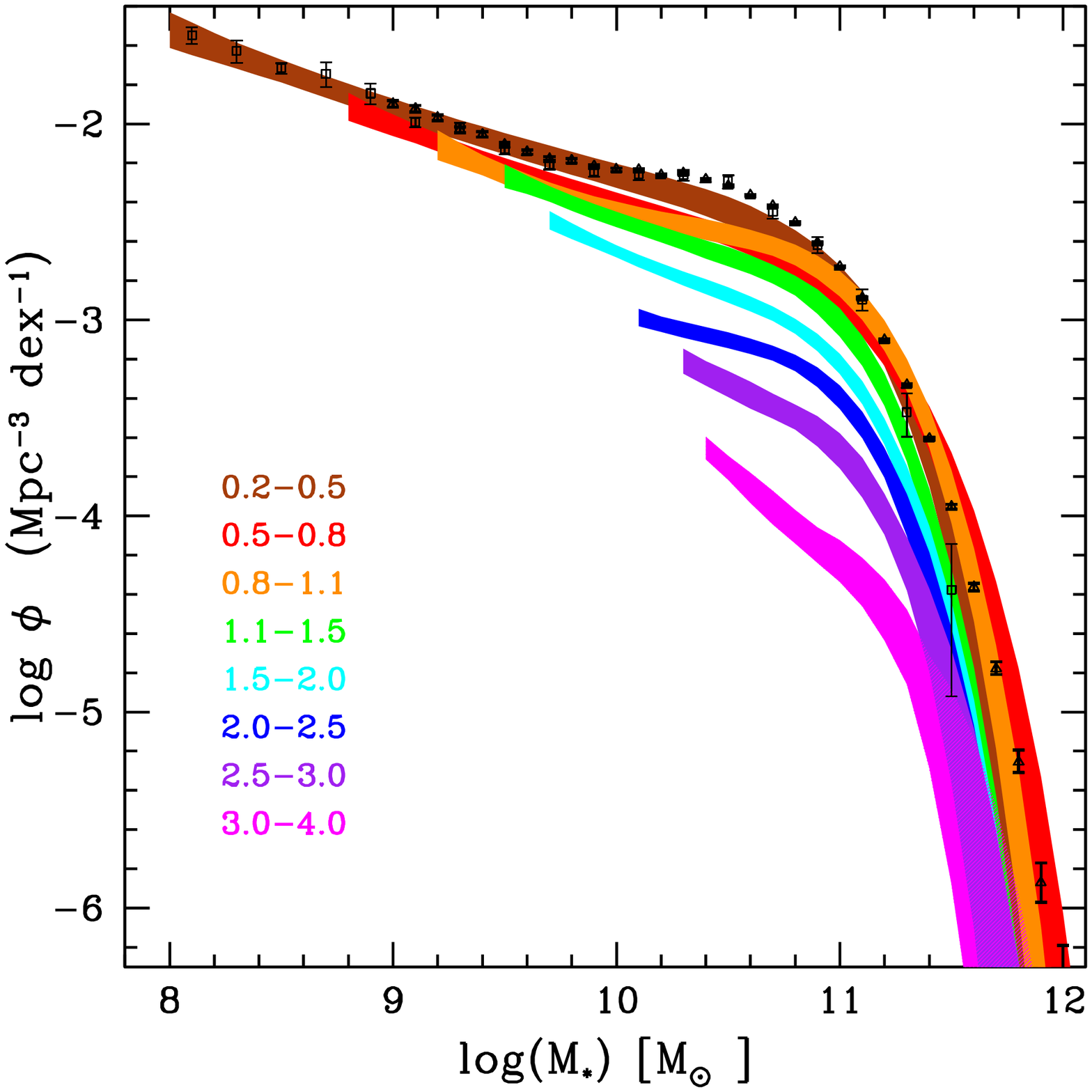,width=0.35\textwidth}
}
\caption{\footnotesize ({\it Left panel}) Redshift evolution of the FUV 
luminosity function at $0 < z < 4$. The colored bands indicate the 68\% 
confidence intervals on the space densities over the observed luminosities 
(uncorrected for dust attenuation), in different redshift ranges as indicated by the legend, 
from Cucciati \etal (2012). Data points, coded by shape, also as indicated in the legend, 
show the FUV luminosity functions for LBGs at mean redshifts 2.3 and 3.05 from Reddy \& Steidel (2009) 
and 3.8 from Bouwens \etal (2007). These luminosity functions use color selection techniques to extend 
the measurements to much fainter luminosities than those measured in the purely spectroscopic 
samples from Cucciati \etal (2012).  The FUV luminosity functions at $2.3 < z < 3.8$ are observed to be quite similar. 
({\it Middle panel}) Redshift evolution of the FIR luminosity function at $0<z<4$ from Gruppioni \etal (2013). 
The bands indicate the 68\% confidence intervals at each redshift, as indicated by the color coding.
({\it Right panel}) Galaxy stellar mass function at $0<z<4$ for a large, deep ($K_s<24$) sample of 220,000 
galaxies, from Ilbert \etal (2013).  Once again, the bands correspond to the 68\% confidence intervals 
at each redshift, including estimated uncertainties in the derived stellar masses. The open triangles 
and squares correspond to the local estimates by Moustakas \etal (2013) and Baldry et al. (2012), respectively. 
}
\label{fig7}
\end{figure}

Figure~\ref{fig7} shows recent determinations of the IRLFs and UVLFs 
at $0 < z < 4$.   The UVLFs in the figure show the observed 
luminosities uncorrected for extinction, and are presented in units of solar luminosities 
for more direct comparison with the IRLFs.  This figure illustrates several points that 
indicate low extinction for galaxies with lower SFRs and a significant contribution from 
low-luminosity galaxies to the global SFRD at high redshift.

\begin{enumerate}

\item Compared with the UVLFs, the IRLFs cut off less steeply at high luminosities.

\item The IRLFs extend to much higher luminosities than the UVLFs at the same redshifts,
as the most actively star-forming galaxies tend to be strongly obscured by dust.

\item There is strong luminosity evolution, particularly for the IRLFs, but also
in the UV, with more modest density evolution.  

\item The UVLFs shown in Figure \ref{fig7} (from Cucciati \etal 2012) exhibit a trend toward steeper faint-end slopes
at higher redshifts, especially for $z > 2$. Although this point lacks universal agreement, most studies do measure 
quite steep UVLF at $z > 2$ and a trend toward bluer UV colors at faint luminosities.

\end{enumerate}

Before {\it Spitzer} and {\it Herschel}, ground-based submillimeter bolometer arrays, especially
SCUBA at JCMT, provided an essential glimpse at dusty star formation at very high redshifts.
As is frequently noted, the negative $K$ correction at submillimeter wavelengths approximately
cancels luminosity distance dimming at $z > 1$.  The flux limits of most submillimeter surveys constrain
individual source detections to ultra- and hyper-luminous galaxies, so that only the tip
of the IRLF is sampled.  In practice, the greatest limitation for deriving LFs or SFRD is identifying
galaxy counterparts to submillimeter sources and measuring their redshifts. This limitation is due mainly
to the large beam-size of single-dish submillimeter observations, but it is also due to the fact
that the optical counterparts are often very faint and sometimes invisible.  
Another consequence of the negative $K$ correction is that substantial redshift
uncertainties translate to only relatively small uncertainties in the bolometric luminosity.
Hence, using radio-identified counterparts and very rough radio-millimetric redshift estimates,
Barger \etal (2000) were able to make plausible estimates of the SFRD from submillimeter sources in 
broad redshift bins. Barger \etal (2012) recently updated their
findings using a complete submillimeter galaxy (SMG) sample with accurate interferometric positions from the 
Submillimeter Array and with
more extensive redshift information, made possible in part thanks to recent advances in millimetric
CO spectroscopy.  Chapman \etal (2005) measured optical spectroscopic redshifts for 
a sample of radio-identified SMGs, and derived the first LF 
at $\langle z \rangle = 0.9$ and $\langle z \rangle = 2.5$.  These estimates were 
recently updated by Wardlow \etal (2011) using (mainly) photometric redshifts for 
another well-defined SMG sample.  These analyses demonstrated a significant contribution 
of dusty SMGs to the cosmic SFRD at high redshift, mainly limited to $z \lesssim 4$. However, 
recent discoveries of substantial numbers of SMGs out to $z = 6.3$
(e.g., Riechers \etal 2013) suggest that the dusty ultraluminous population may be
important even in the first few billion years of the cosmic SFH.

Although direct FIR detection of individual sources at $z \gg 2$ is limited
to the most extremely luminous objects (Figure ~\ref{fig1}), the CIRB provides 
additional constraints on dusty star formation at the highest redshifts
(Pei \etal 1999, Gispert \etal 2000).  FIR 
source counts and the CIRB were reviewed by Lagache \etal (2005) and recent results 
from {\it Herschel} are reviewed by Lutz (2014) in this volume, so we only briefly discuss this topic 
here. Sources directly detected by {\it Herschel} in the deepest
observations resolve $\sim 75$\% of the CIRB at 100 and 160$\,\mu$m (Berta \etal 2011,  
Magnelli \etal 2013), albeit with significant uncertainties at 100$\,\mu$m. Stacking
{\it Herschel} data at the position of sources detected by {\it Spitzer} at 24$\,\mu$m
detects an even larger fraction that, with modest extrapolation, can account for
the entire CIRB at these wavelengths.  At 250 to 500$\,\mu$m, where confusion is
more severe in {\it Herschel} SPIRE data, the directly resolved fractions
are smaller (from 15\% to 5\% at 250 to 500$\,\mu$m), whereas stacking detects 73\% to 55\%.
Again, an extrapolation is consistent with resolving the entire
background (B\'{e}thermin \etal 2012).   Because the deepest {\it Herschel} observations
were carried out in fields such as GOODS and COSMOS with exceptional ancillary data,
it is possible to stack in bins of photometric redshift, to constrain the redshift 
distribution of the CIRB emission.  At higher redshifts, the peak of dust
emission from galaxies shifts to longer FIR and submillimeter wavelengths,
and the fractional contribution of more distant galaxies increases with
the wavelength of the bandpass analyzed. Combining data from {\it Spitzer} through {\it Herschel}
to ground-based submillimeter observations, B\'{e}thermin \etal (2012) estimated that 
$4.7 \pm 2.0$~nW~m$^{-2}$ sr$^{-1}$, or $17^{+11}_{-9}$\% of the integrated CIRB,
is produced by galaxies at $z > 2$.  Although the fraction is small, it still allows 
for a significant amount of dusty star formation to take place at $z > 2$, beyond what 
is seen in directly detected sources.  However, the 24-$\mu$m sources
used for these stacking analyses are also subject to strong $k$--correction dimming
at $z > 2$. Thus, a larger fraction of the high-redshift CIRB may have been missed.

\subsection{Emission Line Surveys}
\label{sec:surveys_emline}

Among the nebular emission lines that have been most frequently used to quantify SFRs
at high redshift, H$\alpha$ is arguably the most useful (see 
Section~\ref{sec:sfr_other}). Lines from other elements, most commonly [OII]~3,727\,\AA,
have also been used to measure the cosmic SFR (e.g., Hogg \etal 1998),
but their more complex dependence on metallicity and ISM conditions as well as their larger extinction
make them problematic. Narrow-band \Lya\ surveys are popular at high
redshift, but \Lya\ is so strongly subject to resonant scattering and extinction that
it must always be regarded as setting only a lower limit on the true SFR. 

H$\alpha$ is accessible with optical observations only at $z < 0.5$, and NIR
observations are needed to follow it out to higher redshifts ($z \lesssim 2.5$).  Recent
technological developments in IR instrumentation have significantly increased the
potential for such measurements.  A new generation of wide-field imagers using mosaics
of IR arrays is now operating on 4-m- and 8-m-class telescopes
(e.g., {UKIRT} WFCAM, {CFHT} WIRCAM, {NOAO} NEWFIRM, {VISTA} VIRCAM, {VLT} HAWK-I), thereby 
significantly increasing the comoving volumes accessible for deep narrow-band imaging.  
NIR multiobject spectrographs are now becoming mature and efficient
(e.g., {\it Subaru} MOIRCS and FMOS, {\it Keck} MOSFIRE, {VLT} KMOS, {LBT} LUCI).  
Slitless grism spectroscopy with WFC3 on the {HST} 
can measure faint H$\alpha$ lines out to $z < 1.5$ for all objects within its field of view.
Each method has advantages and disadvantages.  Narrow-band imaging surveys are tuned to 
specific, narrow redshift ranges and are strongly subject to density variations due 
to line-of-sight clustering effects.  Flux calibration for objects whose redshifts
place emission lines in the wings of the narrow bandpasses can also be problematic.
Only statistical corrections can be made for the flux contribution from [NII] or
for stellar absorption.  Multislit spectroscopy is subject to slit losses that
complicate measurements of integrated line fluxes, and atmospheric absorption and
emission significant limit the accessible redshifts and can introduce complicated
selection effects. (The new {VLT} KMOS spectrograph uses multiple deployable
integral field units, thus eliminating slit-loss concerns, and may prove to be a valuable
tool for H$\alpha$ surveys.) {HST} WFC3 slitless spectroscopy avoids concerns about
the atmosphere and slit losses, but deep observations covering adequately large
solid angles are time-intensive. As of this writing, results on H$\alpha$ LFs from
the largest ongoing programs [WISPS (Atek \etal 2010) and 3D-HST (Brammer \etal 2012)]
have not appeared in the literature to supersede earlier {HST} NICMOS results
(Yan \etal 1999, Hopkins \etal 2000, Shim \etal 2009).  With all methods,
reliable extinction corrections depend on the measurement of the Balmer decrement
(the ratio of H$\alpha$ to H$\beta$ line fluxes).  This is rarely available for narrow-band surveys, but it is 
sometimes available for spectroscopic samples. However, in such cases, the sample selection is often
limited by the weaker H$\beta$ line. Hence, statistical corrections are often adopted.   AGN and LINERs 
can also contribute significantly to the samples of emission line galaxies and can be reliably taken 
into account only using high-quality spectroscopic data to 
measure line-excitation diagnostics.   Most careful studies of the local galaxy
population using spectroscopy from SDSS or GAMA (see below) have done this,
but it is rarely possible at higher redshifts. Thus, most studies have resorted
to statistical corrections or none at all.

Much literature discusses LF and SFRD measurements from H$\alpha$, 
[OII] and [OIII], and we note only selected works here.  Gunawardhana \etal (2013) 
included a thorough and up-to-date compilation of these measurements at low and high 
redshifts.
Using objective prism photographic data, Gallego \etal (1995) presented an important early study of the local H$\alpha$ 
LF. Subsequently, the SDSS provided a vast 
number of spectroscopic redshifts and line-flux measurements, although the small aperture 
size of the spectroscopic fibers requires careful and inevitably uncertain corrections to the 
total emission line flux for each galaxy.   SDSS spectroscopy covers both H$\alpha$ and H$\beta$ and 
can provide a measurement of extinction via the Balmer decrement, although care 
is needed to account for stellar absorption, flux limits, and selection effects. 
Brinchmann \etal (2004) conducted a widely-cited study of local star formation from 
SDSS optical spectroscopy and photometry. They used a full analysis of the emission
and absorption line spectroscopy. Thus, their study was not strictly based on H$\alpha$ alone, although 
the Balmer lines carry significant weight in the SFR determinations.
As noted above, Salim \etal (2007) carried out an independent SDSS analysis based
mainly on photometry including {GALEX} UV measurements but with extensive cross-comparison
to the H$\alpha$ data.  By using photometry, the Salim analysis bypasses uncertainties
inherent in the spectroscopic H$\alpha$ aperture corrections.  The local SFRDs 
(normalized to $z = 0.1$) from the studies by Brinchmann \etal (2004) and Salim \etal (2007) agree
extremely well.  Most recently, Gunawardhana \etal (2013) combined data from the SDSS 
with significantly deeper (but also much smaller) fiber spectroscopy from the GAMA survey,
to probe fainter down the LF and extend the analysis to higher redshifts ($z < 0.35$).   

Optical spectroscopic surveys have measured H$\alpha$ SFRD typically out to $z < 0.4$ (e.g., Tresse \& Maddox 1998, Sullivan \etal 2000, Westra \etal 2010, 
Gunawardhana \etal 2013).  Until recently, NIR spectroscopy 
was measured only for small samples, object by object, typically at $0.5 < z < 1.1$
(Glazebrook \etal 1999, Tresse \etal 2002, Doherty \etal 2006).
In general, nebular line extinction is not directly measured and can only be assumed. 
The same is true for corrections for stellar absorption. Larger-scale and deeper spectroscopic 
surveys from a new generation of multiobject IR spectrographs on 8--10-m
telescopes should be forthcoming.

Several studies have exploited slitless grism spectroscopy of H$\alpha$ at $0.7 < z < 1.9$ 
from the {HST} NICMOS instrument (Yan \etal 1999,  Hopkins \etal 2000, Shim \etal 
2009). These studies are of small but relatively unbiased samples that are observed with
relatively uniform sensitivity over a broad redshift range and without concerns for
flux losses due to a finite spectrograph slit size.  Once again, direct measurements
of extinction are generally unavailable, and the low-resolution slitless spectroscopy
inextricably blends H$\alpha$ with [NII], thus requiring purely statistical corrections.
New, more sensitive grism surveys with {HST} WFC3 that cover  larger solid angles are under way.

The largest number of high-redshift measurements has come from narrow-band imaging
surveys, using wide-field CCD cameras out to $z \approx 0.4$ and with IR arrays
at higher redshifts.   These are generally carried out using filters that fit into
spectral windows relatively unaffected by atmospheric absorption or emission and that correspond
to specific redshifts ($z = 0.24$, 0.40, 0.84, 1.47, and 2.23 are all common).
Using a new generation of wide-field imagers, the current state-of-the-art H$\alpha$ 
surveys include those of Ly \etal (2007, 2011), Hayes \etal (2010),
and Sobral \etal (2013). The latter survey (HiZELS) combines data from {\it Subaru} 
Suprime-Cam ($z = 0.40$) and {UKIRT} WFCAM ($z = 0.84$, 1.47 and 2.23). It covers 
$\sim 2$~deg$^2$ in two survey fields, with deeper but narrower (0.03 deg$^2$) observations 
at $z = 2.23$ from the {VLT} and its HAWK-I imager.  Between 500 and 1750 H$\alpha$ emitters 
are detected in each redshift window, thereby providing good statistics.  These authors measured 
a steady increase in the characteristic luminosity $L^\ast({\rm H}\alpha)$ with
redshift and a faint-end slope $\alpha \approx 1.6$ that is constant with redshift,
and is also similar to that found in most of the UV continuum surveys.

\subsection{Radio Surveys}
\label{sec:surveys_radio}

Centimeter-wavelength radio continuum emission from star-forming galaxies
arises from a combination of flat-spectrum free-free emission, which is prominent
at high frequencies, and steep spectrum synchrotron emission, which dominates
at lower frequencies.   Although the free-free emission should 
be a direct tracer of SFRs, it has been impractical 
to observe this at high redshift. However, the improved high-frequency sensitivity
of the Jansky VLA should open this capability in the future.   The lower-frequency
emission arises mainly as a consequence of SNe. As such, it is also
related to the SFR.  In practice, its calibration as an SFR
measurement is primarily empirical, e.g., based on the tight correlation
observed between radio and FIR emission for {IRAS}-selected galaxies
in the local Universe (e.g., Yun \etal 2001).   This correlation together with insensitivity to dust 
extinction make radio emission an attractive wavelength for studying star formation.  

Problematically, AGN can also produce powerful radio emission.  Locally, 
radio sources with $L_\nu({\rm 1.4~GHz}) > 10^{23}$~W~Hz$^{-1}$ are predominantly AGN.
This radio luminosity corresponds to a SFR $> 94\,\msun$~year$^{-1}$, using the conversion factor from Murphy \etal (2011) 
scaled to a Salpeter IMF.  Such galaxies are extremely rare locally, 
so very powerful radio sources are commonly excluded as likely AGN.
However, at $z > 1$, galaxies with these SFRs (corresponding to 
ULIRGs in terms of their IR luminosities) are common, even ``normal'' 
(see Section \ref{sec:sfrates_ir}). Thus, it is problematic to disregard them 
simply on the basis of their radio emission.   Other considerations are needed 
to distinguish AGN from star-forming radio sources, such as radio spectral index, 
radio morphology, or radio/IR flux ratios, but these are all more difficult 
to measure, especially for very faint sources.  

The local radio LF has been used to estimate the current 
SFRD in many studies (for recent examples, see Machalski \& Godlowski 2000, Condon \etal 2002, Sadler \etal 2002, 
Serjeant \etal 2002, Mauch \& Sadler 2007).  In these studies, the authors carefully employed various criteria to 
separate AGN from star-forming galaxies. Data were often compared with {IRAS} IR measurements and 
excluded objects with a radio excess relative to their IR luminosities.

At $z > 1$, even the deepest VLA surveys have been able to detect galaxies with SFRs only $\gtrsim 100\,\msun$~year$^{-1}$ 
(Figure ~\ref{fig1}) and $\gtrsim 1000\,\msun$~year$^{-1}$ at $z > 3$, although the Jansky VLA is now pushing 
to fainter sensitivities.  Haarsma \etal (2000) were among the first to combine 
very deep VLA observations with spectroscopic and photometric redshift information to derive radio LFs
and the cosmic SFRD at high redshift, primarily at $0.3 < z < 0.8$. However, they also considered one very 
broad bin from $1 < z < 4.4$.  Based on individually detected radio sources, this work was updated in later studies 
that took advantage of more extensive multiwavelength
data to distinguish AGN from star-forming galaxies.  Seymour \etal (2008) identified
AGN based on radio spectral indexes, radio morphology, and radio to NIR and MIR 
flux ratios.  Smol\v{c}i\'{c} \etal (2009) used optical SED criteria to identify 
star-forming galaxies detected in VLA 1.4~GHz data for COSMOS, at more modest redshifts,
$z < 1.3$.  Seymour \etal (2008) assumed pure luminosity evolution for the LF, consistent with the measurements by 
Smol\v{c}i\'{c} \etal (2009) (and earlier by Haarsma \etal 2000).  All these studies found broad agreement between the radio 
SFRD evolution and optical and IR surveys, noting a steep decline 
from $z = 1$ to 0; Seymour \etal (2008) measured a peak SFRD at $z \approx 1.5$, albeit 
with large uncertainties.

Other studies have used radio stacking to probe to fainter luminosities, below
the detection limits for individual sources -- particularly, stacking for NIR 
samples that approximate stellar mass selection.  Here, the assumption is that,
as in the local Universe, radio sources at fainter luminosities will primarily trace
star formation, with relatively little AGN contribution.  Dunne \etal (2009) stacked 
VLA 1.4~GHz and GMRT 610~MHz data for $K$-band-selected sources in bins of redshift
and $K$-band luminosity. They found a linear (but redshift-dependent) relation between
radio and NIR luminosity.  Using a measurement of the evolving $K$-band LF
from the same data set, they then used this radio/NIR ratio to extrapolate
the observations to the total radio luminosity density and SFRD.  Karim \etal (2011)
used a large 1.4-GHz survey of COSMOS and a {\it Spitzer} 3.6-$\mu$m--selected sample
to carry out the most extensive study of this sort to date. Stacking 
in bins of stellar mass and photometric redshift and converting the mean radio fluxes 
to SFRs, they extensively analyzed the SFR--$M_\ast$ correlation,
and used this and the evolving stellar mass function (SMF) (from Ilbert \etal 2010) to
compute the integrated SFRD.  Dunne \etal (2009) measured a steady increase in the SFRD
from $z \approx 0$ to a peak at $z \approx 1.5$ that declined at higher redshift to $z \approx 4$.
By contrast, Karim \etal (2011) found a monotonic decline in the SFRD from $z = 3$ to today.

\subsection{Stellar Mass Density Surveys}
\label{sec:surveys_smd}

As the technology of NIR detectors advanced, so did surveys that used 
NIR photometry to better sample galaxies by stellar mass at both low and high 
redshifts (e.g., Cowie \etal 1996, Gavazzi \etal 1996).  However, it was not 
until the turn of the millennium that authors started to routinely use stellar 
population synthesis models to convert photometry and redshift information 
to stellar masses for large samples of galaxies, especially at high redshift.
Before the era of large, modern redshift surveys such as the SDSS and the 2dFGRS,
several authors made estimates of the local 
baryonic and SMDs (Persic \& Salucci 1992, Fukugita \etal 1998, 
Salucci \& Persic 1999).  This effort accelerated, however, as new spectroscopic
surveys mapped the local Universe.  Cole \etal (2001) used 2dFGRS redshifts
and NIR photometry from 2MASS to measure the local $K$-band LF
more accurately than had previously been possible.  Then, using stellar 
population modeling, they inferred stellar masses from the galaxy colors and 
luminosities and derived the local galaxy stellar mass function (GSMF) and the
local comoving SMD. Bell \etal (2003) did the same using SDSS 
and 2MASS.  Many studies have subsequently derived the GSMF from incrementally 
improving SDSS data releases, using additional ancillary data and a variety of 
methods for stellar population modeling; mass-to-light ratios are sometimes derived from 
photometry and from the SDSS spectra (Panter \etal 2007,  Baldry \etal 2008, 
Li \& White 2009).  Moustakas \etal (2013) incorporated 
photometry from the ultraviolet ({GALEX}) to the MIR ({WISE})
in their analysis of the SDSS GSMF.  Baldry \etal (2012) analyzed a sample 
from GAMA, a wide-area spectroscopic survey extending significantly fainter than 
the SDSS, modeling $M/L$ from optical SDSS photometry.   As the samples have 
grown larger, more elaborate functional forms have been fit to the GSMF, 
including double and even triple Schechter functions, and there is some evidence 
that the GSMF slope at low masses ($< 10^9\,\msun$) may be steeper than was 
previously believed (e.g., $\alpha = -1.47$) (Baldry \etal 2012).   
Other authors have modeled stellar populations for ensembles of galaxies
rather than deriving individual galaxy stellar masses, e.g., by fitting the 
integrated local luminosity densities for the entire local volume
from UV to NIR wavelengths (Baldry \& Glazebrook 2003) or by modeling 
coadded SDSS spectra in bins of luminosity, velocity dispersion, and 4,000-\AA\ break
strength (Gallazzi \etal 2008).

At higher redshifts, Brinchmann \& Ellis (2000) derived stellar masses for 
galaxies at $z < 1$ in fields with both {HST} and NIR imaging. As expected given the declining cosmic SFRD at those 
redshifts, they found relatively little evolution in the integrated mass density at $0.4 < z < 0.9$, 
but a marked evolution in the mass breakdown by morphological type.  Cohen (2002) similarly found no 
significant evolution in the SMD at $0.25 < z < 1.05$.
Because the cosmic SFRD declines steeply with time at $z < 1$, relatively
little new stellar mass accumulates in the late stages of cosmic history.
Moustakas \etal (2013) recently broke new ground with a low-resolution 
prism spectroscopic survey measuring $\sim 40,000$ redshifts for galaxies at 
$0.2 < z < 1$ in five fields with {\it Spitzer} IRAC photometry covering 5.5~deg$^2$.  
Even with such outstanding data, the evolution in the overall SMF at $0 < z < 1$
is nearly imperceptible, but the migration of galaxies from the star-forming
to the quiescent population is confirmed with exquisite detail.

Sawicki \& Yee (1998), Giallongo \etal (1998), Papovich \etal (2001), and Shapley \etal (2001) 
pioneered the use of stellar population models to derive stellar masses for 
LBGs at $z \approx 2$--3. Giallongo et al. (1998) computed comoving SMDs out to $z \approx 4$ for 
galaxies from a relatively bright, optically-selected sample, and measured steep evolution, but did 
not attempt to correct for unobserved galaxies fainter than the limits of their data. Subsequently, Dickinson \etal (2003), 
Fontana \etal (2003), and Rudnick \etal (2003) used the deepest NIR 
imaging then available for the two HDFs, {HST} NICMOS (HDF-N)
and {VLT} ISAAC (HDF-S), together with extensive photometric and spectroscopic 
redshifts to derive the comoving SMD in several redshift bins 
out to $z \approx 3$.  Strong evolution was found over that longer
redshift baseline; the SMD at $z \approx 3$ measured in the range 5 to 15\% of its present-day value, although a somewhat 
broader range of values would be permitted if systematic assumptions about
the galaxy SFHs or stellar metallicities were pushed
well beyond the range of models used for standard analysis.  The SMD reached half its present-day value somewhere between $z = 2$ and 1.
Strong cosmic variance in these small fields was also evident: At $z > 2$, red 
galaxies with high mass-to-light ratios were nearly absent in the HDF-N but were 
found in moderate abundance in the HDF-S. By contrast, the corresponding SMDs
differed substantially. This indicated the importance of surveying 
larger fields and more sightlines, but obtaining IR imaging to satisfactory 
depth over these larger regions of sky has proven to be very challenging and 
has required another decade of effort.

The launch of {\it Spitzer} and the impressive performance of its IRAC camera 
for imaging at 3.6 to 8$\,\mu$m made it possible to measure {rest-frame}
NIR photometry for galaxies at high redshift, and major public survey
imaging campaigns such as GOODS, S-COSMOS, and SWIRE produced widely-accessible
and heavily-used imaging data sets spanning a wide range of area--depth parameter
space, ideal for deriving SMFs and densities at high redshift.
Indeed, NIR imaging has struggled to catch up with IRAC in terms of
comparable area--depth coverage. Despite the vast swaths of telescope time 
that have been invested in obtaining NIR data on popular fields such as GOODS 
and COSMOS, imaging at 1 to 2.5$\,\mu$m still tends to fall short of {\it Spitzer}'s
sensitivity at 3.6 and 4.5$\,\mu$m.  At $z > 4.5$, the ground-based $K$-band 
samples rest-frame ultraviolet wavelengths, and IRAC offers the only viable 
way to measure optical rest-frame light to constrain stellar masses.  

From 2006 onward, most (although not all) papers on SMFs
and densities at high redshift have made use of IRAC data -- often in the 
same survey fields that are repeatedly analyzed.  Among many other papers, Fontana \etal (2006), 
P\'{e}rez-Gonz\'{a}lez \etal (2008), Kajisawa \etal (2009), and Marchesini \etal (2009)
analyzed stellar masses in the GOODS fields (sometimes in combination with other 
data sets) for galaxies out to $z \approx 4$. Arnouts \etal (2007), 
Pozzetti \etal (2010), Ilbert \etal (2010), and Brammer \etal (2011) analyzed
wider-area but shallower surveys (e.g., COSMOS, VVDS-SWIRE, NMBS), generally focusing
on redshifts $z \leq 2$.   Despite differences in their methodologies, the conclusions
of these papers painted a remarkably consistent picture of the evolution of the 
SMF at $0 < z < 3$, with very little change in its shape,
characteristic mass $M^\ast$, or faint-end slope, but with steady evolution
in the characteristic density $\phi^\ast$.   There are indications that the 
faint-end slope of the mass function may steepen at higher redshifts 
(e.g., Kajisawa \etal 2009, Santini \etal 2012). The integrated SMDs
measured in the different analyses generally agreed within factors of 2 at most
redshifts, and a consistent picture of mass-build-up emerged.  

Some of the most recent additions to this literature have taken advantage of 
deeper, wider NIR imaging from the largest-format cameras on 4-m-class
telescopes (Bielby \etal 2012, Ilbert \etal 2013, Muzzin \etal 2013) to map
relatively wide survey areas such as COSMOS or the CFHT Legacy Survey fields to depths
previously reserved for small, deep surveys such as GOODS.  Together with ever
growing spectroscopic surveys and increasingly excellent photometric redshifts,
these have yielded the most statistically robust measurements of the SMFs at $z < 2.5$.
However, several of these surveys repeat analysis
in COSMOS. Some even use essentially the same imaging data sets, so these analyses
are not always robust against cosmic variance.  

{\it Spitzer} IRAC has been essential for deriving stellar masses at $z > 4$, and 
very deep observations are necessary to detect typical galaxies at those redshifts.
For this reason, nearly all analyses of SMDs at $z > 4$ have been 
carried out in GOODS and the HUDF [in a departure from the GOODS-dominated norm,
McLure \etal (2009) stacked relatively shallow IRAC data for LBGs at $z = 5$ and 6
in the UKIDSS Ultra Deep Survey to measure average SEDs and
mass-to-light ratios and, hence, to estimate the SMD] and nearly all have studied UV-selected LBGs, 
for which there are abundant samples. Early analyses of small 
samples of galaxies at $z = 5$ and 6, including estimates of the integrated SMD,
were presented by Yan \etal (2006), Eyles \etal (2007), Stark \etal (2007),
Verma \etal (2007), and Yabe \etal (2009).  These were followed by larger
and more systematic analyses of LBG samples at $4 \geq z \geq 7$ 
(Stark \etal 2009, Gonz\'{a}lez \etal 2011, Lee \etal 2012), all of which used
similar procedures, and found broadly similar results.  In particular,
the derived LBG SMFs have somewhat shallower low-mass slopes 
than do the UVLFs, because $M/L_{UV}$ decreases
at fainter UV luminosities, at least at $z = 4$ and 5 where this could be
measured with some robustness from galaxies with individual IRAC detections
(e.g., Lee \etal 2012). Using {HST} WFC3-selected samples in the HUDF
and GOODS/CANDELS fields, Labb\'{e} \etal (2013) recently extended this
analysis to $z \approx 8$. 

Exceptionally, some studies have used IRAC selection to avoid the potential
bias of UV selection against older or dustier galaxies.  Caputi \etal (2011)
analyzed an IRAC 4.5-$\mu$m-selected sample in the UKIDSS Ultra Deep Survey, thereby computing 
SMDs at $3 \leq z \leq 5$.  The depth of their IRAC data 
limited direct detections to a stellar mass completeness limit $\gtrsim 10^{10.4}\,\msun$.
Their extrapolated mass densities based on Schechter function fits fall
somewhat below those from other LBG-based studies (e.g., Gonz\'{a}lez \etal 2011, 
Lee \etal 2012), but this is likely due to uncertainties in the large extrapolation
required.  Several other authors have analyzed partially or wholly IRAC-selected
candidates for massive galaxies at $z \gtrsim 3.5$ (Wiklind \etal 2008, Mancini \etal 2009, Marchesini \etal 2010, 
Caputi \etal 2012). In some cases, they have estimated comoving SMDs, although generally without fitting SMFs
and often without rigorous analysis of sample completeness.

Several studies have suggested that LBGs at $z \gtrsim 4$ have much stronger 
optical rest-frame nebular line emission (particularly H$\alpha$ and [OIII]) 
than do similar UV-selected galaxies at lower redshifts (Chary \etal 2005, Raiter \etal 2010, Shim \etal 2011, 
Stark \etal 2013, Labb\'{e} \etal 2013).  In most cases this has
been inferred on the basis of {\it Spitzer} IRAC colors that would be unusual for 
pure stellar populations but that can be explained if strong line emission boosts
the IRAC fluxes.  This line emission, if not taken into account, can significantly 
effect derived stellar population parameters and generally leads to overestimated
stellar masses.  For LBG samples at $z = 4$ to 8, Stark \etal (2013) and 
Labb\'{e} \etal (2013) estimated that average stellar masses derived from 
models without nebular lines should be reduced by factors from 10 to 70\%, 
with the effect increasing at higher redshifts. Although the photometric evidence
for this strong nebular emission is compelling, it will be vitally important
for {JWST} spectroscopy to confirm and quantify its effects.

\subsection{The State of the Art, and What's Wrong with It}
\label{sec:stateoftheart}

\subsubsection{Local measurements}

To be statistically meaningful, measurements of the current SFR or SMD require surveys covering a large fraction of the sky.
Salim \etal (2007) and Robotham \& Driver (2011) used most or all of the {GALEX} 
Medium Imaging Survey data, covering $\sim 1,000$~deg$^2$ overlapping the SDSS and
2dFGRS spectroscopic surveys, and there is little prospect for improving the UV
data in the near future.  Only refinements in the analysis can be expected, 
e.g., incorporating improved photometric data at optical or NIR wavelengths, 
or further joint analysis with spectroscopic stellar population measurements. As discussed 
below (Section \ref{sec:sfrd}) (Figure ~\ref{fig8}), there is significant
disagreement in the literature about the net FUV extinction correction at $z \approx 0$.
Wide-area spectroscopic emission line surveys (e.g., from SDSS or GAMA) (Brinchmann \etal 2004, 
Gunawardhana \etal 2013) are limited by uncertain aperture corrections to line fluxes,
whereas narrow-band imaging surveys have yet to cover enough galaxies over a wide enough area
and are usually limited by the absence of direct measurements of extinction from the Balmer 
decrement, the [NII] contribution to H$\alpha$ measurements, or the contribution of AGN
emission.  There is still room
for progress in combined narrow-band plus spectroscopic data for large local samples.
The local FIRLF has not been drastically revised since the
final {IRAS} analyses (Sanders \etal 2003, Takeuchi \etal 2003);  additional {AKARI} 
data did not drastically change earlier results (Goto \etal 2011a,b, Sedgwick \etal 2011).
The biggest remaining uncertainties pertain to the faint-end slope, where measurements
vary significantly from $\alpha = -1.2$ to $-1.8$ (or, somewhat implausibly, 
even -2.0) (e.g., Goto \etal 2011b).  Analysis of the widest-area FIR surveys
from {\it Herschel}, such as H-ATLAS (570~deg$^2$) (Eales \etal 2010) may help with this.
The present uncertainties lead to a difference of a factor of at least 2 to 3 in the
local FIR luminosity density. Nevertheless, as previously noted, in today's
relatively ``dead'' epoch of cosmic star formation, a significant fraction of the FIR
emission from ordinary spiral galaxies may arise from dust heated by intermediate-age
and older stellar populations, not newly formed OB stars. Hence, it is not necessarily
the best measure of the SFR.  At higher redshifts, when the cosmic-specific SFR was much 
larger, new star formation should dominate dust heating, making the IR emission a more robust global tracer.

Local measurements of the SMD have relied mainly on purely optical
data (e.g., SDSS photometry and spectroscopy) or on relatively shallow NIR 
data from 2MASS.  There may still be concerns about missing light, surface brightness
biases, etc., in the 2MASS data (e.g., Bell \etal 2003), and deeper very-wide-field 
NIR data would be helpful.  All-sky MIR data from WISE may be valuable and have been used 
by Moustakas \etal (2013), but without extensive analysis
specifically focused on this topic.  Deeper NIR data covering a significant 
fraction of the sky, either from new wide-field ground-based NIR telescopes with large 
apertures or from space-based surveys with {EUCLID} or {WFIRST}, would make 
a valuable new contribution.  Otherwise, as for UV SFRs,
the most likely improvements will come from refined stellar population analyses, rather than from new data.

\subsubsection{$0 < z < 1$}

During the decline and fall of cosmic star formation, from $z \approx 1$ to 0, the greatest
weakness of current measurements is that they have generally covered small sky areas
and small comoving volumes over few independent sightlines. Hence, they 
may be subject to significant cosmic variance uncertainties.  Fields such as GOODS,
which have been analyzed many times, are too small to offer precision measurements in
fine redshift bins at $z < 1$.  Even the 2~deg$^2$ COSMOS field subtends less than 
100~Mpc at $z < 1$ and can be spanned by large-scale structure; as a single sightline,
it is subject to density fluctuations.  Although very good data for measuring 
the SFRD or SMD at $z < 1$ exist in many fields, relatively little information has 
been analyzed thoroughly, in part because intensive effort on spectroscopic
(or even photometric) redshifts has been applied to only the few, most famous fields.
Sometimes even fields such as AEGIS, which has outstanding spectroscopy and deep 
{GALEX}, {\it Spitzer} and {\it Herschel} data, have not been fully exploited 
for this purpose.

For example, very deep {GALEX} data exist for several of the most famous survey fields,
but the one widely-cited analysis of the UV luminosity density at $z < 1$ 
(Schiminovich \etal 2005) uses only $\sim 1,000$ sources with redshifts in 
a single 0.5-deg$^2$ field.  Expanded analysis of comparably deep {GALEX} data
in COSMOS, AEGIS, and several other survey fields with existing, extensive 
spectroscopy is long overdue.  There are no opportunities to collect more
{GALEX} data, but deep $U$-band imaging measuring somewhat longer rest-frame UV
wavelengths may be quite adequate for many purposes.  Such data exist or could
be obtained with wide-field imagers, but the best analysis to date 
(Cucciati \etal 2012), using $\sim 11,000$ spectroscopic redshifts, 
is limited to a single 0.6-deg$^2$ sightline [indeed, the same field analyzed 
by Schiminovich \etal (2005) with {GALEX}].  Much more work can be done to improve
this situation, with relatively limited new observational effort and often using
data that already exist.

Similarly, most analyses of MIR and FIR data from {\it Spitzer} and {\it Herschel} data 
have used data from at most three independent sightlines (e.g., Magnelli \etal 2009, Rodighiero \etal 2010, 
Gruppioni \etal 2013), nearly always combining the two
GOODS fields with one shallower but wider data set (e.g., COSMOS or the VVDS-SWIRE 
field).  The widest-area analysis to date is that of Rujopakarn \etal (2010), which 
used 24-$\mu$m-selected sources at $z < 0.65$ from the 9-deg$^2$ Bo\"{o}tes 
survey.  In practice, more data over more sightlines exist.  Large consortium 
surveys such as the PEP and HerMES Herschel Guaranteed Time programs have
mapped many fields, often with an elegant hierarchy of different areas and depths,
but these have not yet been exploited and combined into a single, definitive
analysis.  Typically, this is because the ancillary imaging and spectroscopy needed
to identify IR source counterparts and to determine their redshifts is 
available or adequate in only a few fields (hence, the repeated analyses of GOODS
and COSMOS).  

Many of these same comments apply to SMD estimates at $z < 1$. In this case, the state of the art has recently been improved by Moustakas \etal (2013),
who analyzed five independent fields with a combined solid angle of 5.5~deg$^2$
with (relatively shallow) IRAC photometry and (low-resolution) spectroscopic redshifts.
This is the best combination of area, depth, number of sightlines, and redshift quality 
for any study of the SMD at $z < 1$ and is also superior to any data
used to date for SFRD studies at similar redshifts. Deeper ground-based NIR data 
were used by Bielby \etal (2012), who analyzed four 
fields covering 2.1~deg$^2$, and by several studies of the single 2-deg$^2$ COSMOS 
field (e.g., Ilbert \etal 2013).  New wide-area surveys such as VISTA VIDEO 
(Jarvis 2012) (ground-based NIR, three fields, 12 deg$^2$) SERVS 
(Mauduit \etal 2012) ({\it Spitzer} IRAC, five fields, 18~deg$^2$, overlapping VIDEO)
will provide excellent new data to improve mass function estimates, but only 
if adequate redshift information and supporting optical photometry are available.

\subsubsection{$1 < z < 4$}

At $z > 1$, deep surveys are needed to probe typical ($L^\ast$ and fainter)
luminosities and to directly detect the majority of cosmic star formations. 
In principle, many fields have suitable data. However, in practice, a few survey 
fields have been re-analyzed many times, in part because they have the best spectroscopic 
and photometric redshift measurements and in part because they have the richest 
multiwavelength data. As such, they are magnets for studies of all kinds. 
In the UV, the 1,500-\AA\ rest-frame is easily probed at $z > 1.5$ using ground-based
or {HST} optical imaging.  At $z \approx 2$ and 3, the surveys of LBGs
by Steidel and collaborators (e.g., Reddy \& Steidel 2009) cover many independent 
sightlines. They offer excellent control of cosmic variance and have outstanding 
spectroscopic calibration.  UV selection is biased against dusty star formation,
but for pure measurements of the UVLF, their surveys are arguably
definitive.  The best direct measurements of dust emission from {\it Spitzer} and {\it Herschel}
are limited to fewer fields, and still fewer have the very deep data needed to probe 
galaxies near $L^\ast$.  As described above, most analysis has revolved around GOODS
and COSMOS, and even in GOODS, the {\it Herschel} data only barely reach $L^\ast$ at $z = 2$.
The faint-end slope of the IRLF is not directly constrained
by individually detected sources at $z > 2$, leading to potentially large uncertainties
in any extrapolation to the bolometric IR luminosity density.  With no more sensitive
FIR space missions on the horizon, there is little prospect for wide-area surveys
to fainter flux limits.  ALMA can reach fainter submillimeter continuum limits and
detect dust continuum from ordinary galaxies at very high redshifts, but only for
very small solid angles. Mosaics of hundreds or even thousands of pointings would be 
needed to survey a field the size of GOODS.  Well-designed observations targeting 
intelligently selected galaxy samples, rather than panoramic mapping, may be required.

Stellar masses below the characteristic mass $M^\ast$ can be probed using the best 
ground-based NIR data, and using {\it Spitzer} IRAC even with modest integration 
times. Thus, more fields have been analyzed, although GOODS and COSMOS still 
tend to dominate the literature.  The extended {\it Spitzer} warm mission generated 
a wealth of valuable data that has yet to be fully exploited.
In practice, there is uncannily good agreement between most determinations of the 
SMF and density at $1 < z < 4$, and it is not clear that new
surveys are needed, rather than more sophisticated analysis of the existing data.
However, the CANDELS {HST} WFC3 survey reaches fainter multiband NIR 
fluxes in fields that already have the deepest IRAC data and, hence, provides an
important opportunity for measuring photometric redshifts and stellar masses for
galaxies fainter than the limits of most studies to date.  This should provide
a better constraint on the slope of the SMF at low masses.

\subsubsection{$z > 4$}

At $z \geq 4$, the large majority of UV-based SFR and SMD 
measurements have been derived from {HST} data, mainly in the GOODS fields and 
the HUDF plus its deep parallel fields (also located in and around GOODS-South).  
Although ground-based imaging can select galaxies at $z \geq 4$, in practice the {HST} 
surveys have gone deeper, especially at the very red optical wavelengths ($I$- and $z$-bands)
and the NIR wavelengths needed to select galaxies at the highest redshifts;
only with these data can LF measurements probe significantly
fainter than $L^\ast$.  GOODS also has the deepest {\it Spitzer} IRAC data, essential for 
deriving stellar masses at $z > 4$, where even the $K$-band samples rest-frame UV 
wavelengths.  Although the comoving volumes in the GOODS fields are significant at 
these large redshifts (nearly $10^6$ comoving Mpc$^3$ at $z = 4$ for the combined 
GOODS fields), one may still worry about clustering and cosmic variance, given only 
two sightlines.   The CANDELS program (Grogin \etal 2011, Koekemoer \etal 2011)
is obtaining multiband optical and NIR {HST} imaging over five fields, each
similar in size, including the two GOODS fields. This program is supplemented by very deep
IRAC imaging from SEDS (Ashby \etal 2013) and S-CANDELS (G. Fazio, research in progress) 
and will help with cosmic variance control and improved statistics.

At $6 < z < 8$, the HUDF data become essential to get any handle on the faint-end slope 
of the LF, and one must worry about the very small field size.   
At $z > 8$, the HUDF (or lensing clusters) are needed to detect convincing candidates. 
Hence, nearly all the literature on the SFRD or stellar masses
at these redshifts consists of serial and parallel analyses of the same HUDF data
sets, as well as recent work from CLASH (Zheng \etal 2012, Coe \etal 2013).
Here, the new Hubble Frontier Fields should be important.  This program, to be
executed from 2013- to 2016, will obtain multiband optical and NIR {HST} imaging with near-HUDF depth
along with ultradeep {\it Spitzer} IRAC data for six massive galaxy clusters. Their lensing potentials 
will magnify the faintest high-redshift background galaxies.  In addition, there will be six new ``blank'' parallel fields 
that will improve statistics for unlensed sources and provide cosmic variance control 
for analyses that now depend on the HUDF (and its satellites) alone.  

At $z > 4$, galaxies detected in ground-based NIR data (and at $z > 3$ for {HST} observations in the reddest WFC3-IR filters)
are observed at rest-frame UV wavelengths.  Hence, even with CANDELS, the HUDF, and the Frontier Fields, {HST} selection will be biased against quiescent
or dusty galaxies.   Massive galaxies with huge SFRs that are
detected at submillimeter wavelengths but invisible even out to the $K$-band
have been detected. Redshifts $z > 4$ have been confirmed from CO measurements
(e.g., Daddi \etal 2009), but it is unknown how much they might contribute
to the SFR or SMD. IRAC selection 
should, in principle, provide a less biased census (e.g., Mancini \etal 2009, Huang \etal 2011, Caputi \etal 
2011, 2012), but spectroscopic confirmation or even photometric redshift
estimates for this population may prove to be very difficult before the launch of the {JWST}.

\section{FROM OBSERVATIONS TO GENERAL PRINCIPLES}
\label{sec:obs_to_para}

Equation \ref{chem_ev} was first used by Lanzetta \etal (1995) to study the chemical evolution of the 
damped \Lya\ absorption systems, where one infers the comoving rate of star formation from the 
observed cosmological mass density of \HI. Pei \& Fall (1995) then generalized it to models with inflows and outflows.
Madau \etal (1996, 1998b) and Lilly \etal (1996) developed a different method where data from galaxy 
surveys were used to infer the SFRD $\psi(t)$ directly.  
This new approach relies on coupling the equations of chemical evolution to the spectrophotometric properties of the 
cosmic volume under consideration. The specific luminosity density at time $t$ of a ``cosmic stellar population'' characterized by an
SFRD $\psi(t)$ and a metal-enrichment law $Z_\ast(t)$ is given by the convolution integral
\begin{equation}
\rho_{\rm \nu}(t)=\int_0^t \psi(t-\tau){\cal L}[\tau,Z_\ast(t-\tau)]d\tau,
\label{eq:rhofuv}
\end{equation}
where ${\cal L}_\nu[\tau,Z_\ast(t-\tau)]$ is the specific luminosity density radiated per unit initial stellar mass by 
a SSP of age $\tau$ and metallicity $Z_\ast(t-\tau)$. The theoretical calculation of ${\cal L}_\nu$ requires stellar 
evolutionary tracks, isochrones, and stellar atmosphere libraries. As an illustrative example of this technique, we provide in this section
a current determination of the SFH of the Universe and discuss a number of possible implications. 

\subsection{Star-Formation Density}
\label{sec:sfrd}

Rather than trying to be exhaustive, we base our modeling below on a limited number of contemporary 
(mostly post-2006) galaxy surveys (see Table 1).  For the present purpose, we consider only surveys that have measured 
SFRs from rest-frame FUV (generally 1,500\,\AA) or MIR and FIR measurements.
Other surveys of nebular line or radio emission are also important, but they provide more limited or indirect 
information as discussed in previous sections (Sections \ref{sec:surveys_emline} and \ref{sec:surveys_radio}).
For the IR measurements, we emphasize 
surveys that make use of FIR data from {\it Spitzer} or {\it Herschel}, rather than relying on 
MIR (e.g., {\it Spitzer} 24-$\mu$m) measurements alone, owing to the complexity and lingering 
uncertainty over the best conversions from MIR luminosity to SFR, particularly 
at high redshift or high luminosity.  In a few cases, we include older measurements when they are the
best available, particularly for local luminosity densities from {IRAS} or {GALEX}, or {GALEX}-based
measurements at higher redshifts that have not been updated since 2005.

For rest-frame FUV data, we use local {GALEX} measurements by Wyder \etal (2005) and Robotham \& Driver (2011)
and also include the 1,500-\AA\ {GALEX} measurements at $z < 1$ from Schiminovich \etal (2005).
We use the FUV luminosity densities of Cucciati \etal (2012) at $0.1 < z < 4$, noting that for $z < 1$
these are extrapolations from photometry at longer UV rest-frame wavelengths.  At $1 \lta  z \lta 3$) 
we also use FUV luminosity densities from Dahlen \etal (2007) and Reddy \& Steidel (2009).  
At redshifts $4 \le z \le 8$, there are a plethora of HST-based studies, with some groups of authors 
repeatedly re-analyzing samples in GOODS and the HUDF as new and improved data have accumulated.  
We restrict our choices to a few of the most recent analyses, taking best-fit Schechter parameters 
$(\phi^\ast, L^\ast, \alpha)$ from Bouwens \etal (2012b) and Schenker \etal (2013).  For the 
present analysis, we stop at $z = 8$ and do not consider estimates at higher redshifts.

For local IR estimates of the SFRD, we use {IRAS} LFs from Sanders \etal (2003) 
and Takeuchi \etal (2003).  At $0.4 < z < 2.3$, we include data from Magnelli \etal (2009, 2011), 
who used stacked {\it Spitzer} 70-$\mu$m measurements for 24-$\mu$m-selected sources.
We also use the {\it Herschel} FIRLFs of Gruppioni \etal (2013) 
and Magnelli \etal (2013).  Although both groups analyze data from the GOODS fields, Gruppioni \etal (2013) incorporate
wider/shallower data from COSMOS. By contrast, Magnelli \etal (2013) include the deepest 100- and 160-$\mu$m
data from GOODS-{\it Herschel}, extracting sources down to the faintest limits using 
24-$\mu$m prior positions.

All the surveys used here provide best-fit LF parameters -- generally Schechter
functions for the UV data, but other functions for the IR measurements, such as double power laws 
or the function by Saunders \etal (1990).  These allow us to integrate the LF down 
to the same {relative} limiting luminosity, in units of the characteristic luminosity $L^\ast$.  
We adopt an integration limit $L_{\rm min}=0.03\,L^\ast$ when computing the luminosity density 
$\rho_{\rm FUV}$ or $\rho_{\rm IR}$.  For the case of a Schechter function, this integral is
\begin{equation}
\rho_{\rm FUV}(z)= \int_{0.03L^\ast}^\infty L\phi(L,z){\rm d}L = \Gamma(2+\alpha,0.03)\phi^\ast L^\ast.
\end{equation}
Here $\alpha$ denotes the faint-end slope of the Schechter parameterization, and $\Gamma$ is the incomplete gamma function.
The integrated luminosity density has a strong dependence on $L_{\rm min}$ at high redshift, where the faint-end slope 
is measured to be very steep, i.e., $\alpha=-2.01\pm 0.21$ at $z\sim 7$ and $\alpha=-1.91\pm 0.32$ at $z\sim 8$ (Bouwens \etal 2011b). 
Slopes of $\alpha\lta -2$ lead to formally divergent luminosity densities. Our choice of a limiting luminosity that is 3.8 magnitudes 
fainter than $L^\ast$, although it samples a significant portion of the faint-end of the FUV LF, requires only a mild extrapolation (1.3 mag) from 
the deepest HST WFC3/IR observations ($\sim 2.5$ mag beyond $L^\ast$ at $z\sim 5-8$) of the HUDF (Bouwens \etal 2011b).  For the IR data, 
we use analytic or numerical integrations depending on the LF form
adopted by each reference, but the same faint-end slope considerations apply.  (Note, however, that some authors use logarithmic
slopes for IRLFs, which differ from the linear form used in the standard Schechter formula by $\Delta \alpha = +1$.)

\begin{figure}[ht]
\vspace{0cm}
\centerline{\psfig{figure=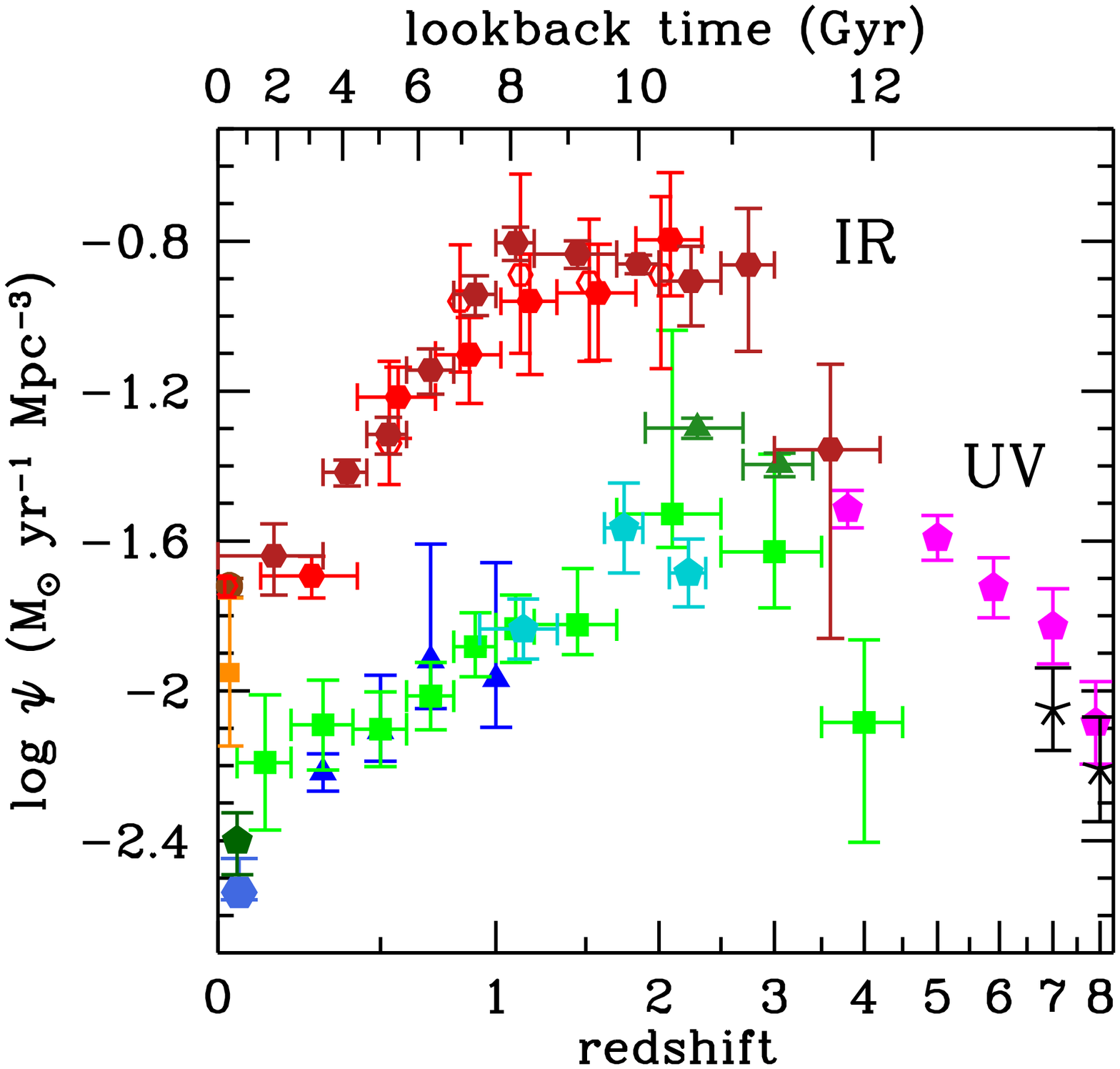,width=0.52\textwidth}
\psfig{figure=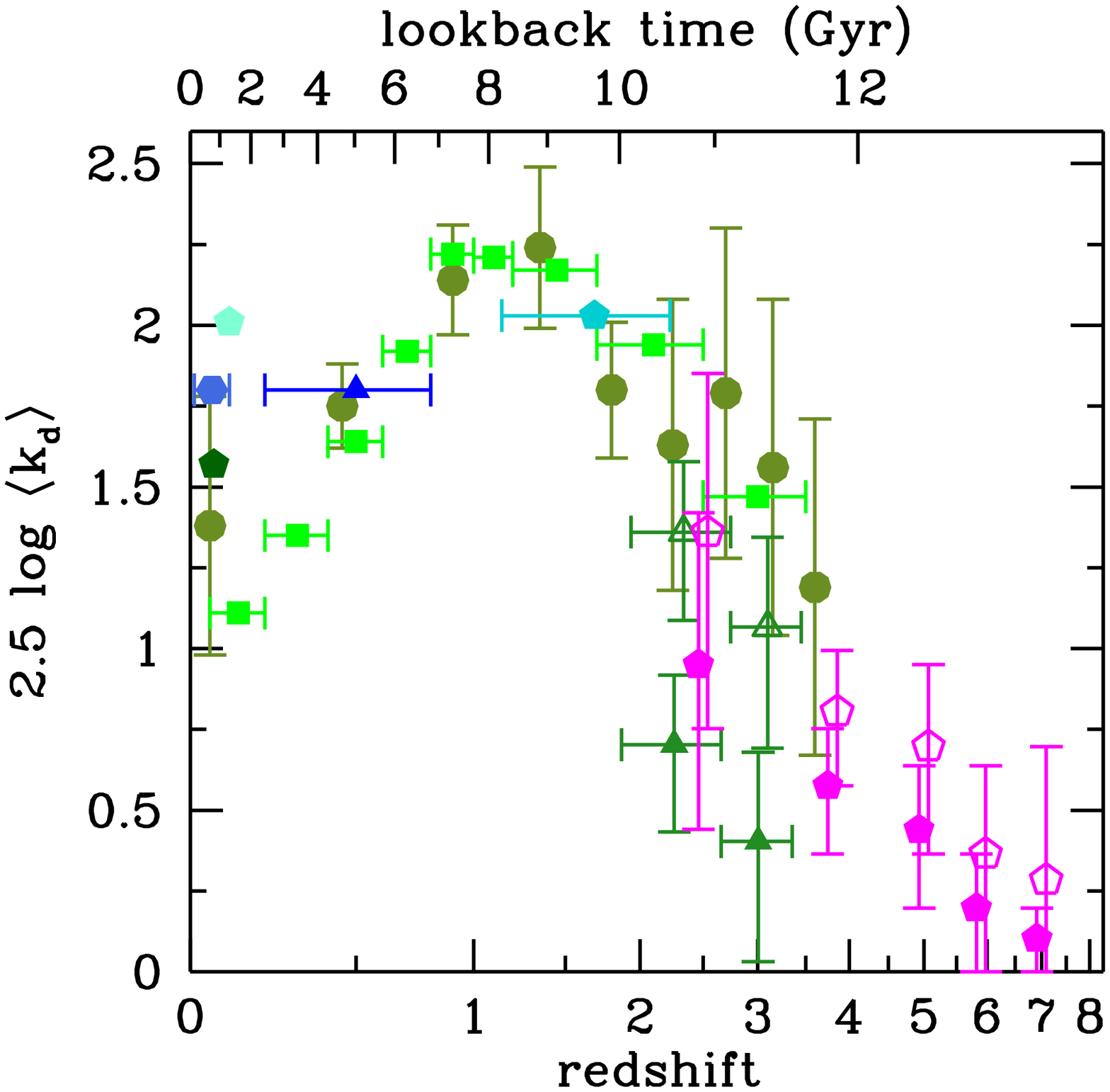,width=0.52\textwidth}}
\vspace{-0cm}
\caption{\footnotesize 
({\it Left panel}) SFR densities in the FUV (uncorrected for dust attenuation) and in the FIR. 
The data points with symbols are given in Table 1.  All UV and IR luminosities have been converted to 
instantaneous SFR densities using the factors ${\cal K}_{\rm FUV}=1.15 \times 10^{-28}$ 
and ${\cal K}_{\rm IR}=4.5 \times 10^{-44}$ (cgs units) valid for a Salpeter IMF.  
({\it Right panel}) Mean dust attenuation in magnitudes as a function of redshift. Most
of the data points shown are based on ultraviolet spectral slopes or stellar population model 
fitting.  The symbol shapes and colors correspond to the data sets cited in Table 1,
with the addition of Salim \etal (2007) ({\it cyan pentagon}).  Two versions of the 
attenuation factors are shown for UV-selected galaxies at $2 < z < 7$ (Reddy \& Steidel 2009,  
Bouwens \etal 2012a) (offset slightly in the redshift axis for clarity):  one integrated over 
the observed population ({\it open symbols}), the other extrapolated down down to $L_{\rm FUV} = 0$
({\it filled symbols}).  Data points from Burgarella \etal (2013) ({\it olive green dots})
are calculated by comparing the integrated FIR and FUV luminosity densities in redshift bins, 
rather than from the UV slopes or UV-optical SEDs.
}
\label{fig8}
\end{figure}

Multiplying the integrated FUV and IR comoving luminosity densities by the conversion
factors ${\cal K}_{\rm FUV}$ (Section \ref{sec:sfrates_uv}) and ${\cal K}_{\rm IR}$ (Section \ref{sec:sfrates_ir}),
we obtain measurements of the ``observed'' UV and IR SFRDs (shown in Figure~\ref{fig8}). Here, the FUV measurements are {uncorrected} for dust 
attenuation.  This illustrates the now well-known result that most of the energy from star-forming galaxies 
at $0 < z < 2$ is absorbed and reradiated by dust;  only a minority fraction emerges directly 
from galaxies as UV light.   The gap between the UV and IR measurements increases with
redshift out to at least $z \approx 1$ and then may narrow from $z = 1$ to 2.
Robust measurements of the FIR luminosity density are not yet available at $z > 2.5$.

Clearly, a robust determination of dust attenuation is essential to transform FUV 
luminosity densities into total SFRDs.  Figure \ref{fig8} shows measurements 
of the effective dust extinction, $\langle k_d\rangle$, as a function of redshift.
This is the multiplicative factor needed to correct the observed FUV luminosity density 
to the intrinsic value before extinction, or equivalently, 
$\langle k_d\rangle=\rho_{\rm IR}/\rho_{\rm FUV}+1$ (e.g., Meurer \etal 1999).  
For most of the data shown in Figure \ref{fig8},
the attenuation has been estimated from the UV spectral slopes of star-forming galaxies using the 
attenuation--reddening relations from Meurer \etal (1999) or Calzetti \etal (2000) or occasionally from 
stellar population model fitting to the full UV--optical SEDs of galaxies integrated over the observed 
population (e.g., Salim \etal 2007, Cucciati \etal 2012).  
Robotham \& Driver (2011) used the empirical attenuation correction of Driver \etal (2008).  
We note that the estimates of UV attenuation in the local Universe span a broad range,
suggesting that more work needs to be done to firmly pin down this quantity
(and perhaps implying that we should be cautious about the estimates at higher redshift).
Several studies of UV-selected galaxies at $z \geq 2$ (Reddy \& Steidel 2009, Bouwens \etal 2012a, Finkelstein 
\etal 2012b) have noted strong trends for less luminous galaxies as having bluer UV spectral slopes and, hence, lower 
inferred dust attenuation.  Because the faint-end slope of the far-UV luminosity function (FUVLF) is so steep 
at high redshift, a large fraction of the reddened FUV luminosity density is emitted by galaxies much 
fainter than $L^\ast$; this extinction--luminosity trend also implies that the net extinction for the 
entire population will be a function of the faint integration limit for the sample.  
In Figure \ref{fig8}, the points from Reddy \& Steidel (2009) (at $z = 2.3$ and 3.05) and from 
Bouwens \etal (2012a) (at $2.5 \leq z \leq 7$) are shown for two faint-end integration limits: These are 
roughly down to the observed faint limit of the data, $M_{\rm FUV} < -17.5$ to $-17.7$ for the different 
redshift subsamples and extrapolated to $L_{\rm FUV} = 0$.  The net attenuation for the brighter limit, 
which more closely represents the sample of galaxies actually observed in the study, is significantly 
larger than for the extrapolation -- nearly two times larger for the Reddy \& Steidel (2009) samples, 
and by a lesser factor for the more distant objects from Bouwens \etal (2012a).  In our analysis 
of the SFRDs, we have adopted the mean extinction factors inferred by each 
survey to correct the corresponding FUV luminosity densities.

Adopting a different approach, Burgarella \etal (2013) measured total UV attenuation from the ratio
of FIR to observed (uncorrected) FUV luminosity densities (Figure ~\ref{fig8}) as a function of redshift, using FUVLFs 
from Cucciati \etal (2012) and {\it Herschel} FIRLFs from Gruppioni \etal (2013).
At $z < 2$, these estimates agree reasonably well with the measurements inferred from the UV slope or from SED fitting.
At $z > 2$, the FIR/FUV estimates have large uncertainties owing to the similarly large uncertainties
required to extrapolate the observed FIRLF to a total luminosity density.
The values are larger than those for the UV-selected surveys, particularly when compared with the
UV values extrapolated to very faint luminosities.  Although galaxies with lower
SFRs may have reduced extinction, purely UV-selected samples at high redshift may also be 
biased against dusty star-forming galaxies.  As we noted above, a robust census for star-forming galaxies at $z \gg 2$ 
selected on the basis of dust emission alone does not exist, owing to the sensitivity limits of past and present FIR and submillimeter observatories.
Accordingly, the total amount of star formation that is missed from UV surveys at such high redshifts
remains uncertain.

\begin{figure}[ht]
\vspace{0.cm}
\centerline{\psfig{figure=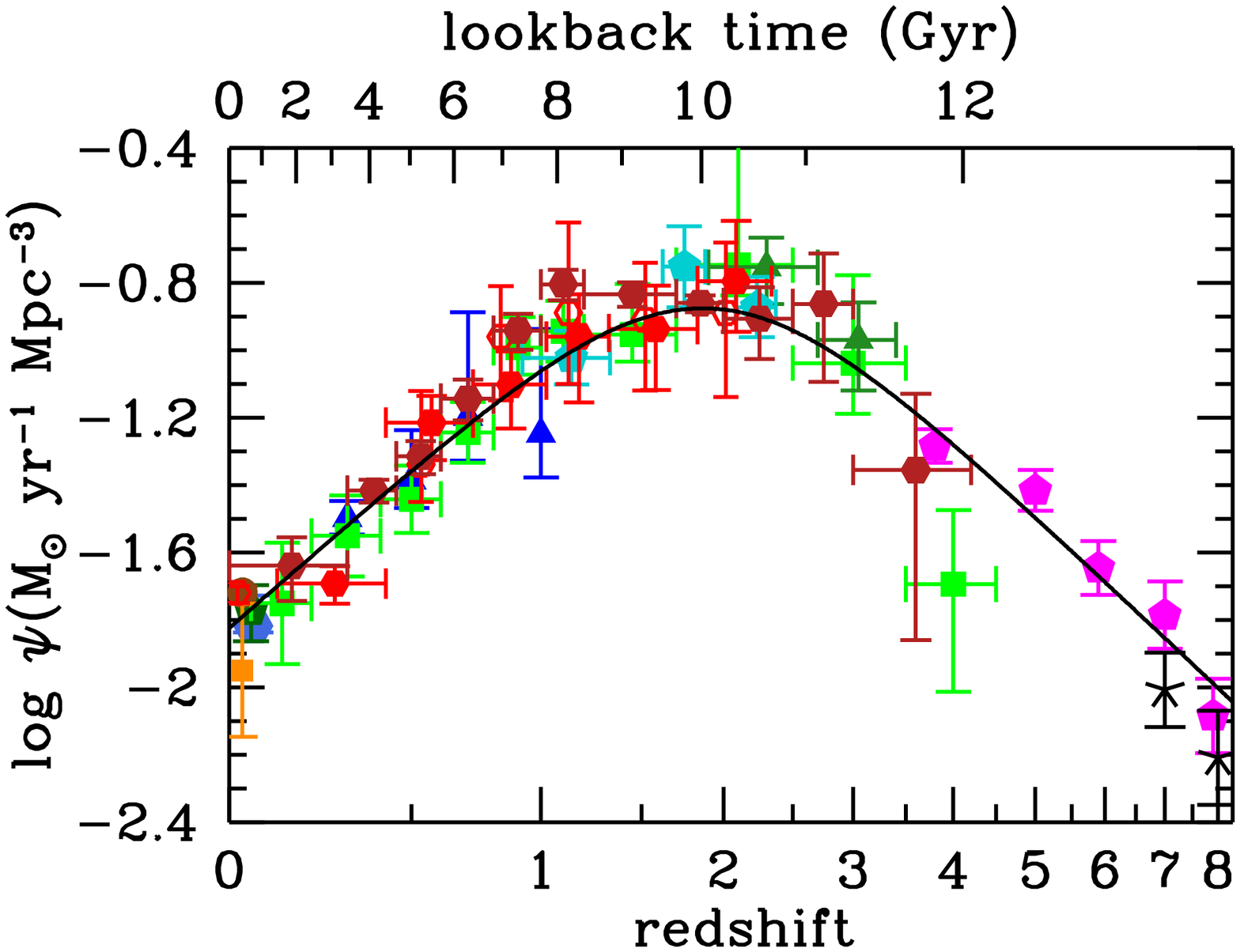,width=0.65\textwidth}
\psfig{figure=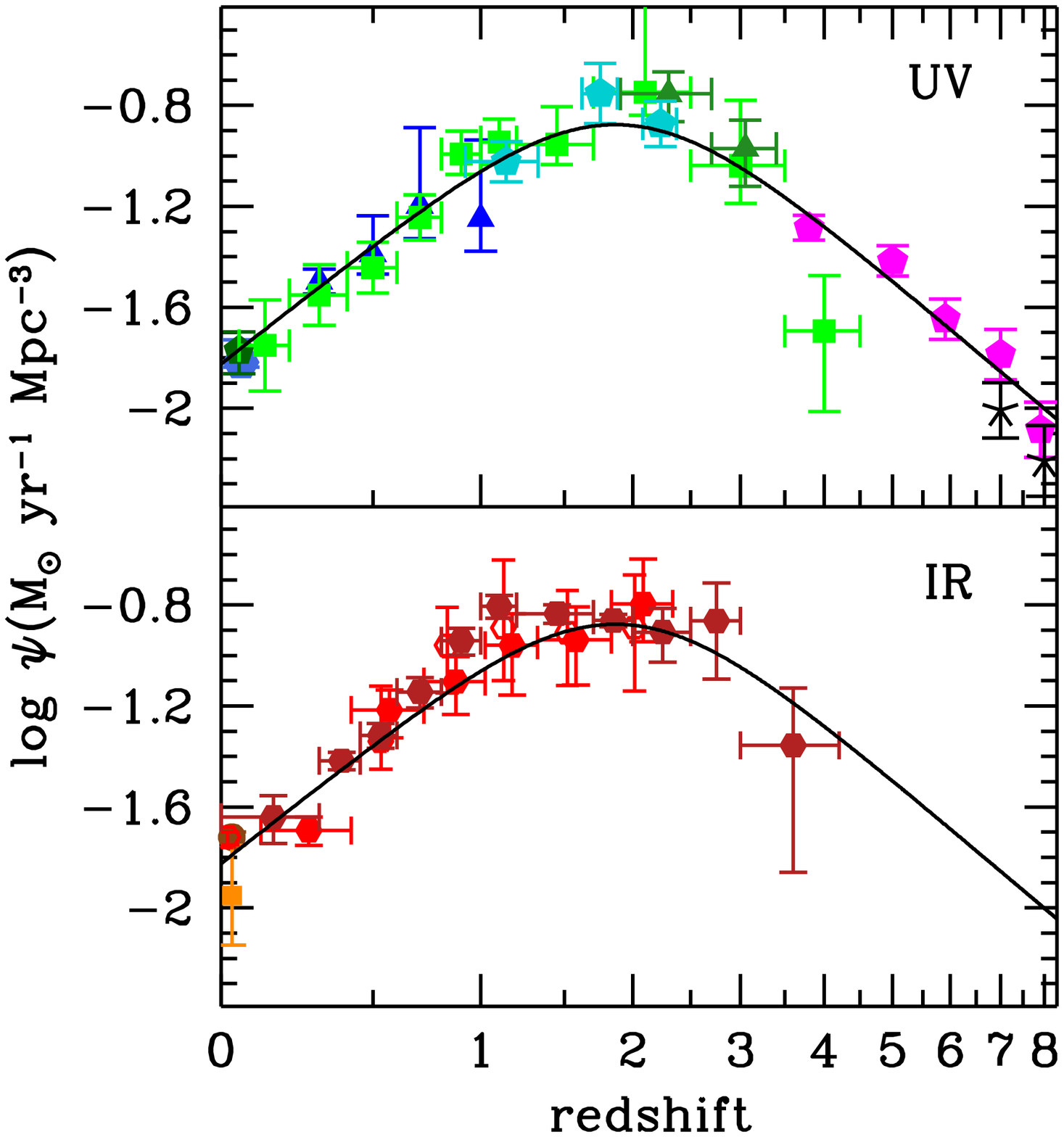,width=0.6\textwidth,height=0.6\textwidth}}
\vspace{-0.cm}
\caption{\footnotesize The history of cosmic star formation from ({top right panel}) FUV, 
({bottom right panel}) IR, and ({left panel}) FUV+IR rest-frame measurements. 
The data points with symbols are given in Table 1. 
All UV luminosities have been converted to instantaneous SFR densities using the 
factor ${\cal K}_{\rm FUV}=1.15 \times 10^{-28}$ (see Equation \ref{eq:KFUV}), valid for a Salpeter IMF. 
FIR luminosities (8--1,000$\,\mu$m) have been converted to instantaneous SFRs using the factor
${\cal K}_{\rm IR}=4.5 \times 10^{-44}$ (see Equation \ref{eq:KIR}), also valid for a Salpeter IMF. 
The solid curve in the three panels plots the best-fit SFRD in Equation \ref{eq:sfrd}. 
}
\label{fig9}
\end{figure}

Figure~\ref{fig9} shows the cosmic SFH 
from UV and IR data following the above prescriptions, as well as the best-fitting function
\begin{equation}
\psi(z)=0.015\,{(1+z)^{2.7}\over 1+[(1+z)/2.9]^{5.6}}\,\sfrd. 
\label{eq:sfrd}
\end{equation}
These state-of-the-art surveys provide a remarkably consistent picture of the cosmic SFH:
a rising phase, scaling as $\psi(z) \propto (1+z)^{-2.9}$ at $3 \lta z \lta 8$, slowing and peaking at some
point probably between $z = 2$ and 1.5, when the Universe was $\sim 3.5$ Gyr old, followed by a gradual
decline to the present day, roughly as  $\psi(z) \propto (1+z)^{2.7}$. The comoving SFRD at redshift 7 was 
approximately the same as that measured locally. The increase in $\psi(z)$ from $z \approx 8$ to 3 appears to have 
been steady, with no sharp drop at the highest redshifts, although there is now active debate in the literature about
whether that trend continues or breaks at redshifts 9 and beyond (Coe \etal 2013, Ellis \etal 2013, Oesch \etal 2013).
Although we have adopted a fitting function  that is a double power-law in $(1+z)$, we note that the SFRD data at $z < 1$  
can also be fit quite well by an exponential  decline with cosmic time and an e-folding timescale of 3.9 Gyr. Compared with
the recent empirical fit to the SFRD by Behroozi \etal (2013),  the function in Equation \ref{eq:sfrd} reaches its 
peak at a slightly higher redshift, with a lower maximum value of $\psi$ and with slightly shallower rates of change at both
lower and higher redshift, and produces 20\% fewer stars by $z=0$.

We also note that each published measurement has its 
own approach to computing uncertainties on the SFRD and takes different random and systematic factors into account, and we have made no attempt to 
rationalize these here.  Moreover, the published studies integrate their measurements down to different luminosity limits.   We have instead adopted 
a fixed threshold of $0.03\,L^\ast$ to integrate the published LFs, and given the covariance on the measurements and uncertainties of 
LF parameters, there is no simple way for us to correct the published uncertainties to be appropriate for our adopted integration limit. 
Therefore, we have simply retained the fractional errors on the SFRD measurements published by each author without modification to provide 
an indication of the relative inaccuracy derived by each study. These should not be taken too literally, especially when there
is significant difference in the faint-end slopes of LFs reported in different studies, which can lead to large differences
in the integrated luminosity density.  Uncertainties in the faint-end slope and the resulting extrapolations are not always fully
taken into account in published error analyses, especially when LFs are fit at high redshift by fixing the slope to some value
measured only at lower redshift.

\begin{figure}[ht]
\vspace{-1cm}
\centerline{\psfig{figure=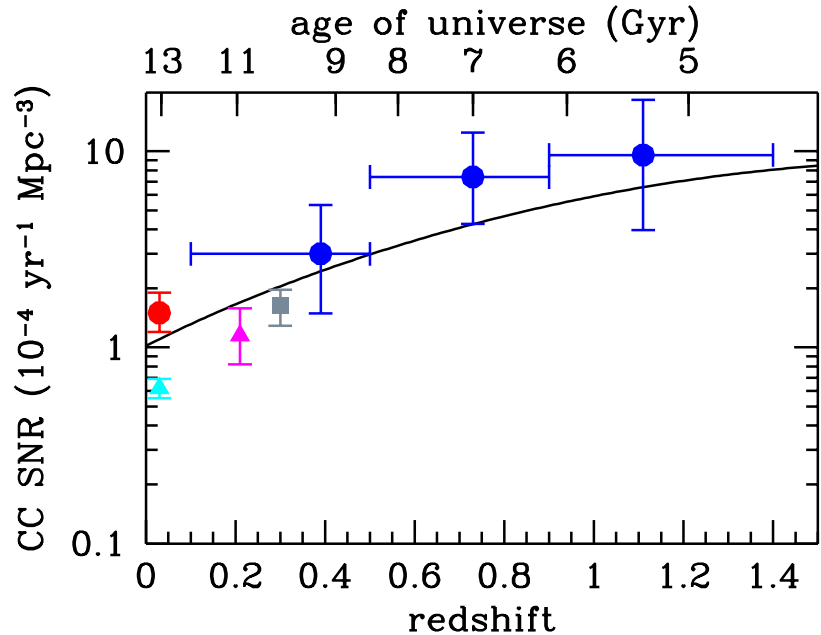,width=0.9\textwidth}}
\vspace{-1cm}
\caption{\footnotesize The cosmic core-collapse SN rate. The data points are taken from 
Li \etal (2011) ({\it cyan triangle}), Mattila \etal (2012) ({\it red dot}), Botticella \etal (2008) ({\it magenta triangle}), 
Bazin \etal (2009) ({\it gray square}), and Dahlen \etal (2012) ({\it blue dots}).  
The solid line shows the rates predicted from our fit to the cosmic star-formation history. The local overdensity in star formation may 
boost the local rate within 10--15 Mpc of Mattila \etal (2012). 
}
\label{fig10}
\end{figure}

\subsection{Core-Collapse Supernova Rate}

Because core-collapse supernovae (CC SNe) (i.e., Type II and Ibc SNe) originate from massive, short-lived stars, the rates of these events 
should reflect ongoing star formation and offer an independent determination of the cosmic star formation and metal production rates at different 
cosmological epochs (e.g., Madau \etal 1998a, Dahlen \etal 2004). While poor statistics and dust obscuration are major liming factors for using   
CC SNe as a tracer of the SFH of the Universe, most derived rates are consistent with each other and increase with lookback time between 
$z=0$ and $z\sim 1$ (see Figure \ref{fig10}). The comoving volumetric SN rate is determined by multiplying Equation \ref{eq:sfrd}   
by the efficiency of forming CC SNe
\begin{equation}
R_{\rm CC}(z)=\psi(z)\times {\int_{m_{\rm min}}^{m_{\rm max}}\phi(m){\rm d}m \over  \int_{m_l}^{m_u}m\phi(m){\rm d}m}\equiv \psi(z)\times k_{\rm CC},   
\label{eq:CC}
\end{equation}
where the number of stars that explode as SNe per unit mass is $k_{\rm CC}=0.0068$ M$_{\odot}^{-1}$ for a Salpeter IMF, $m_{\rm min}=8\,\msun$, 
and $m_{\rm max}=40\,\msun$. The predicted cosmic SN rate is shown in Figure \ref{fig10} and appears to be in good agreement with the data. 
The IMF dependence in Equation \ref{eq:CC} is largely canceled out by the IMF dependence of the derived SFRD 
$\psi(z)$, as the stellar mass range probed by SFR indicators is comparable to the mass range of stars exploding as CC SNe. Recent comparisons between 
SFRs and CC SN rates have suggested a discrepancy between the two rates: The numbers of CC SNe detected are too low by a factor of approximately
(Horiuchi \etal 2011). Our revised cosmic SFH does not appear to show such systematic discrepancy (see also Dahlen \etal 2012).

\begin{figure}[ht]
\vspace{-1cm}
\centerline{\psfig{figure=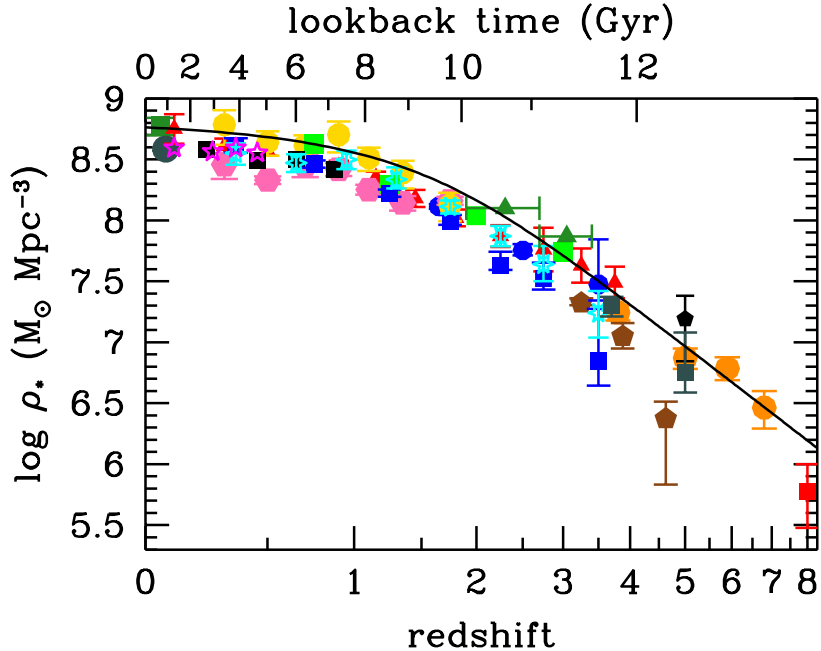,width=0.9\textwidth}}
\vspace{-1cm}
\caption{\footnotesize  The evolution of the stellar mass density. The data points with symbols are given 
in Table 2. The solid line shows the global stellar mass density obtained by integrating the best-fit instantaneous 
star-formation rate density $\psi(z)$ (Equations \ref{eq:rhostar} and \ref{eq:sfrd}) with a return fraction $R=0.27$. 
}
\label{fig11}
\end{figure}

Observations show that at least some long-duration gamma-ray bursts (GRBs) happen simultaneously with CC SNe, 
but neither all SNe, nor even all SNe of Type Ibc produce GRBs (for a review, see Woosley \& Bloom 2006). In principle,
the rate of GRBs of this class could provide a complementary estimate of the SFRD 
(e.g., Porciani \& Madau 2001), but it is only a small fraction ($<$1\% after correction for beaming) of 
the CC SN rate (Gal-Yam \etal 2006), suggesting that GRBs are an uncommon chapter in the evolution of massive 
stars requiring special conditions that are difficult to model. Recent studies of the GRB-SFR 
connection have claimed that GRBs do not trace the SFR in an unbiased way, and are more frequent per unit stellar mass 
formed at early times (Kistler \etal 2009, Robertson \& Ellis 2012, Trenti \etal 2012). 

\subsection{Stellar Mass Density}

Figure \ref{fig11} shows a compilation (see also Table 2) of recent (mostly post 2006) measurements of 
the SMD as a function of redshift (for a compilation of older data see Wilkins \etal 2008a).  We show local 
SDSS-based SMDs from Gallazzi \etal (2008), Li \& White (2009), and Moustakas \etal (2013).  Moustakas \etal (2013) 
also measured SMFs at $0.2 < z < 1$. However, at $z > 0.5$ their mass completeness limit is
larger than $10^{9.5}\,\msun$, so we have used their points only below that redshift.  At higher redshifts (as 
in Moustakas \etal 2013), nearly all the modern estimates incorporate {\it Spitzer} IRAC photometry;
we include only one recent analysis (Bielby \etal 2012) that does not but that otherwise uses excellent deep, wide-field NIR 
data in four independent sightlines. We also include measurements at $0.1 < z \lta 4$
from Arnouts \etal (2007), P\'{e}rez-Gonz\'{a}lez \etal (2008), Kajisawa \etal (2009), Marchesini \etal (2009),
Reddy \& Steidel (2009), Pozzetti \etal (2010), Ilbert \etal (2013), and Muzzin \etal (2013).
We show measurements for the IRAC-selected sample of Caputi \etal (2011) at $3 \leq z \leq 5$ and for
UV-selected LBG samples at $4 < z < 8$ by Yabe \etal (2009), Gonz\'{a}lez \etal (2011), Lee \etal (2012), 
and Labb\'{e} \etal (2013).

When needed, we have scaled from a Chabrier IMF 
to a Salpeter IMF by multiplying the stellar masses by a factor of 1.64 (Figure \ref{fig4}).
At high redshift, authors often extrapolate their SMFs beyond the observed range by fitting a Schechter 
function.  Stellar mass completeness at any given redshift is rarely as well defined as luminosity completeness, 
given the broad range of $M/L$ values that galaxies can exhibit.  Unlike the LFs used for the SFRD calculations
where we have tried to impose a consistent faint luminosity limit (relative to $L^\ast$) for integration,
in most cases we have simply accepted whatever low-mass limits or integral values that the various authors reported.  Many authors 
found that the characteristic mass $M^\ast$ appears to change little for $0 < z < 3$ 
(e.g., Fontana \etal 2006, Ilbert \etal 2013) and is roughly $10^{11}\,\msun$ (Salpeter).  Therefore, a low-mass 
integration limit similar to that which we used for the LFs ($L_{\rm min} = 0.03 L^\ast$) would 
correspond to $M_{\rm min} \approx 10^{9.5}\,\msun$ in that redshift range.  A common but by no means universal low-mass 
integration limit used in the literature is $10^8\,\msun$.  Generally, SMFs have
flatter low-mass slopes than do UVLFs (and sometimes IRLFs), so the lower-mass limit 
makes less difference to the SMDs than it does to the SFRDs.

Our model predicts an SMD that is somewhat high ($\sim 0.2$ dex on average, or 60\%) compared with many, but not all, 
of the data at $0 < z \lesssim 3$.  At $0.2 < z < 2$, our model matches the SMD measurements for the {\it Spitzer} 
IRAC-selected sample of Arnouts \etal (2007), but several other modern measurements in this redshift range from 
COSMOS (Pozzetti \etal 2010, Ilbert \etal 2013, Muzzin \etal 2013) fall below our curve.  Carried down to 
$z = 0$, our model is somewhat high compared with several, but not all (Gallazzi \etal 2008), 
local estimates of the SMD (e.g., Cole \etal 2001, Baldry \etal 2008, Li \& White 2009).

Several previous analyses (Hopkins \& Beacom 2006, but see the erratum; Hopkins \& Beacom 2008; Wilkins \etal 2008a) 
have found that the instantaneous SFH over-predicts the SMD by larger factors, up to 0.6 dex at redshift 3.  We find 
little evidence for such significant discrepancies, although there does appear to be a net offset over 
a broad redshift range.  Although smaller, a $\sim 60$\% effect should not be disregarded.   One can imagine
several possible causes for this discrepancy;   we consider several of them here.  

Star formation rates may be overestimated, particularly at high redshift during the peak era of galaxy growth.
For UV-based measurements, a likely culprit may be the luminosity-weighted dust corrections, which could be 
too large, although it is often asserted that UV data are likely to underestimate SFRs in very 
dusty, luminous galaxies.  IR-based SFRs may be overestimated and indeed {were} overestimated for 
some high-redshift galaxies in earlier {\it Spitzer} studies, although this seems less likely now in the era 
of deep {\it Herschel} FIR measurements.  It seems more plausible that the SFRs inferred for individual galaxies
may be correct on average but that the luminosity function extrapolations could be too large. Many authors 
adopt fairly steep ($\alpha \geq -1.6$) faint-end slopes to both the UVLFs and IRLFs for distant galaxies.  
For the UVLFs, the best modern data constrain these slopes quite well, but in the IR current measurements 
are not deep enough to do so.  However, although these extrapolations may be uncertain, the good agreement between 
the current best estimates of the UV- and IR-based SFRDs at $0 < z < 2.5$ (Figure ~\ref{fig9})
does not clearly point to a problem in either one.

Instead, stellar masses or their integrated SMD may be systematically underestimated.  This is not
implausible, particularly for star-forming galaxies, where the problem of recent star formation
``outshining'' older high $M/L$ stars is well known (see Section \ref{sec:sfrd}).  By analyzing
mock catalogs of galaxies drawn from simulations with realistic (and complex) SFHs, 
Pforr \etal (2012) found that the simplifying assumptions that are typically
made when modeling stellar masses for real surveys generally lead to systematically underestimated
stellar masses at all redshifts.  That said, other systematic effects can work in 
the opposite direction and lead to mass overestimates, e.g., the effects of TP-AGB stars on the 
red and NIR light if these are not correctly modeled (Maraston~2005).  
A steeper low-mass slope to the GSMF could also increase the total SMD.  This has been
suggested even at $z = 0$, where mass functions have been measured with seemingly great precision
and dynamic range (e.g., Baldry \etal 2008).  At high redshift, most studies to date have found relatively 
flat low-mass SMF slopes, but galaxy samples may be incomplete (and photometric redshift estimates poor)
for very faint, red, high-$M/L$ galaxies if they exist in significant numbers.  Some recent SMF
determinations using very deep {HST} WFC3 observations have found steeper SMF slopes at $z > 1.5$
(e.g., Santini \etal 2012), and new measurements from extremely deep NIR surveys such as  
CANDELS are eagerly anticipated.   That said, it seems unlikely that the SMF slope at {low}
redshift has been underestimated enough to account for a difference of 0.2~dex in the SMD.

Recent evidence has suggested that strong nebular line emission can significantly affect
broadband photometry for galaxies at high redshift, particularly $z > 3.8$, where H$\alpha$
(and, at $z > 5.3$, [OIII]) enter the {\it Spitzer} IRAC bands (Shim \etal 2011).
Therefore, following Stark \etal (2013), we have divided the SMD of Gonz\'{a}lez \etal (2011) 
at $z\simeq 4, 5, 6,$ and 7 in Figure \ref{fig11} by the factor 1.1, 1.3, 1.6, and 2.4, 
respectively, to account for this effect.  Although considerable uncertainties in these corrections
remain, such downward revision to the inferred early SMD improves consistency 
with expectations from the time-integrated SFRD. 

Alternatively, some authors have considered how changing the IMF may help reconcile SMD($z$)
with the time-integrated SFRD($z$) (e.g., Wilkins \etal 2008b).  Generally, a more top-heavy or 
bottom-light IMF will lead to larger luminosities per unit SFR, hence smaller SFR/$L$ conversion
factors $K$ (Section \ref{sec:sfrates}).   Mass-to-light ratios for older stellar populations will
also tend to be smaller, but not necessarily by the same factor.  Although we have 
used a Salpeter IMF for reference in this review, an IMF with a low-mass turnover (e.g., Chabrier
or Kroupa) will yield a larger mass return fraction $R$ and proportionately lower final 
stellar masses for a given integrated past SFH, by the factor 
$(1 - R_1) / (1 - R_2) = 0.81$, where $R_1$ and $R_2$ = 0.41 and 0.27 for the Chabrier
and Salpeter IMFs, respectively (Section \ref{sec:chemev}).  The apparent offset between the SMD
data and our integrated model $\psi(z)$ can be reduced further to only $\sim 0.1$~dex 
without invoking a particularly unusual IMF.  Given the remaining potential 
for systemic uncertainties in the measurements of SFRDs and SMDs, it seems premature to 
tinker further with the IMF, although if discrepancies remain after further improvements
in the measurements and modeling then this topic may be worth revisiting.

\begin{figure}[ht]
\centerline{\psfig{figure=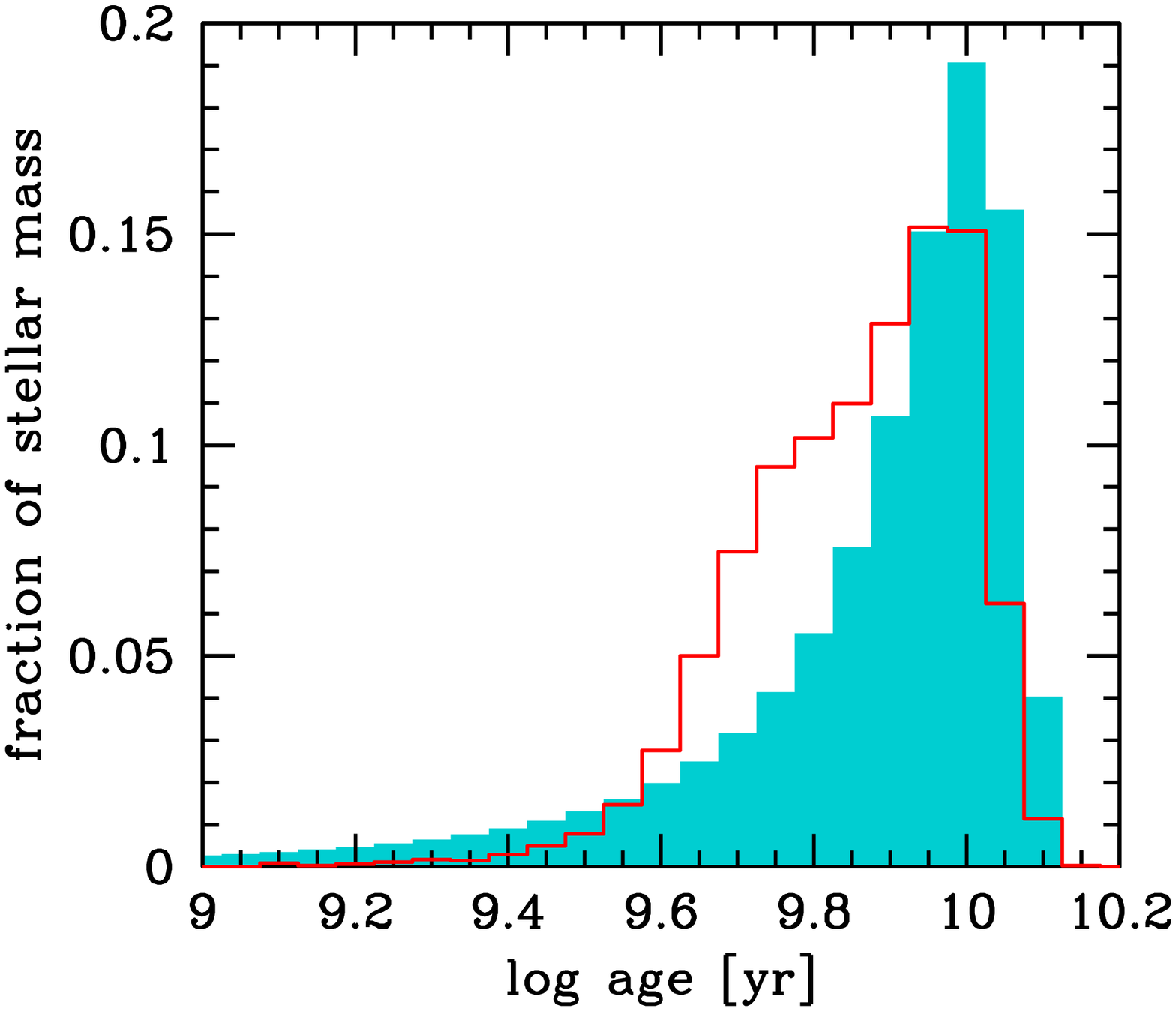,width=0.49\textwidth}
\psfig{figure=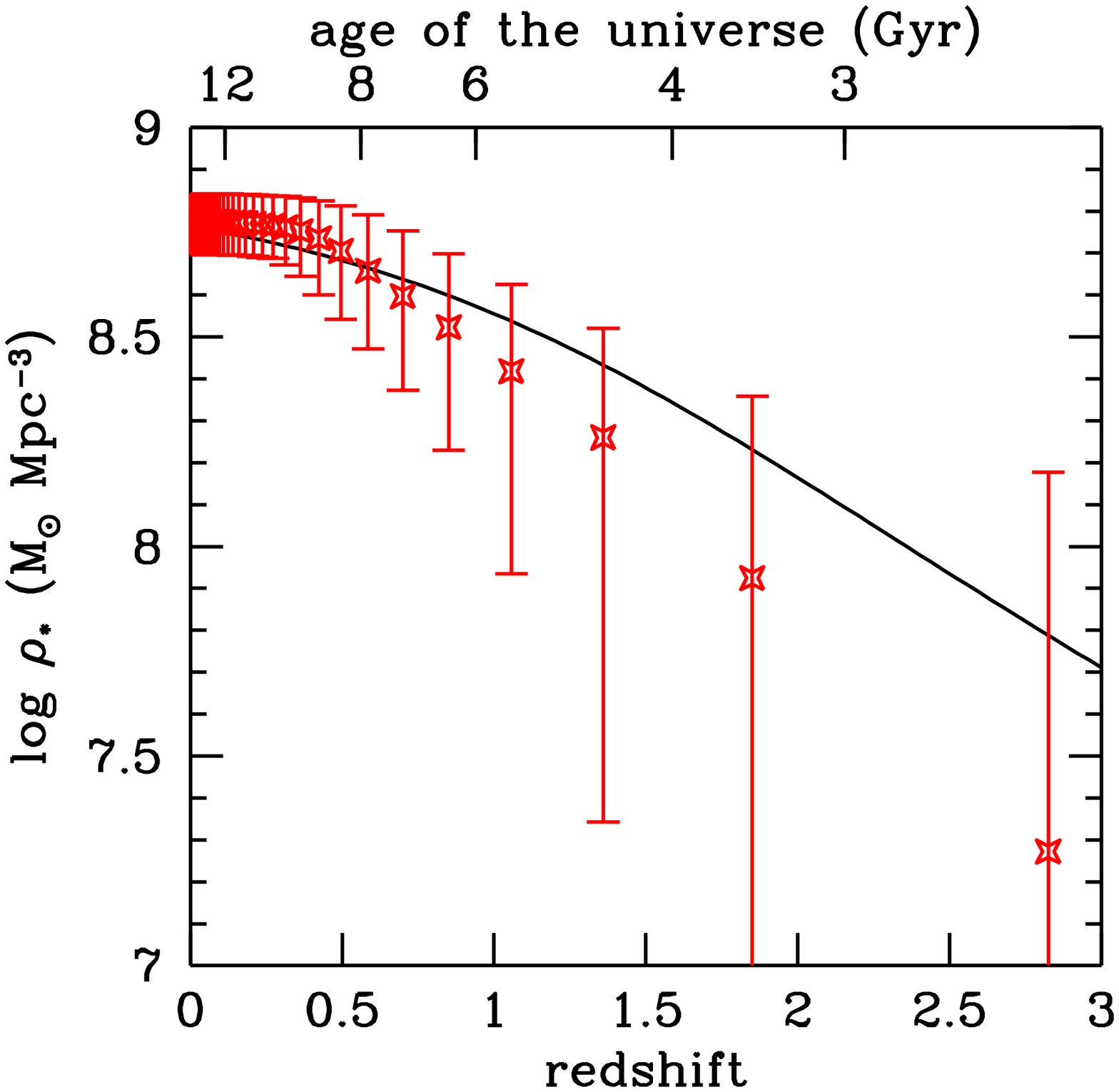,width=0.49\textwidth}}
\vspace{-0.0cm}
\caption{\footnotesize ``Stellar archaeology'' with the SDSS. ({\it Left panel}) Normalized distribution of stellar mass in the local
Universe as a function of age. The red histogram shows estimates from the SDSS (Gallazzi \etal 2008). The measured (mass-weighted) 
ages have been corrected by adding the lookback time corresponding to the redshift at which the galaxy is observed. The turquoise shaded histogram
shows the distribution generated by integrating the instantaneous star-formation rate density in Equation \ref{eq:rhostar}.
({\it Right panel}) Evolution of the SMD with redshift. The points, shown with systematic error bars, are derived from
the analysis of SDSS data by Gallazzi \etal (2008), assuming a Salpeter IMF in the range $0.1-100\,\msun$.
The solid line shows the mass assembly history predicted by integrating our best-fit star-formation history. 
}
\label{fig12}
\end{figure}

\subsection{Fossil Cosmology}

In concordance with estimates from the cosmic SFH, the measurements of the SMD discussed above 
imply that galaxies formed the bulk ($\gta$ 75\%) of their stellar mass at $z<2$. Stars formed in galaxies before 11.5 Gyr are predicted to contribute 
only 8\% of the total stellar mass today. An important consistency check for all these determinations could come from studies 
of the past SFH of the Universe from its present contents. This ``fossil cosmology'' approach has benefited 
from large spectroscopic surveys in the local Universe such as the 2dFGRS (Colless \etal 2001) and the SDSS 
(York \etal 2000), which provide detailed spectral information for hundreds of thousands of galaxies.
Using a sample of $1.7\times 10^5$ galaxies drawn from the SDSS DR2, and comparing the spectrum of each galaxy to a library of 
templates by Bruzual \& Charlot (2003) (the comparison was based on five spectral absorption features, namely D4,000, H$\beta$, 
and H$\delta_{\rm A}$ $+$ H$\gamma_{\rm A}$ as age-sensitive indexes, 
and [Mg$_2$Fe] and [MgFe]' as metal-sensitive indexes), Gallazzi \etal (2008) have constructed a distribution of stellar mass as a function of 
age (for a similar analysis on the SDSS DR3 sample, see also Panter \etal 2007). In Figure \ref{fig12}, we compare this distribution 
with the one predicted by our best-fit cosmic SFH.  The latter implies a mass-weighted mean stellar age,
\begin{equation}
\langle t_{\rm age}\rangle=t_0-\int_0^{t_0} t\psi(t)dt\left[\int_0^{t_0} \psi(t)dt\right]^{-1},
\end{equation}
equal to $\langle t_{\rm age}\rangle =8.3$ Gyr. Both distributions have a peak at $8-10$ Gyr and decline rapidly at younger ages, with the peak age 
corresponding to the formation redshift, $z\sim 2$, where the cosmic star-formation density reaches a maximum. The SDSS distribution, however, appears to be skewed 
toward younger ages. This is partly caused by a bias toward younger populations in the SDSS ``archaeological'' approach, 
where individual galaxies are assigned only an average (weighted by mass or light) age that is closer to the last significant episode of star formation.  
Such bias appears to be reflected in Figure \ref{fig12} where the mass assembly history predicted by our model SFH is compared with that 
inferred by translating the characteristic age of the stellar populations measured by Gallazzi \etal (2008) into a characteristic redshift of formation. 
The agreement is generally good, although the SDSS distribution would predict later star formation and more rapid SMD growth at $z < 2$ 
and correspondingly less stellar mass formed at $z > 2$.  The present-day total SMD derived by Gallazzi \etal (2008) is 
$(6.0\pm 1.0) \times 10^8\,\mdens$ (scaled up from a Chabrier to a Salpeter IMF), in excellent agreement with $\rho_\ast=5.8 \times 10^8\,\mdens$ 
predicted by our model SFH. This stellar density corresponds to a stellar baryon fraction of only 9\% (5\% for a Chabrier IMF).

\begin{figure}[ht]
\vspace{-2cm}
\centerline{\psfig{figure=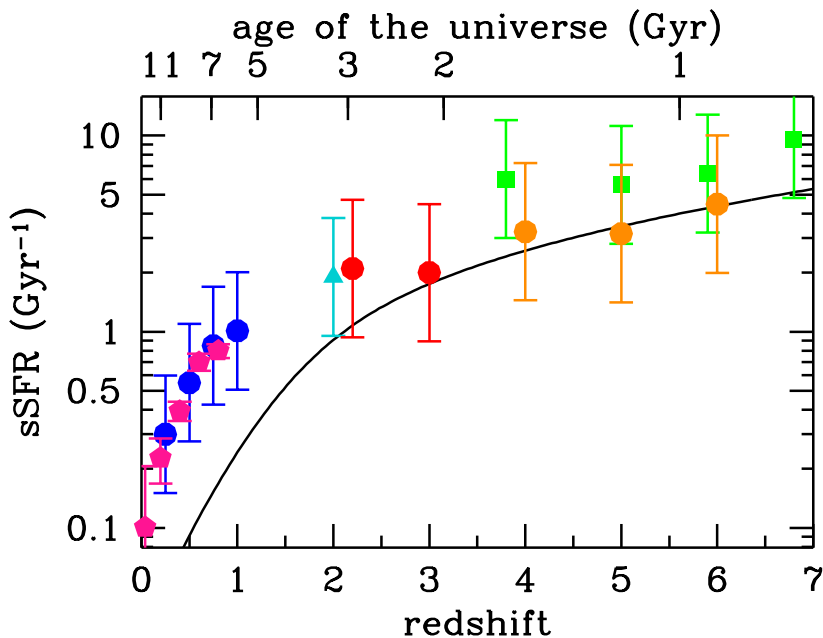,width=0.9\textwidth}}
\vspace{-1cm}
\caption{\footnotesize The mean specific star-formation rate (sSFR$\equiv {\rm SFR}/M_\ast$) for galaxies with estimated stellar masses
in the range $10^{9.4}-10^{10}\,\msun$. The values are from the literature: Daddi \etal (2007) ({\it cyan triangle}), 
Noeske \etal (2007) ({\it blue dots}), Damen \etal (2009) ({\it magenta pentagons}), Reddy \etal (2009) ({\it red dots}), 
Stark \etal (2013) ({\it green squares}), and Gonz\'{a}lez \etal (2014) ({\it orange dots}). The error bars correspond to systematic 
uncertainties.  The high-redshift points from Stark \etal (2013) and Gonz\'{a}lez \etal (2014) have been corrected upward owing 
to the effect of optical emission lines on the derived stellar masses, using their ``fixed H$\alpha$ EW'' model (Stark \etal 2013) and 
``RSF with emission lines'' model (Gonz\'{a}lez \etal 2014). The curve shows the predictions from our best-fit star-formation history. 
}
\label{fig13}
\end{figure}

\subsection{The Global Specific Star Formation Rate}

In recent years, there has been considerable interest in the sSFR (sSFR$\equiv {\rm SFR}/M_\ast$) of galaxies with 
different masses at different times in the history of the Universe. The sSFR describes the fractional growth rate of stellar mass in
a galaxy or, equivalently, the ratio of current to past star formation.  The inverse of the sSFR is the characteristic stellar mass 
doubling time (Guzman \etal 1997).  At $0 < z < 2$ and quite possibly at higher redshifts, most star-forming galaxies follow
a reasonably tight relation between SFR and $M_\ast$, whose normalization (e.g., the mean sSFR at some fiducial mass) 
decreases steadily with cosmic time (decreasing redshift) at least from $z = 2$ to the present (Brinchmann \etal 2004, Daddi \etal 2007,
Elbaz \etal 2007, Noeske \etal 2007). A minority population of starburst galaxies exhibits elevated sSFRs, whereas quiescent or 
passive galaxies lie below the SFR--$M_\ast$ correlation.  For the ``main sequence'' of star-forming galaxies, most studies find that the 
average sSFR is a mildly declining function of stellar mass (e.g., Karim \etal 2011).  This implies that more massive 
galaxies completed the bulk of their star formation earlier than that did lower-mass galaxies (Brinchmann \& Ellis 2000, Juneau \etal 2005), 
a ``downsizing'' picture first introduced by Cowie \etal (1996).  Dwarf galaxies continue to undergo major episodes of activity.
The tightness of this SFR--$M_\ast$ correlation has important implications for how star formation is regulated 
within galaxies, and perhaps for the cosmic SFH itself.  Starburst galaxies, whose SFRs are significantly
elevated above the main-sequence correlation, contribute only a small fraction of the global SFRD at $z \leq 2$
(Rodighiero \etal 2011, Sargent \etal 2012).  Instead, the evolution of the  cosmic SFR is primarily
due to the steadily-evolving properties of main-sequence disk galaxies.

Figure \ref{fig13} compares the sSFR (in Gyr$^{-1}$) for star-forming galaxies with estimated stellar masses in the range $10^{9.4}-10^{10}\,\msun$
from a recent compilation by Gonz\'{a}lez \etal (2014), with the predictions from our best-fit SFH. At $z < 2$, the globally averaged 
sSFR ($\equiv \psi/\rho_\ast$) declines more steeply than does that for the star-forming population, as star formation is ``quenched'' for an increasingly large
fraction of the galaxy population.  These passive galaxies are represented in the global sSFR, but not in the sSFR of the star-forming 
``main sequence.'' Previous derivations showed a nearly constant sSFR of $\sim 2\,$Gyr$^{-1}$ for galaxies in the redshift range $2<z<7$, 
suggesting relatively inefficient early star formation and exponential growth in SFRs and stellar masses with cosmic time. 
Recent estimates of reduced stellar masses were derived after correcting for nebular emission in broad bandphotometry and 
appear to require some evolution in 
the high-$z$ sSFR (Stark \etal 2013, Gonz\'{a}lez \etal 2014). At these epochs, the global sSFR decreases with increasing cosmic
time $t$ as sSFR $\sim 4/t$ Gyr$^{-1}$, a consequence of the power-law scaling of our SFRD, $\psi(t)\propto t^{1.9}$. 

\subsection{Cosmic Metallicity}
\label{sec:enrich}

According to Equation {\ref{eq:rhoZ}, the sum of the heavy elements stored in stars and in the gas phase at any given time, $Z\rho_g 
+\langle Z_\ast\rangle \rho_\ast$, is equal to the total mass of metals produced over cosmic history, $y\rho_\ast$. It can be useful to 
express this quantity relative to the baryon density, 
\begin{equation}
Z_b(z)\equiv {y\rho_\ast(z)\over \rho_b}, 
\end{equation}
where $\rho_b=2.77\times 10^{11}\Omega_bh^2\,\mdens$. The evolution of the ``mean metallicity of the Universe'', $Z_b$, predicted by our model SFH 
is plotted in Figure \ref{fig14}. The global metallicity is $Z_b\simeq 0.09\ (y/\zsun)$ solar at the present epoch (note that this is the same value 
derived by Madau \etal 1998b). It drops to $Z_b\simeq 0.01 (y/\zsun)$ solar at $z=2.5$, i.e., the star-formation activity we believe to have taken 
place between the Big Bang and $z=2.5$ (2.5 Gyr later) was sufficient to enrich the Universe as a whole to a metallicity of $\sim 1$\% 
solar (for $y\simeq \zsun$). Note that the metal production term $y\rho_\ast$ (and therefore $Z_b$) depends only weakly on the 
IMF (at a fixed luminosity density): Salpeter-based mass-to-light ratios are 1.64 times higher than those based on Chabrier. 
This is counterbalanced by Salpeter-based net metal yields that are a factor of $\sim 2$ lower than those based on Chabrier (see Section \ref{sec:chemev}). 

\begin{figure}[ht]
\vspace{-1.5cm}
\centerline{\psfig{figure=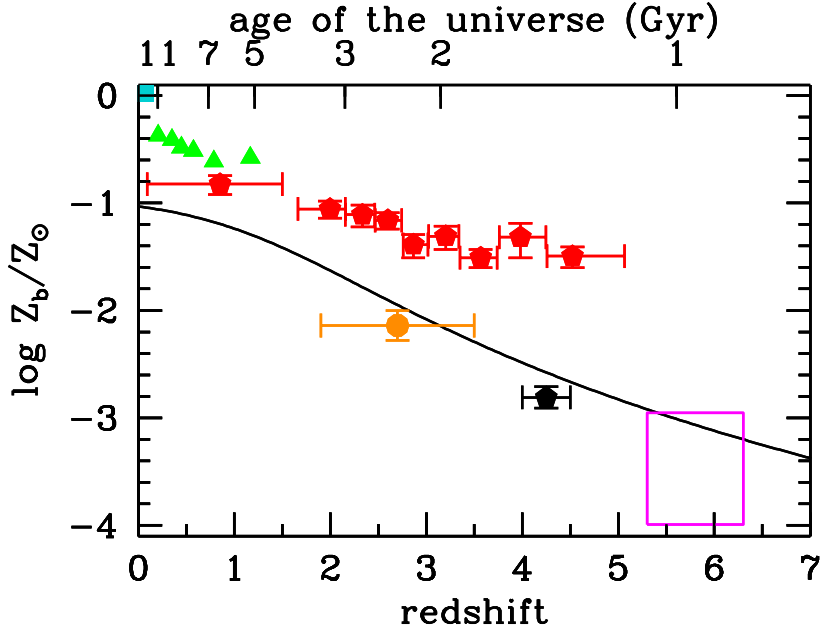,width=0.9\textwidth}}
\vspace{-1cm}
\caption{\footnotesize Mean metallicity of the Universe (in solar units): ({\it solid curve}) mass of heavy elements ever produced per cosmic 
baryon from our model SFH, for an assumed IMF-averaged yield of $y=0.02$; ({\it turquoise square}) mass-weighted stellar metallicity 
in the nearby Universe from the SDSS (Gallazzi \etal 2008); ({\it green triangles)} mean iron abundances in the central regions of galaxy clusters 
(Balestra \etal 2007); ({\it red pentagons}) column density-weighted metallicities of the 
damped \Lya\ absorption systems (Rafelski \etal 2012); ({\it orange dot}) metallicity of the IGM as probed by \OVI\ absorption in 
the \Lya\ forest (Aguirre \etal 2008); ({\it black pentagon}) metallicity of the IGM as probed by \CIV\ absorption (Simcoe 2011); 
({\it magenta rectangle}) metallicity of the IGM as probed by \CIV\ and \CII\ absorption (Ryan-Weber \etal 2009, Simcoe \etal 2011, Becker \etal 2011). 
}
\label{fig14}
\end{figure}

Figure \ref{fig14} also shows the metallicity of a variety of astrophysical objects at different epochs. The mass-weighted average stellar metallicity in the local 
Universe, $\langle Z_\ast(0)\rangle=(1.04\pm 0.14)\,\zsun$ (Gallazzi \etal 2008), is plotted together with the metallicity of three different gaseous components 
of the distant Universe: {\it a)} galaxy clusters, the largest bound objects for which chemical enrichment can be thoroughly studied, and perhaps the best example 
in nature of a ``closed box''; {\it b)} the damped \Lya\ absorption systems that originate in galaxies and dominate the neutral-gas content of the 
Universe; and {\it c)} the highly ionized circumgalactic and intergalactic gas that participates in the cycle of baryons in and out 
of galaxies in the early Universe.

The iron mass in clusters is several times larger than could have been produced by CC SNe if stars formed with a standard IMF, a discrepancy that 
may indicate an IMF in clusters that is skewed toward high-mass stars (e.g., Portinari \etal 2004) and/or to enhanced iron production by 
Type Ia SNe (e.g., Maoz \& Gal-Yam 2004). The damped \Lya\ absorption systems are detected in absorption (i.e., have no luminosity bias) and their large optical 
depths at the Lyman limit eliminates the need for uncertain ionization corrections to deduce metal abundances. Their metallicity is  
determined with the highest confidence from elements such as O, S, Si, Zn, and Fe and decreases with increasing redshift down to 
$\approx 1/600$ solar to $z\sim 5$ (e.g., Rafelski \etal 2012). The enrichment of the circumgalactic medium, as probed in absorption by 
\CIII, \CIV, \SiIII, \SiIV, \OVI, and other transitions, provides us with a record of past star formation and of the impact of galactic 
winds on their surroundings. Figure \ref{fig14} shows that the ionization-corrected metal abundances from \OVI\ absorption at $z\sim 3$ (Aguirre \etal 
2008) and \CIV\ absorption at $z\sim 4$ (Simcoe 2011) track well the predicted mean metallicity of the Universe, i.e. that these systems are an unbiased 
probe of the cosmic baryon cycle.

The Universe at redshift 6 remains one of the most challenging observational frontiers, as the high opacity of the \Lya\ forest inhibits detailed 
studies of hydrogen absorption along the line-of-sight to distant quasars. In this regime, the metal lines that fall longwards of the \Lya\ 
emission take on a special significance 
as the only tool at our disposal to recognize individual absorption systems, whether in galaxies or the IGM, and probe cosmic enrichment 
following the earliest episodes of star formation. Here, we used recent surveys of high- and low-ionization intergalactic absorption to estimate the  
metallicity of the Universe at these extreme redshifts. According to Simcoe \etal (2011) (see also Ryan-Weber \etal 2009), the comoving mass 
density of triply ionized carbon over the redshift range $5.3-6.4$ is (expressed as a fraction of the critical density) $\Omega_{\rm CIV}=(0.46\pm 0.20)\times 10^{-8}$.
Over a similar redshift range, \CII\ absorption yields $\Omega_{\rm CII}=0.9\times 10^{-8}$ (Becker \etal 2011). The total carbon metallicity
by mass, $Z_{\rm C}$, at $\langle z\rangle=5.8$ implied by these measurements is   
\begin{equation}
Z_{\rm C}={\Omega_{\rm CIV}+\Omega_{\rm CII}\over \Omega_b}\times {{\rm C}\over \CII+\CIV}\simeq 3\times 10^{-7}\,{{\rm C}\over \CII+\CIV},
\label{eq:ZC}
\end{equation}
where $(\CII+\CIV)/$C is the fraction of carbon that is either singly or triply ionized. In Figure \ref{fig14}, we plot $Z_{\rm C}$ 
in units of the mass fraction of carbon in the Sun, $Z_{\rm C\odot}=0.003$ (Asplund \etal 2009). The lower bound to the rectangle  
centered at redshift 5.8 assumes no ionization correction, i.e., $(\CII+\CIV)/$C$=1$. To derive the upper bound, we adopt 
the conservative limit (\CII+\CIV)/C $\ge 0.1$; this is the minimum fractional abundance reached by \CII+\CIV\ under the most favorable 
photoionization balance conditions at redshift 6. [To obtain this estimate, we have computed photoionization models based on 
the CLOUDY code (Ferland \etal 1998) assuming the UV radiation background at $z=6$ of Haardt \& Madau (2012) and a range of gas 
overdensities $0<\log \delta <3$.] If the ionization state of the early metal-bearing IGM is such that most 
of the C is either singly or triply ionized, then most of the heavy elements at these epochs appear to be ``missing'' compared with 
the expectations based on the integral of the cosmic SFH (Ryan-Weber \etal 2009, Pettini 2006). Conversely, if \CII\ and \CIV\ are only trace ion 
stages of carbon, then the majority of the heavy elements produced by stars 1 Gyr after the Big Bang ($z=6$) may have been detected already.

A simple argument can also be made against the possibility that our best-fit SFH significantly overpredicts the cosmic metallicity at 
these early epoch. The massive stars that explode as Type II SNe and seed the IGM with metals are also the sources of nearly all of the 
Lyman-continuum (LyC) photons produced by a burst of star formation. It is then relatively straightforward to link a given IGM metallicity 
to the minimum number of LyC photons that must have been produced up until that time. The close correspondence 
between the sources of metals and photons makes the conversion from one to the other largely independent of the details of the 
stellar IMF (Madau \& Shull 1996). Specifically, the energy emitted in hydrogen ionizing radiation per baryon, $E_{\rm ion}$, is related
to the average cosmic metallicity by
\begin{equation}
E_{\rm ion}=\eta m_pc^2 Z_b, 
\label{eq:eta}
\end{equation}
where $m_pc^2=938$ MeV is the rest mass of the proton and $\eta$ is the efficiency of conversion of the heavy element rest mass into LyC radiation. 
For stars with $Z_\ast=\zsun/50$, one derives $\eta=0.014$ (Schaerer 2002, Venkatesan \& Truran 2003). An average energy of 22 eV per LyC photon 
together with our prediction of $Z_b=7\times 10^{-4}\,\zsun$ (assuming a solar yield) at redshift 6 imply that approximately 8 LyC photons per baryon 
were emitted by early galaxies prior to this epoch. Although at least one photon per baryon is needed for reionization to occur, this is a reasonable value
because the effect of hydrogen recombinations in the IGM and within individual halos will likely boost the number of photons required. A global 
metallicity at $z=6$ that was much lower than our predicted value would effectively create a deficit of UV radiation and leave the 
reionization of the IGM unexplained. In Section \ref{sec:first} below we link the production of LyC photons to stellar mass, show that the 
efficiency of LyC production decreases with increasing $Z_\ast$, and discuss early star formation and the epoch of reionization in more detail. 

\subsection{Black Hole Accretion History}

Direct dynamical measurements show that most local massive galaxies host a quiescent massive black hole in their nuclei.
Their masses correlate tightly with the stellar velocity dispersion of the host stellar bulge, as manifested 
in the $M_{\rm BH}-\sigma_\ast$ relation of spheroids (Ferrarese \& Merritt 2000, Gebhardt \etal 2000).
It is not yet understood whether such scaling relations were set in primordial structures and maintained throughout cosmic
time with a small dispersion or which physical processes established such correlations in the first place. 
Nor it is understood whether the energy released during the luminous quasar phase has a global impact on the host, 
generating large-scale galactic outflows and quenching star formation (Di Matteo \etal 2005), or just modifies gas 
dynamics in the galactic nucleus (Debuhr \etal 2010). 

Here, we consider a different perspective on the link between the assembly of the stellar component of galaxies 
and the growth of their central black holes.  The cosmic mass accretion history of massive black holes can be inferred using 
Soltan's argument (Soltan 1982), which relates the quasar bolometric luminosity density to the rate at which 
mass accumulates into black holes, 
\begin{equation}
\dot \rho_{\rm BH}(z)={1-\epsilon\over \epsilon c^2} \int L\phi(L,z)dL, 
\end{equation}
where $\epsilon$ is the efficiency of conversion of rest-mass energy into radiation.  In practice,
bolometric luminosities are typically derived from observations of the AGN emission at X-ray,
optical or IR wavelengths, scaled by a bolometric correction. In Figure \ref{fig15}, 
several recent determinations of the massive black hole mass growth rate are compared with the SFRD 
(Equation ~\ref{eq:sfrd}). Also shown is the accretion history derived from the hard X-ray LF 
of Aird \etal (2010), assuming a radiative efficiency $\epsilon = 0.1$ and a constant 
bolometric correction of 40 for the observed 2-10 KeV X-ray luminosities.  This accretion rate peaks
at lower redshift than does the SFRD and declines more rapidly from $z \approx 1$ to 0.   However, 
several authors have discussed the need for luminosity-dependent bolometric corrections, which 
in turn can affect the derived accretion history (e.g., Marconi \etal 2004, Hopkins \etal 2007, 
Shankar \etal 2009).   Moreover, although the hard X-ray LF includes unobscured as well as moderately obscured 
sources that may not be identified as AGN at optical wavelengths, it can miss Compton thick AGN, 
which may be identified in other ways, particularly using IR data.  Delvecchio \etal (2014) 
have used deep {\it Herschel} and {\it Spitzer} survey data in GOODS-S and COSMOS to identify AGN 
by SED fitting.  This is a potentially powerful method but depends on reliable decomposition of the 
IR emission from AGN and star formation.  

Black hole mass growth rates derived from the bolometric AGN LFs of Shankar \etal (2009) and 
Delvecchio \etal (2014) are also shown in Figure \ref{fig15}.  These more closely track the 
cosmic SFH, peaking at $z \approx 2$, and suggest that star formation and black hole 
growth are closely linked at all redshifts (Boyle \& Terlevich 1998, 
Silverman \etal 2008).  However, the differences between accretion histories published in the recent 
literature would caution that it is premature to consider this comparison to be definitive.

\begin{figure}[ht]
\vspace{-1.5cm}
\centerline{\psfig{figure=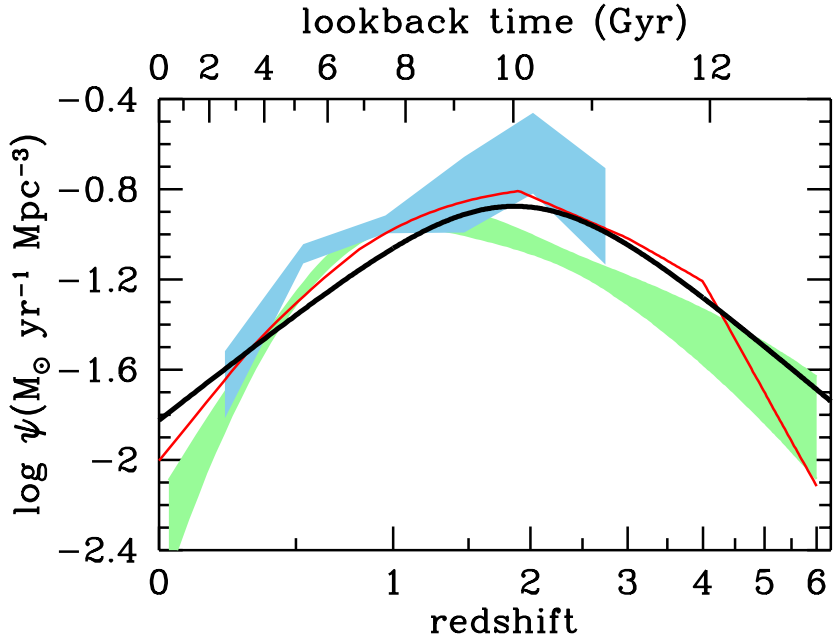,width=0.9\textwidth}}
\vspace{-1cm}
\caption{\footnotesize Comparison of the best-fit star formation history ({\it thick solid curve}) 
with the massive black hole accretion history from X-ray [{\it red curve} (Shankar \etal 2009); 
{\it light green shading} (Aird \etal 2010)] and infrared ({\it light blue shading}) (Delvecchio \etal 2014)
data.  The shading indicates the $\pm 1\sigma$ uncertainty range on the total bolometric luminosity density. 
The radiative efficiency has been set to $\epsilon=0.1$.  The comoving rates of black hole 
accretion have been scaled up by a factor of 3,300 to facilitate visual comparison to the star-formation history.
}
\label{fig15}
\end{figure}

\subsection{First Light and Cosmic Reionization} \label{sec:first} 

Fundamental to our understanding of how the Universe evolved to its present state is the epoch of ``first light,'' the first 
billion years after the Big Bang when the collapse of the earliest baryonic objects -- the elementary building blocks for the 
more massive systems that formed later -- determined the ``initial conditions'' of the cosmological structure formation process.
The reionization in the all-pervading IGM -- the transformation of neutral hydrogen into an ionized state -- is a landmark 
event in the history of the early Universe. Studies of \Lya\ absorption in the spectra of distant quasars show that the IGM is 
highly photoionized out to redshift $z\gta 6$ (for a review, see Fan \etal 2006), whereas polarization data from the {\it Wilkinson 
Microwave Anisotropy Probe} constrain the redshift of any sudden reionization event to be significantly higher, 
$z=10.5 \pm 1.2$ (Jarosik \etal 2011). It is generally thought that the IGM is kept ionized by the integrated UV emission from 
AGN and star-forming galaxies, but the relative contributions of these sources as a function of epoch are poorly
known (e.g., Madau \etal 1999, Faucher-Gigu\`ere \etal 2008, Haardt \& Madau 2012, Robertson \etal 2013). 

\begin{figure}[ht]
\centerline{\psfig{figure=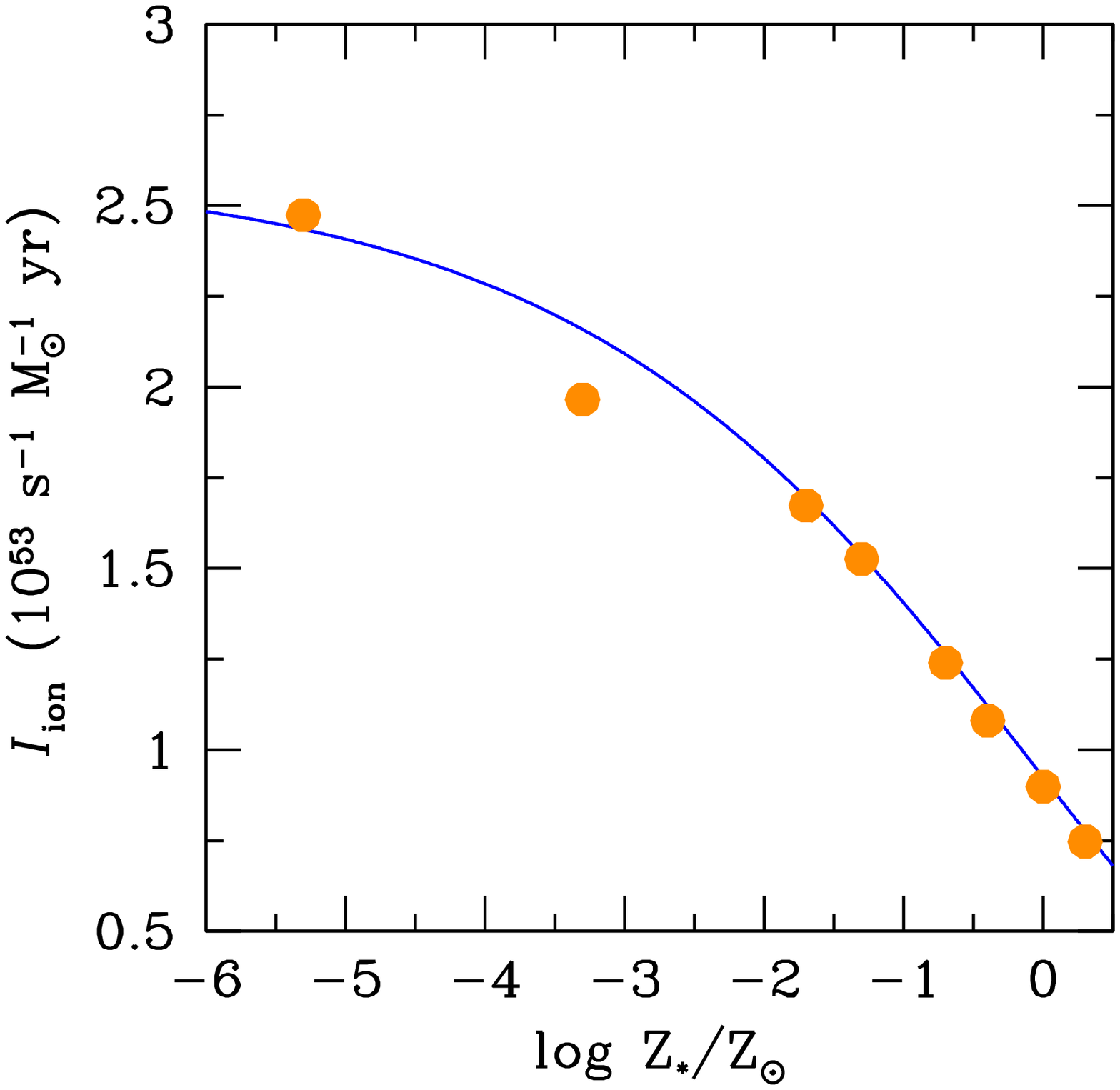,width=0.49\textwidth}
\psfig{figure=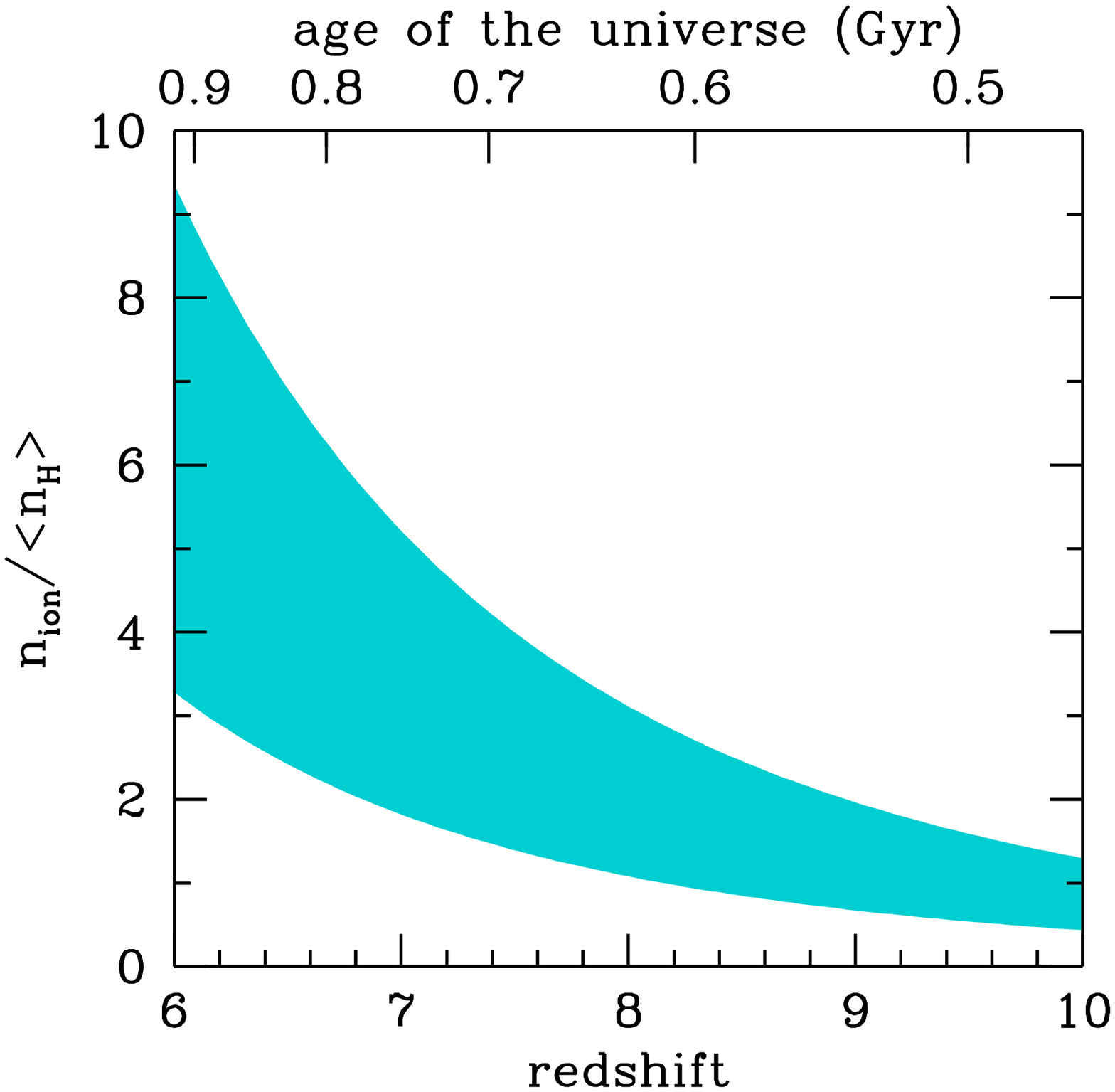,width=0.49\textwidth}}
\vspace{-0.0cm}
\caption{\footnotesize ({\it Left panel}) Metallicity dependence of the ionizing photon yield, ${I}_{\rm ion}$, for a stellar 
population with a Salpeter IMF (initial mass function) and constant star-formation rate. The points show the values given 
in table 4 of Schaerer (2003), computed for 
a Salpeter IMF in the 1-100$\,\msun$ range, and divided by a mass conversion factor of 2.55 to rescale to a mass range 
0.1-100$\,\msun$.  The solid curve shows the best-fitting function, 
$\log {I}_{\rm ion}=[(0.00038Z_\ast^{0.227}+0.01858)]^{-1}-\log(2.55)$. 
({\it Right panel}) Number of H-ionizing photons emitted per hydrogen atom since the Big Bang, $n_{\rm ion}/\langle n_{\rm H}
\rangle$, according to our best-fit star formation history. The shaded area is delimited by stellar populations with 
metallicities $Z_\ast=\zsun$ (lower boundary) and $Z_\ast=0$ (upper boundary).
}
\label{fig16}
\end{figure}

Establishing whether massive stars in young star-forming galaxies were responsible for cosmic reionization requires a determination of the early history 
of star formation, of the LyC flux emitted by a stellar population, and of the fraction of hydrogen-ionizing photons 
that can escape from the dense sites of star formation into the low-density IGM. We can use stellar population synthesis 
to estimate the comoving volumetric rate at which photons above 1 Ryd are emitted from star-forming galaxies as  
\begin{equation}
 {\dot n}_{\rm ion}=I_{\rm ion} \times \psi(t), 
\label{eq:rhoi}
\end{equation}
where ${\dot n}_{\rm ion}$ is expressed in units of $\ndotunits$ and $\psi$ in units of $\sfrd$.
The LyC photon yield $I_{\rm ion}$ is plotted in Figure \ref{fig16} for our reference Salpeter IMF and for a wide 
range of metallicities spanning from extremely metal poor to metal-rich stars (Schaerer 2003). The yield 
increases with decreasing metallicity by more than a factor of 3. 
Integrating Equation \ref{eq:rhoi} over time and dividing by the mean comoving hydrogen density $\langle 
n_{\rm H}\rangle=1.9\times 10^{-7}$ cm$^{-3}$, we can write the total number of stellar LyC photons emitted per hydrogen atom 
since the Big Bang as 
\begin{equation}
{\cal N}_{\rm ion}(z) \equiv {{n}_{\rm ion}\over\langle n_{\rm H}\rangle}= I_{\rm ion}\times 7.8\times 10^{-60} \times \rho_\ast(z),
\end{equation}
where $\rho_\ast(z)$ is the cosmic SMD. [Within purely stellar radiation or energetic X-ray photons, either the total number of 
ionizing photons produced or the total radiated energy, respectively, is what matters for reionization.
This is because, in a largely neutral medium, each photoionization produces a host of secondary collisional ionizations, 
with approximately one hydrogen secondary ionization for every 37 eV of energy in the primary photoelectron (Shull \& van Steenberg 1985).
As the medium becomes more ionized, however, an increasing fraction of this energy is deposited as heat.] 
Figure \ref{fig16} depicts the quantity ${\cal N}_{\rm ion}$ at $z>6$ according to our best-fit SFH for the range of stellar 
metallicities $0<Z_\ast<\zsun$. Cosmological reionization requires at least one LyC photon  
per hydrogen atom {escaping} into the intergalactic space, i.e., ${\cal N}_{\rm ion}\langle f_{\rm esc}\rangle>1$. The exact number depends on the 
rate of radiative recombinations in a clumpy IGM. Here, the escape fraction $\langle f_{\rm esc}\rangle$ is the angle-averaged, absorption 
cross-section-weighted, and luminosity-weighted fraction of ionizing photons that leaks into the IGM from the dense star-forming regions within galaxies. 

Figure \ref{fig16} shows the well-known result that, even if the emission of LyC photons at early cosmic times was dominated by extremely low-metallicity 
stars, escape fractions in excess of 20\% would be required to reionize the Universe 
by redshift 6-7 (e.g., Bolton \& Haehnelt 2007, Ouchi \etal 2009, Bunker \etal 2010, Haardt \& Madau 2012, Finkelstein \etal 2012b). 
Although the mechanisms regulating the escape 
fraction of ionizing radiation from galaxies and its dependence on cosmic time and luminosity are unknown, these leakage values are higher 
than typically inferred from observations of LBGs at $z\sim 3$ (e.g., Nestor \etal 2013, Vanzella \etal 2012, and references therein).

The ``reionization budget'' is made even tighter by two facts. First, the volume-averaged hydrogen recombination timescale  in the IGM,
\begin{equation}
\langle t_{\rm rec}\rangle=(\chi \langle n_{\rm H}\rangle \alpha_B\,C_{\rm IGM})^{-1}\simeq 3.2~ {\rm Gyr}\left({1+z\over 
7}\right)^{-3}C_{\rm IGM}^{-1}, \label{eq:trec}
\end{equation}
is $\sim $60\% of the expansion timescale $H^{-1}=H_0^{-1}[\Omega_M(1+z)^3+\Omega_\Lambda]^{-1/2}$ at $z=10$, i.e., close to two ionizing photon
per baryon are needed to keep the IGM ionized. Here, $\alpha_B$ is the recombination coefficient to the excited states of hydrogen, $\chi=1.08$ accounts for the presence of
photoelectrons from singly ionized helium, and $C_{\rm IGM}$ is the clumping factor of ionized hydrogen. The above estimate assumes 
a gas temperature of $2\times 10^4$ K and a clumping factor for the IGM, $C_{\rm IGM}=1+43\,z^{-1.71}$, that is equal to the clumpiness 
of gas below a threshold overdensity of 100 found at $z\ge 6$ in a suite of cosmological hydrodynamical simulations (Pawlik \etal 2009; 
see also Shull \etal 2012).
Second, the rest-frame UV continuum properties of very high redshift galaxies appear to show little evidence for the ``exotic'' stellar populations (e.g., 
extremely sub-solar metallicities or top-heavy IMFs) that would significantly boost the LyC photon yield 
(e.g., Finkelstein \etal 2012a, Dunlop \etal 2013). 
Although improved knowledge of the SFRD beyond redshift 9 may help to better characterize this photon shortfall, the prospects 
of a direct observational determination of the escape fraction of ionizing photons leaking into the IGM at these early epochs look rather bleak, as the
Universe becomes opaque to LyC radiation above redshift 4.

\section{CONCLUDING REMARKS}
\label{sec:conclusion}

The cosmic history of star formation is one of the most fundamental observables in astrophysical cosmology. 
We have reviewed the range of complementary techniques and theoretical tools that are allowing astronomers to
map the transformation of gas into stars, the production of heavy elements, and the reionization of the Universe
from the cosmic dark ages to the present epoch. Under the simple assumption of a universal IMF, there is reasonable
agreement between the global SMD inferred at any particular time and the time integral of all the preceding 
instantaneous star-formation activity, although modest offsets may still point toward systematic uncertainties.
A consistent picture is emerging, whereby the SFRD peaked $\sim $3.5 Gyr after the Big Bang, and dropped 
exponentially at $z<1$ with an e-folding timescale of 3.9 Gyr.  The Universe was a much more active place 
in the past: Stars formed at a peak rate approximately nine times higher than is seen today. Approximately 
25\% of the present-day SMD formed at $z > 2$, before the peak of the SFRD, 
and another 25\% formed since $z = 0.7$, i.e., roughly over the last half of the Universe's age.
From the peak of the SFRD at $z \approx 2$ to the present day, and perhaps earlier as well, 
most stars formed in galaxies that obey a relatively tight SFR--$M_\ast$ correlation, and
only a small fraction formed in starbursts with significantly elevated specific SFRs.
The smooth evolution of this dominant main-sequence galaxy population suggests that the 
evolution of the cosmic SFH is primarily determined by a balance between
gas accretion and feedback processes, both closely related to galaxy mass, and that stochastic
events such as merger-driven starbursts play a relatively minor role.  
The growth histories of the stellar component of galaxies and their central black holes are similar in 
shape, suggesting broad co-evolution of black holes and their host galaxies.
The rise of the mean metallicity of the 
Universe to 0.001 solar by redshift six, 1 Gyr after the Big Bang, appears to have been 
accompanied by the production of fewer than 10 hydrogen LyC photons per baryon, 
indicating a rather tight budget for cosmological reionization.  The SFRD
at $z \approx 7$ was approximately the same as that of today, at $z \approx 0$, but only 1\% of today's 
SMD was formed during the epoch of reionization.

As far as the observations and data are concerned, there is still room for improvement in both
SFRD and SMD measurements at virtually every redshift, from the local Universe to the epoch of
reionization (Section \ref{sec:stateoftheart}).  That said, it would be somewhat surprising if new {measurements}
changed the picture dramatically at $z < 1$;  it is more likely that stellar population modeling, e.g., for
deriving stellar masses or SFRs, could still change the details of the picture during
the decline and fall of the cosmic SFH. Indeed, at all redshifts, limitations
of our methods for interpreting light as mass may play a significant, even dominant, role in the
error budget for the analyses described in this review.  The peak era of cosmic star formation has been
extensively mapped, and yet even with the current data (Figure ~\ref{fig9}), it is still hard to accurately 
pinpoint the redshift of maximum SFRD within a range $\Delta z = 1$.   Our fitting function (Equation ~\ref{eq:sfrd}) 
places this peak at $z \approx 1.85$, which is plausible but still uncertain.   Uncertainties in the faint-end slope 
of the IRLF and in extinction corrections for the UVLF
still dominate at this peak era of cosmic star formation. Although evidence seems to point clearly 
to a steady increase in the SFRD from $z = 8$ to $z \approx 2$, our direct knowledge of dust-obscured
star formation at these redshifts is, for the most part, limited to the rarest and most ultraluminous
galaxies, leaving considerable uncertainty about how much SFRD we may be missing in the UV census
of that early phase of galaxy evolution.   At $z > 4$, our galaxy surveys have been strongly biased 
toward UV-bright galaxies, and may underestimate both SFRDs and SMDs.
Even for UV-selected galaxies, the measurements at $z \geq 8$ are very new and likely uncertain, 
unsupported by spectroscopic confirmation to date.  In addition to measuring redshifts, spectroscopy 
from the {JWST} will help clarify basic issues about nebular line emission and the degree to which 
it has affected the photometric analyses that have been carried out to date.  

Painstaking though all this vast effort has been, it does miss a crucial point. It says little 
about the inner workings of galaxies, i.e., their ``metabolism'' and the basic process of ingestion (gas infall and 
cooling), digestion (star formation), and excretion (outflows). Ultimately, it also says little about the mapping from dark 
matter halos to their baryonic components. Its roots are in optical-IR astronomy, statistics, stellar populations, and 
phenomenology, rather than in the physics of the ISM, self-regulated accretion and star formation, stellar feedback, 
and SN-driven galactic winds. It provides a benchmark against which to compare semi-analytic modeling and hydrodynamical 
simulations of galaxy formation, but it offers little guidance in identifying the smaller-scale basic mechanisms that determine 
the rate of conversion of gas into stars and lead to the grandiose events in the history of the Universe described in this review.

A variety of physical processes are thought to shape the observed distribution of galaxy properties, 
ranging from those responsible for galaxy growth (e.g., star formation and galaxy merging) to those that regulate 
such growth (e.g., energetic feedback from SNe, AGN, and the UV radiation background). However, many of these processes 
likely depend primarily on the mass of a galaxy's dark matter halo. Relating the stellar masses and SFRs of galaxies to the masses and 
assembly histories of their parent halos is a crucial piece of the galaxy formation and evolution puzzle. 
With the accumulation of data from large surveys and from cosmological numerical simulations, several statistical methods have been 
developed over the past decade to link the properties of galaxies to the underlying dark matter structures (e.g. Berlind \& Weinberg 2002, 
Yang \etal 2003, Vale \& Ostriker 2004). One of them,  the ``abundance matching" technique, assumes in its simplest form a unique 
and monotonic relation between galaxy light and halo mass, and it reproduces galaxy clustering as a function 
of luminosity over a wide range in redshift (e.g. Conroy \etal 2006, Guo \etal 2010, Moster \etal 2010). Modern versions of this approach 
(Moster \etal 2013, Behroozi \etal 2013) have shown that {\it a)} halos of mass $\sim 10^{12}\,\msun$ are the most efficient at forming stars at 
every epoch, with baryon conversion efficiencies of 20-40\%  that fall rapidly at both higher and lower masses; {\it b)} in halos similar to that 
of the Milky Way, approximately half of the central stellar mass is assembled after redshift 0.7; and {\it c)} in low-mass halos, the accretion of satellites 
contributes little to the assembly of their central galaxies, whereas in massive halos more than half of the central stellar mass is formed ``ex-situ." 
These studies represent promising advances, albeit with serious potential shortcomings (e.g. Guo \& White 2013, Zentner \etal 2013). The assumption 
of a monotonic relation between stellar mass and the mass of the host halo is  likely incorrect in detail, and it predict only numerically 
converged properties on scales that are well resolved in simulations. The matching procedure requires minimal assumptions and avoids an 
explicit treatment of the physics of galaxy formation. As such it provides relatively little new insight into this physics. In the version of this technique 
by Behroozi \etal (2013), for example, the cosmic SFH is reproduced by construction.
 
As of this writing, a solid interpretation of the cosmic SFH from first principles is still missing (for a recent review, see Mac Low 2013). 
Generically, one expects that star formation may be limited at early times by the build-up of dark matter halos and quenched at low 
redshift as densities decline from Hubble expansion to the point where gas cooling is inhibited. These two regimes 
could then lead to a peak in the SFH at intermediate redshifts (Hernquist \& Springel 2003). A decade ago, hydrodynamical simulations predicted 
that the peak in star-formation activity should occur at a much higher redshift, $z\gta 5$, than is actually observed (Springel \& Hernquist 2003, 
Nagamine \etal 2004). Theoretical modeling has been unable to correctly forecast the evolution of the SFRD because of the large range of 
galaxy masses that contribute significantly to cosmic star formation and the difficulty in following the feedback of energy into the 
ISM and circumgalactic medium from stellar radiation, SN explosions, and accreting massive black holes. Gas cooling in an 
expanding Universe is an intrinsically unstable process because cooling acts to increase the density of the gas, which in turn 
increases the cooling rate. Systems collapsing at low redshift have low mean densities and long cooling times, whereas systems collapsing 
at higher redshifts have higher mean densities and cool catastrophically. Without feedback processes that transfer energy to the 
ISM  and reheat it, one is faced with the classical overcooling problem -- the unphysical cooling of hot gas in the poorly resolved 
inner regions of galaxies -- and with the consequent overproduction of stars at early times. And yet, a completely satisfactory treatment of feedback in hydrodynamical 
simulations that capture large cosmological volumes remains elusive, as these mechanisms operate on scales too small to be resolved and 
must therefore be incorporated via ad-hoc recipes that are too simplistic to capture the complex subgrid physics involved (e.g., Schaye \etal 2010). 

In-depth knowledge of the mechanisms responsible for suppressing  star formation in small halos (e.g., Governato \etal 2010, 
Krumholz \& Dekel 2012, Kuhlen \etal 2012), more powerful supercomputers, better algorithms as well as more robust numerical implementations of stellar feedback (e.g., Agertz \etal 2013) 
all now appear as crucial prerequisites for predicting more realistic SFHs. Newer and deeper observations from the ground and space 
should improve our measurements of the galaxy population and its integrated properties, especially at and beyond the current redshift frontier where data 
remains sparse. It seems likely, however, that the most important contribution of new surveys and better modeling will be toward a detailed 
understanding of the physics of galaxy evolution, not simply its demographics.

\section{DISCLOSURE STATEMENT}

The authors are not aware of any affiliations, memberships, funding, or financial holdings that might be perceived as affecting the objectivity of this review.

\section{ACKNOWLEDGMENTS}

This review has benefited from many discussions with and the help of J. Aird, R. Chary, C. Conroy, O. Cucciati, D. Elbaz, S. Faber, H. Ferguson, A. Gallazzi,
V. Gonz\'{a}lez, A. Klypin, K.-S. Lee, D. Maoz, P.  Oesch, M. Pettini, J. Pforr, L. Pozzetti, J. Primack, J. X. Prochaska, M. Rafelski, 
A. Renzini, B. Robertson, M. Schenker, and D. Stark. Support for this work was provided by the National Science Foundation (NSF) through grant OIA-1124453,
by NASA through grant NNX12AF87G (P.M.), and by NOAO, which is operated by the Association of Universities for Research in Astronomy, 
under a cooperative agreement with the NSF (M.D.).

\vfill\eject

\begin{table*}[t]
\begin{threeparttable}
\centering
\footnotesize 
\caption*{\enspace Table 1: Determinations of the cosmic star formation rate density
from UV data (top group) and IR data (bottom group) used in this review.}\label{tab:SFRD}
\begin{tabular*}{\hsize}{@{\extracolsep{\fill}}lcccl}
\\[-5pt]
\hline
\hline
\multicolumn{1}{l}{Reference} &
\multicolumn{1}{c}{Redshift range} &
\multicolumn{1}{c}{$A_{\rm FUV}^{a}$} &
\multicolumn{1}{c}{$\log \psi^{b}$} & 
\multicolumn{1}{l}{Symbols used in Figure \ref{fig9}}\\
\multicolumn{1}{l}{} &
\multicolumn{1}{c}{} &
\multicolumn{1}{c}{[mag]} &
\multicolumn{1}{c}{[M$_\odot$ year$^{-1}$ Mpc$^{-3}$]} &  
\multicolumn{1}{l}{} \\
\hline
\\[-5pt]
Wyder \etal (2005)                & 0.01-0.1 & 1.80  & $-1.82^{+0.09}_{-0.02}$  & blue-gray hexagon \\

Schiminovich \etal (2005)         & 0.2-0.4  & 1.80  & $-1.50^{+0.05}_{-0.05}$  & blue triangles \\
                                  & 0.4-0.6  & 1.80  & $-1.39^{+0.15}_{-0.08}$  & \\
                                  & 0.6-0.8  & 1.80  & $-1.20^{+0.31}_{-0.13}$  & \\
                                  & 0.8-1.2  & 1.80  & $-1.25^{+0.31}_{-0.13}$  & \\

Robotham \& Driver (2011)         & 0.05  & 1.57  & $-1.77^{+0.08}_{-0.09}$ & dark green pentagon \\

Cucciati \etal (2012)             & 0.05-0.2 & 1.11 & $-1.75^{+0.18}_{-0.18}$ & green squares \\   
                                  & 0.2-0.4  & 1.35 & $-1.55^{+0.12}_{-0.12}$ & \\ 
                                  & 0.4-0.6  & 1.64 & $-1.44^{+0.10}_{-0.10}$ & \\ 
                                  & 0.6-0.8  & 1.92 & $-1.24^{+0.10}_{-0.10}$ & \\ 
                                  & 0.8-1.0  & 2.22 & $-0.99^{+0.09}_{-0.08}$ & \\
                                  & 1.0-1.2  & 2.21 & $-0.94^{+0.09}_{-0.09}$ & \\
                                  & 1.2-1.7  & 2.17 & $-0.95^{+0.15}_{-0.08}$ & \\ 
                                  & 1.7-2.5  & 1.94 & $-0.75^{+0.49}_{-0.09}$ & \\
                                  & 2.5-3.5  & 1.47 & $-1.04^{+0.26}_{-0.15}$ & \\
                                  & 3.5-4.5  & 0.97 & $-1.69^{+0.22}_{-0.32}$ & \\

Dahlen \etal (2007)               & 0.92-1.33 & 2.03  & $-1.02^{+0.08}_{-0.08}$ & turquoise pentagons \\
                                  & 1.62-1.88 & 2.03  & $-0.75^{+0.12}_{-0.12}$ & \\
                                  & 2.08-2.37 & 2.03  & $-0.87^{+0.09}_{-0.09}$ & \\   

Reddy \& Steidel (2009)           & 1.9-2.7   & 1.36 & $-0.75^{+0.09}_{-0.11}$ & dark green triangles \\
                                  & 2.7-3.4   & 1.07 & $-0.97^{+0.11}_{-0.15}$ & \\ 

Bouwens \etal (2012a),(2012b)     & 3.8   & 0.58 & $-1.29^{+0.05}_{-0.05}$ & magenta pentagons \\
                                  & 4.9   & 0.44 & $-1.42^{+0.06}_{-0.06}$ & \\
                                  & 5.9   & 0.20 & $-1.65^{+0.08}_{-0.08}$ & \\
                                  & 7.0   & 0.10 & $-1.79^{+0.10}_{-0.10}$ & \\
                                  & 7.9   & 0.0  & $-2.09^{+0.11}_{-0.11}$ & \\  

Schenker \etal (2013)             & 7.0   & 0.10 & $-2.00^{+0.10}_{-0.11}$ & black crosses \\
                                  & 8.0   & 0.0  & $-2.21^{+0.14}_{-0.14}$ & \\ 
\hline
\hline
\\[-5pt]
Sanders \etal (2003)              & 0.03   &  \textemdash  & $-1.72^{+0.02}_{-0.03}$ & brown circle \\ 

Takeuchi \etal (2003)             & 0.03  &  \textemdash & $-1.95^{+0.20}_{-0.20}$ & dark orange square \\ 

Magnelli \etal (2011)             & 0.40-0.70 & \textemdash & $-1.34^{+0.22}_{-0.11}$ & red open hexagons \\ 
                                  & 0.70-1.00 & \textemdash & $-0.96^{+0.15}_{-0.19}$ & \\ 
                                  & 1.00-1.30 & \textemdash & $-0.89^{+0.27}_{-0.21}$ & \\ 
                                  & 1.30-1.80 & \textemdash & $-0.91^{+0.17}_{-0.21}$ & \\ 
                                  & 1.80-2.30 & \textemdash & $-0.89^{+0.21}_{-0.25}$ & \\ 

Magnelli \etal (2013)             & 0.40-0.70 & \textemdash & $-1.22^{+0.08}_{-0.11}$ & red filled hexagons \\ 
                                  & 0.70-1.00 & \textemdash & $-1.10^{+0.10}_{-0.13}$ & \\ 
                                  & 1.00-1.30 & \textemdash & $-0.96^{+0.13}_{-0.20}$ & \\ 
                                  & 1.30-1.80 & \textemdash & $-0.94^{+0.13}_{-0.18}$ & \\ 
                                  & 1.80-2.30 & \textemdash & $-0.80^{+0.18}_{-0.15}$ & \\ 

Gruppioni \etal (2013)            & 0.00-0.30 & \textemdash & $-1.64^{+0.09}_{-0.11}$ & dark red filled hexagons \\ 
                                  & 0.30-0.45 & \textemdash & $-1.42^{+0.03}_{-0.04}$ & \\
                                  & 0.45-0.60 & \textemdash & $-1.32^{+0.05}_{-0.05}$ & \\
                                  & 0.60-0.80 & \textemdash & $-1.14^{+0.06}_{-0.06}$ & \\
                                  & 0.80-1.00 & \textemdash & $-0.94^{+0.05}_{-0.06}$ & \\
                                  & 1.00-1.20 & \textemdash & $-0.81^{+0.04}_{-0.05}$ & \\
                                  & 1.20-1.70 & \textemdash & $-0.84^{+0.04}_{-0.04}$ & \\
                                  & 1.70-2.00 & \textemdash & $-0.86^{+0.02}_{-0.03}$ & \\
                                  & 2.00-2.50 & \textemdash & $-0.91^{+0.09}_{-0.12}$ & \\
                                  & 2.50-3.00 & \textemdash & $-0.86^{+0.15}_{-0.23}$ & \\
                                  & 3.00-4.20 & \textemdash & $-1.36^{+0.23}_{-0.50}$ & \\
\hline
\end{tabular*}
\begin{tablenotes}
\item[a] In our notation, $A_{\rm FUV}\equiv -2.5\log_{10} \langle k_d\rangle$.
\item[b] All our star-formation rate densities are based on the integration of the best-fit luminosity function parameters 
down to the same {\it relative} limiting luminosity, in units of the characteristic luminosity $L^\ast$, of 
$L_{\rm min}=0.03\,L^\ast$. A Salpeter initial mass function has been assumed. 
\end{tablenotes}
\end{threeparttable}
\vspace{0.cm}
\end{table*}

\begin{table*}[t]
\footnotesize 
\begin{threeparttable}
\centering
\caption*{\enspace Table 2: Determinations of the Cosmic Stellar Mass Density used in this review.}\label{tab:SMD}
\begin{tabular*}{\hsize}{@{\extracolsep{\fill}}lccl}
\\[-5pt]
\hline
\hline
\multicolumn{1}{l}{Reference} &
\multicolumn{1}{c}{Redshift range} &
\multicolumn{1}{c}{$\log \rho_\ast^{a}$} &
\multicolumn{1}{l}{Symbols used in Figure \ref{fig11}} \\ 
\multicolumn{1}{l}{} &
\multicolumn{1}{c}{} &
\multicolumn{1}{c}{[M$_\odot$ Mpc$^{-3}$]} & 
\multicolumn{1}{l}{} \\
\hline
\\[-5pt]
Li \& White (2009)              & 0.07& $8.59^{+0.01}_{-0.01}$ & gray dot \\
Gallazzi \etal (2008)           & 0.005-0.22& $8.78^{+0.07}_{-0.08}$ & dark green square \\

Moustakas \etal (2013)          & 0.0-0.2 & $8.59^{+0.05}_{-0.05}$ & magenta stars \\
                                & 0.2-0.3 & $8.56^{+0.09}_{-0.09}$ & \\
                                & 0.3-0.4 & $8.59^{+0.06}_{-0.06}$ & \\
                                & 0.4-0.5 & $8.55^{+0.08}_{-0.08}$ & \\

Bielby \etal (2012)\,$^{b}$     & 0.2-0.4 & $8.46^{+0.09}_{-0.12}$ & pink filled hexagons\\
                                & 0.4-0.6 & $8.33^{+0.03}_{-0.03}$ & \\
                                & 0.6-0.8 & $8.45^{+0.08}_{-0.1}$  & \\
                                & 0.8-1.0 & $8.42^{+0.05}_{-0.06}$ & \\
                                & 1.0-1.2 & $8.25^{+0.04}_{-0.04}$ & \\
                                & 1.2-1.5 & $8.14^{+0.06}_{-0.06}$ & \\
                                & 1.5-2.0 & $8.16^{+0.32}_{-0.03}$ & \\

Perez-Gonz\'{a}lez \etal (2008)     & 0.0-0.2 & $8.75^{+0.12}_{-0.12}$ & red triangles \\
                                & 0.2-0.4 & $8.61^{+0.06}_{-0.06}$ & \\
                                & 0.4-0.6 & $8.57^{+0.04}_{-0.04}$ & \\
                                & 0.6-0.8 & $8.52^{+0.05}_{-0.05}$ & \\
                                & 0.8-1.0 & $8.44^{+0.05}_{-0.05}$ & \\
                                & 1.0-1.3 & $8.35^{+0.05}_{-0.05}$ & \\
                                & 1.3-1.6 & $8.18^{+0.07}_{-0.07}$ & \\
                                & 1.6-2.0 & $8.02^{+0.07}_{-0.07}$ & \\
                                & 2.0-2.5 & $7.87^{+0.09}_{-0.09}$ & \\
                                & 2.5-3.0 & $7.76^{+0.18}_{-0.18}$ & \\
                                & 3.0-3.5 & $7.63^{+0.14}_{-0.14}$ & \\
                                & 3.5-4.0 & $7.49^{+0.13}_{-0.13}$ & \\

Ilbert \etal (2013)             & 0.2-0.5 & $8.55^{+0.08}_{-0.09}$ & cyan stars \\
                                & 0.5-0.8 & $8.47^{+0.07}_{-0.08}$ & \\
                                & 0.8-1.1 & $8.50^{+0.08}_{-0.08}$ & \\
                                & 1.1-1.5 & $8.34^{+0.10}_{-0.07}$ & \\
                                & 1.5-2.0 & $8.11^{+0.05}_{-0.06}$ & \\
                                & 2.0-2.5 & $7.87^{+0.08}_{-0.08}$ & \\
                                & 2.5-3.0 & $7.64^{+0.15}_{-0.14}$ & \\
                                & 3.0-4.0 & $7.24^{+0.18}_{-0.20}$ & \\

Muzzin \etal (2013)             & 0.2-0.5 & $8.61^{+0.06}_{-0.06}$ & blue squares \\
                                & 0.5-1.0 & $8.46^{+0.03}_{-0.03}$ & \\
                                & 1.0-1.5 & $8.22^{+0.03}_{-0.03}$ & \\
                                & 1.5-2.0 & $7.99^{+0.05}_{-0.03}$ & \\
                                & 2.0-2.5 & $7.63^{+0.11}_{-0.04}$ & \\
                                & 2.5-3.0 & $7.52^{+0.13}_{-0.09}$ & \\
                                & 3.0-4.0 & $6.84^{+0.43}_{-0.20}$ & \\

Arnouts \etal (2007)            & 0.3 & $8.78^{+0.12}_{-0.16}$ & yellow dots\\
                                & 0.5 & $8.64^{+0.09}_{-0.11}$ & \\
                                & 0.7 & $8.62^{+0.08}_{-0.10}$ & \\
                                & 0.9 & $8.70^{+0.11}_{-0.15}$ & \\
                                & 1.1 & $8.51^{+0.08}_{-0.11}$ & \\
                                & 1.35 & $8.39^{+0.10}_{-0.13}$ & \\
                                & 1.75 & $8.13^{+0.10}_{-0.13}$ & \\
\hline
\end{tabular*}
\end{threeparttable}
\end{table*}

\begin{table*}[t]
\small 
\begin{threeparttable}
\centering
\caption*{\enspace Table 2: (cont.)}
\begin{tabular*}{\hsize}{@{\extracolsep{\fill}}lccl}
\\[-5pt]
\hline
\hline
\multicolumn{1}{l}{Reference} &
\multicolumn{1}{c}{Redshift range} &
\multicolumn{1}{c}{$\log \rho_\ast^{a}$} &
\multicolumn{1}{l}{Symbols used in Figure \ref{fig11}} \\ 
\multicolumn{1}{l}{} &
\multicolumn{1}{c}{} &
\multicolumn{1}{c}{[M$_\odot$ Mpc$^{-3}$]} & 
\multicolumn{1}{l}{} \\
\hline
\\[-5pt]
Pozzetti \etal (2010)           & 0.1-0.35   & $8.58$ & black squares \\
                                & 0.35-0.55  & $8.49$ & \\
                                & 0.55-0.75  & $8.50$ & \\
                               & 0.75-1.00  & $8.42$  & \\

Kajisawa \etal (2009)           & 0.5-1.0 & $8.63$ & green squares \\
                                & 1.0-1.5 & $8.30$ & \\
                                & 1.5-2.5 & $8.04$ & \\
                                & 2.5-3.5 & $7.74$ & \\

Marchesini \etal (2009)         & 1.3-2.0 & $8.11^{+0.02}_{-0.02}$ & blue dots \\
                                & 2.0-3.0 & $7.75^{+0.05}_{-0.04}$ & \\
                                & 3.0-4.0 & $7.47^{+0.37}_{-0.13}$ & \\

Reddy \etal (2012)              & 1.9-2.7 & $8.10^{+0.03}_{-0.03}$ & dark green triangles \\
                                & 2.7-3.4 & $7.87^{+0.03}_{-0.03}$ & \\

Caputi \etal (2011)             & 3.0-3.5  & $7.32^{+0.04}_{-0.02}$ & brown pentagons \\
                                & 3.5-4.25 & $7.05^{+0.11}_{-0.10}$ & \\
                                & 4.25-5.0 & $6.37^{+0.14}_{-0.54}$ & \\

Gonz\'{a}lez \etal (2011)$^c$     & 3.8 & $7.24^{+0.06}_{-0.06}$ & orange dots \\
                                  & 5.0 & $6.87^{+0.08}_{-0.09}$ & \\
                                  & 5.9 & $6.79^{+0.09}_{-0.09}$ & \\
                                  & 6.8 & $6.46^{+0.14}_{-0.17}$ & \\

Lee \etal (2012)$^c$              & 3.7 & $7.30^{+0.07}_{-0.09}$ & gray squares \\
                                  & 5.0 & $6.75^{+0.33}_{-0.16}$ & \\

Yabe \etal (2009)                 & 5.0 & $7.19^{+0.19}_{-0.35}$ & small black pentagon\\
Labb\'{e} \etal (2013)                & 8.0 & $5.78^{+0.22}_{-0.30}$ &  red square \\

\hline
\end{tabular*}
\begin{tablenotes}
\item[a] All the stellar mass densities have been derived assuming a Salpeter initial mass function.
\item[b] Stellar mass densities were computed by averaging over the four fields studied by Bielby \etal (2012).
\item[c] Following Stark \etal (2013), the mass densities of Gonz\'{a}lez \etal (2011) and Lee \etal (2012) at $z\simeq 4, 5, 6,$ and 7
have been reduced by the factor 1.1, 1.3, 1.6, and 2.4, respectively, to account for contamination by nebular emission lines.
\end{tablenotes}
\end{threeparttable}
\end{table*}

\end{document}